\documentclass{aa}
\usepackage{tabularx}
\usepackage{amsmath,amssymb}
\usepackage{color}
\usepackage{xcolor}
\usepackage{multirow}
\usepackage{CJK}
\usepackage{multirow} 
\usepackage{color, colortbl}
\usepackage{array,ragged2e,longtable}
\usepackage{txfonts}
\usepackage{hyperref}

\begin{document}
\begin{CJK*}{UTF8}{gbsn}

\title{Ten Supernova-rise in Binary Driven Gamma-ray Bursts}

\author{
R.~Ruffini\inst{\ref{ICRANet}\ref{ICRA}\ref{INAF}\ref{Nizza}}
\and
C.~L.~Bianco\inst{\ref{ICRANet}\ref{ICRA}\ref{IAPS}\ref{Nizza}}
\and
Liang~Li (李亮)\inst{\ref{ICRANet}\ref{ICRA}\ref{Teramo}}
\and
M.~T.~Mirtorabi\inst{\ref{ICRANet}\ref{Iran}}
\and
R.~Moradi\inst{\ref{ICRANet}\ref{ICRA}\ref{Teramo}\ref{key}}
\and
F.~Rastegarnia\inst{\ref{ICRANet}\ref{Iran}}
\and
J.~A.~Rueda\inst{\ref{ICRANet}\ref{ICRA}\ref{ICRANet-Ferrara}\ref{Ferrara}\ref{IAPS}}
\and
S.~R.~Zhang (张书瑞)\inst{\ref{ICRANet}\ref{ICRANet-Ferrara}\ref{school}\ref{CAS}}
\and
Y.~Wang (王瑜)\inst{\ref{ICRA}\ref{ICRANet}\ref{Teramo}}
}

\institute{
ICRANet, Piazza della Repubblica 10, I-65122 Pescara, Italy\email{yu.wang@icranet.org, liang.li@icranet.org,  ruffini@icra.it}\label{ICRANet}
\and
ICRA, Dip. di Fisica, Sapienza Universit\`a  di Roma, Piazzale Aldo Moro 5, I-00185 Roma, Italy\label{ICRA}
\and
INAF, Viale del Parco Mellini 84, 00136 Rome, Italy\label{INAF}
\and
Universit\'e de Nice Sophia-Antipolis, Grand Ch\^ateau Parc Valrose, Nice, CEDEX 2, France\label{Nizza}
\and
INAF, Istituto di Astrofisica e Planetologia Spaziali, Via Fosso del Cavaliere 100, 00133 Rome, Italy\label{IAPS}
\and
INAF, Osservatorio Astronomico d'Abruzzo, Via M. Maggini snc, I-64100, Teramo, Italy\label{Teramo}
\and
Department of Fundamental Physics, Faculty of Physics, Alzahra University, Tehran, Iran\label{Iran}
\and
Key Laboratory of Particle Astrophysics, Institute of High Energy Physics, Chinese Academy of Sciences, Beijing 100049, People's Republic of China\label{key}
\and
ICRANet-Ferrara, Dip. di Fisica e Scienze della Terra, Universit\`a degli Studi di Ferrara, Via Saragat 1, I--44122 Ferrara, Italy\label{ICRANet-Ferrara}
\and
Dip. di Fisica e Scienze della Terra, Universit\`a degli Studi di Ferrara, Via Saragat 1, I--44122 Ferrara, Italy\label{Ferrara}
\and
School of Astronomy and Space Science, University of Science and Technology of China, Hefei 230026, China\label{school}
\and
CAS Key Laboratory for Research in Galaxies and Cosmology, Department of Astronomy, University of Science and Technology of China, Hefei 230026, China\label{CAS}
}
 
\abstract{
The observation of a gamma-ray burst (GRB) associated with a supernova (SN) coincides remarkably with the energy output from a binary system comprising a very massive carbon-oxygen (CO) core and an associated binary neutron star (NS) by the Binary-Driven Hypernova (BdHN) model. The dragging effect in the late evolution of such systems leads to co-rotation, with binary periods on the order of minutes, resulting in a very fast rotating core and a binary NS companion at a distance of $\sim 10^5$ km. Such a fast-rotating CO core, stripped of its hydrogen and helium, undergoes gravitational collapse and, within a fraction of seconds, leads to a supernova (SN) and a newly born, fast-spinning neutron star ($\nu$NS), we name the emergence of the SN and the $\nu$NS as the SN-rise and $\nu$NS-rise. Typically, the SN energies range from $10^{51}$ to $10^{53}$ erg. We address this issue by examining 10 cases of Type-I BdHNe, the most energetic ones, in which SN accretion onto the companion NS leads to the formation of a black hole (BH). In all ten cases, the energetics of the SN events are estimated, ranging between $0.18$ and $12 \times 10^{52}$ erg. Additionally, in all 8 sources at redshift $z$ closer than $4.61$, a clear thermal blackbody component has been identified, with temperatures between $6.2$ and $39.99$ keV, as a possible signature of pair-driven SN. The triggering of the X-ray afterglow induced by the $\nu$NS-rise are identified in three cases at high redshift where early X-ray observations are achievable, benefits from the interplay of cosmological effects. 
}


\maketitle

%%%%%%%%%%%%%%%%%%%%%%%%%%%%%%%%%%%%%%%%%%%%%%%%%%%%%%%%%%%%%%%%%%%%
%%%%%%%%%%%%%%%%%%%%%%%%%%%%%%%%%%%%%%%%%%%%%%%%%%%%%%%%%%%%%%%%%%%%
\section{Introduction} \label{sec:intro}
%%%%%%%%%%%%%%%%%%%%%%%%%%%%%%%%%%%%%%%%%%%%%%%%%%%%%%%%%%%%%%%%%%%%
%%%%%%%%%%%%%%%%%%%%%%%%%%%%%%%%%%%%%%%%%%%%%%%%%%%%%%%%%%%%%%%%%%%%

Supernovae (SNe) and gamma-ray bursts (GRBs) are among the most violent stellar events in the universe due to their luminosity and energy. These cataclysmic phenomena not only reshape their immediate cosmic neighborhood but also offer vital clues about the life cycle of stars, the interstellar medium, and the dynamics of galaxy evolution. A particularly intriguing subclass of these events involves the simultaneous occurrence of a Type Ic supernova (SN Ic) and a long-duration gamma-ray burst \citep{2023ApJ...955...93A, 2006ARA&A..44..507W}, a scenario that challenges our understanding of stellar evolution and explosion mechanisms.

Historically, simultaneous observation of SN Ic and GRBs has been rare, in total around 30 events up to now, with the first case being GRB 980425 / SN 1998bw \citep{1998Natur.395..670G}. These events have prompted intense observational campaigns and theoretical efforts to understand the connection between supernovae and GRBs. The BdHN model, in particular, provides a framework to interpret these observations, suggesting that the co-evolution of a CO core and a neutron star (NS) in a close binary system can lead to such dual explosions \citep{2012ApJ...758L...7R, 2014ApJ...793L..36F, 2015ApJ...812..100B, 2015PhRvL.115w1102F, 2016ApJ...833..107B, 2017PhRvD..96b4046C, 2018ApJ...852..120B, 2019ApJ...871...14B,2021IJMPD..3030007R}.

The BdHN process, which includes the formation of a Type Ic supernova (SN Ic) and the associated gamma-ray burst (GRB), is initiated by the gravitational collapse of the carbon-oxygen (CO) core of a massive star. This event's early detection, specifically the first emergence of the supernova linked to the CO core collapse (termed as the SN–rise), is referred to as Episode I.

This initial episode possesses a lower luminosity, ranging from $10^{51}$ to $10^{52} \text{~erg s}^{-1}$ , compared to subsequent episodes. It precedes the remaining episodes by a time interval varying from a few seconds to around 100 seconds. The specific characteristics of this event depend on numerous factors, such as the GRB's energy, the distance to the source, and notably the functionality of the multi-wavelength detectors at the unpredictable moment when the gravitational collapse occurs.

In the previous paper \citep{2021MNRAS.504.5301R}, we suggested that this episode one signal might have been present in three specific GRBs (GRB 160625B , GRB 221009A and GRB 220101A). In this article, we extend our study to cover 10 sources, here indicated by their time of appearance, and perform a comprehensive spectral analysis to further confirm and investigate this phenomenon.

The article is structured as follows: Section 2 presents a  review of the physical background underlying the BdHN model, including three types of BdHNe and the seven episodes of BdHN. Section 3 describes the observational data and the analysis of each GRB, we focus on the thermal component found in the SN-rise. Section 4 summarizes this study and outlines potential directions for future research.

%%%%%%%%%%%%%%%%%%%%%%%%%%%%%%%%%%%%%%%%%%%%%%%%%%%%%%%%%%%%%%%%%%%%
%%%%%%%%%%%%%%%%%%%%%%%%%%%%%%%%%%%%%%%%%%%%%%%%%%%%%%%%%%%%%%%%%%%%
\section{Binary Driven Hypernova}  \label{sec:bdhn}
%%%%%%%%%%%%%%%%%%%%%%%%%%%%%%%%%%%%%%%%%%%%%%%%%%%%%%%%%%%%%%%%%%%%
%%%%%%%%%%%%%%%%%%%%%%%%%%%%%%%%%%%%%%%%%%%%%%%%%%%%%%%%%%%%%%%%%%%%

The BdHN suggests that GRBs originate from a binary system comprising a carbon-oxygen star and a neutron star. When the carbon-oxygen core collapses, it triggers a hypernova, resulting in a fast-spinning new neutron star at the center. Depending on the orbital separation of the binary system, different types of BdHN events occur, characterized by varying energy outputs and mechanisms.

\begin{enumerate}
    \item \textbf{BdHN I}: Occurs in systems with very short orbital periods (approximately 4-5 minutes) and involves extremely high energies ranging from $10^{52}$ to $10^{54}$ ergs. The high energy is due to the accretion of supernova ejecta onto a companion NS, leading to the formation of a BH. BdHN I is typically associated with hypernova (HN) with energy around $10^{52}$ ergs \citep{2015ApJ...798...10R}.

    \item \textbf{BdHN II}: Characterized by longer orbital periods (about 20 minutes) and lower energy outputs ranging from $10^{50}$ to $10^{52}$ ergs. In these systems, the NS does not undergo collapse into a black hole, given the comparatively slower accretion rates. This type still leads to significant energetic outputs but at a scale less than that of BdHN I \citep{2021IJMPD..3030007R}.

    \item \textbf{BdHN III}: Involves even longer orbital periods, up to several hours, and the lowest energy range, below $10^{50}$ ergs. The accretion rate is minimal, preventing any significant alteration to the neutron star, and likely does not lead to black hole formation. BdHN III events typically occur in systems where the supernova explosion disrupts the binary, with the energy mainly contributed by interactions between the supernova ejecta and the neutron star \citep{2022ApJ...936..190W}.
\end{enumerate}

The BdHN model outlines seven distinct emission episodes. These episodes cover a range of phenomena from the initial SN-rise to later afterglow emissions that follow the major burst even. Each type of BdHN (I, II, III) exhibits a subset of these seven episodes. Seven episodes include:
\begin{enumerate}
    \item \textbf{SN-rise (Episode I)}: This episode involves the gravitational collapse of the Carbon-Oxygen core, leading to a SN explosion and the formation of a new neutron star (νNS) \citep{2021MNRAS.504.5301R}.

    \item \textbf{νNS-rise and SN Ejecta Accretion (Episode II)}: Following the SN, the supernova ejecta begins to accrete onto the newly formed νNS and the existing neutron star \citep{2022ApJ...936..190W}.

    \item \textbf{Ultra-high-energy Prompt Emission (UPE) Phase - BH Overcritical (Episode III)}: Hypercritical accretion can cause the existing neutron star in the binary system to accumulate sufficient mass to form a BH. The formation of the BH triggers ultra-relativistic prompt emission, due to the overcritical field near the BH \citep{2015ApJ...798...10R,2021PhRvD.104f3043M}.

    \item \textbf{BH GeV Emission - Undercritical (Episode IV)}: After the UPE phase, the environment near the newly formed BH transitions to a state where high-energy GeV emissions are observable, driven by synchrotron radiation from charged particles accelerated in the undercritical magnetic fields near the BH \citep{2021MNRAS.504.5301R}.

    \item \textbf{BH Echoes (Episode V)}: Following the BH formation, the environment stabilizes somewhat, allowing for observable emissions known as ``BH echoes'' which are interactions of the emitted radiation with surrounding matter \citep{2019ApJ...883..191R}.

    \item \textbf{Multiwavelength Afterglow (Episode VI)}: This episode involves the extended emission of X-rays, optical, and radio waves as the ejected material from the supernova interacts with the interstellar medium \citep{2015ApJ...798...10R,2020ApJ...893..148R}. 

    \item \textbf{The optical SN Emission (Episode VII)}: this episode involves the optical emission from the supernova ejecta powered by the decay of nickel to cobalt \citet{2019ApJ...874...39W,2023ApJ...955...93A}. 
\end{enumerate}

A detailed analysis of 24 Type Ic SNe that are spectroscopically well-identified and associated with long GRBs are analysed in \citet{2023ApJ...955...93A}. The SNe display consistent peak luminosities and timing relative to the onset of their associated GRBs. This consistency occurs despite the wide range of energies and redshifts among the GRBs, suggesting a predictable underlying mechanism dictated by the dynamics of the binary systems in the BdHN model.

%%%%%%%%%%%%%%%%%%%%%%%%%%%%%%%%%%%%%%%%%%%%%%%%%%%%%%%%%%%%%%%%%%%%
%%%%%%%%%%%%%%%%%%%%%%%%%%%%%%%%%%%%%%%%%%%%%%%%%%%%%%%%%%%%%%%%%%%%
\section{Analysis of Supernova-rise}  \label{sec:sn-rise}
%\section{General Information of Observations}  \label{sec:general}
%%%%%%%%%%%%%%%%%%%%%%%%%%%%%%%%%%%%%%%%%%%%%%%%%%%%%%%%%%%%%%%%%%%%
%%%%%%%%%%%%%%%%%%%%%%%%%%%%%%%%%%%%%%%%%%%%%%%%%%%%%%%%%%%%%%%%%%%%

\subsection{GRB~090423}

GRB 090423 was detected on April 23, 2009, and it has been classified as one of the most distant cosmic explosions ever observed. With a redshift of z = 8.2, it happened when the universe was just about 630 million years old (or roughly $4\%$ of its current age). 

\begin{figure}
\centering
\includegraphics[width=\hsize,clip]{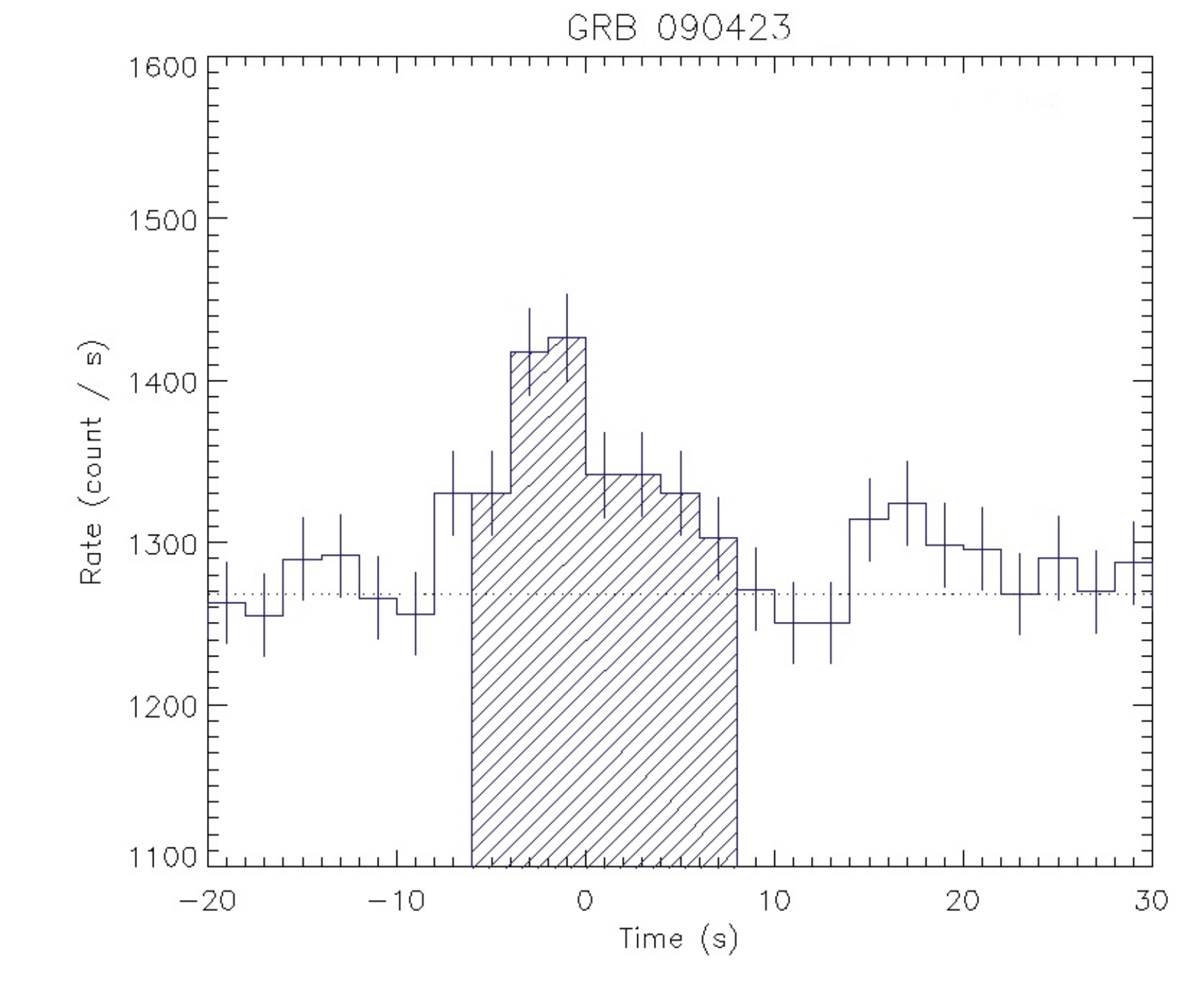} 
\includegraphics[width=\hsize,clip]{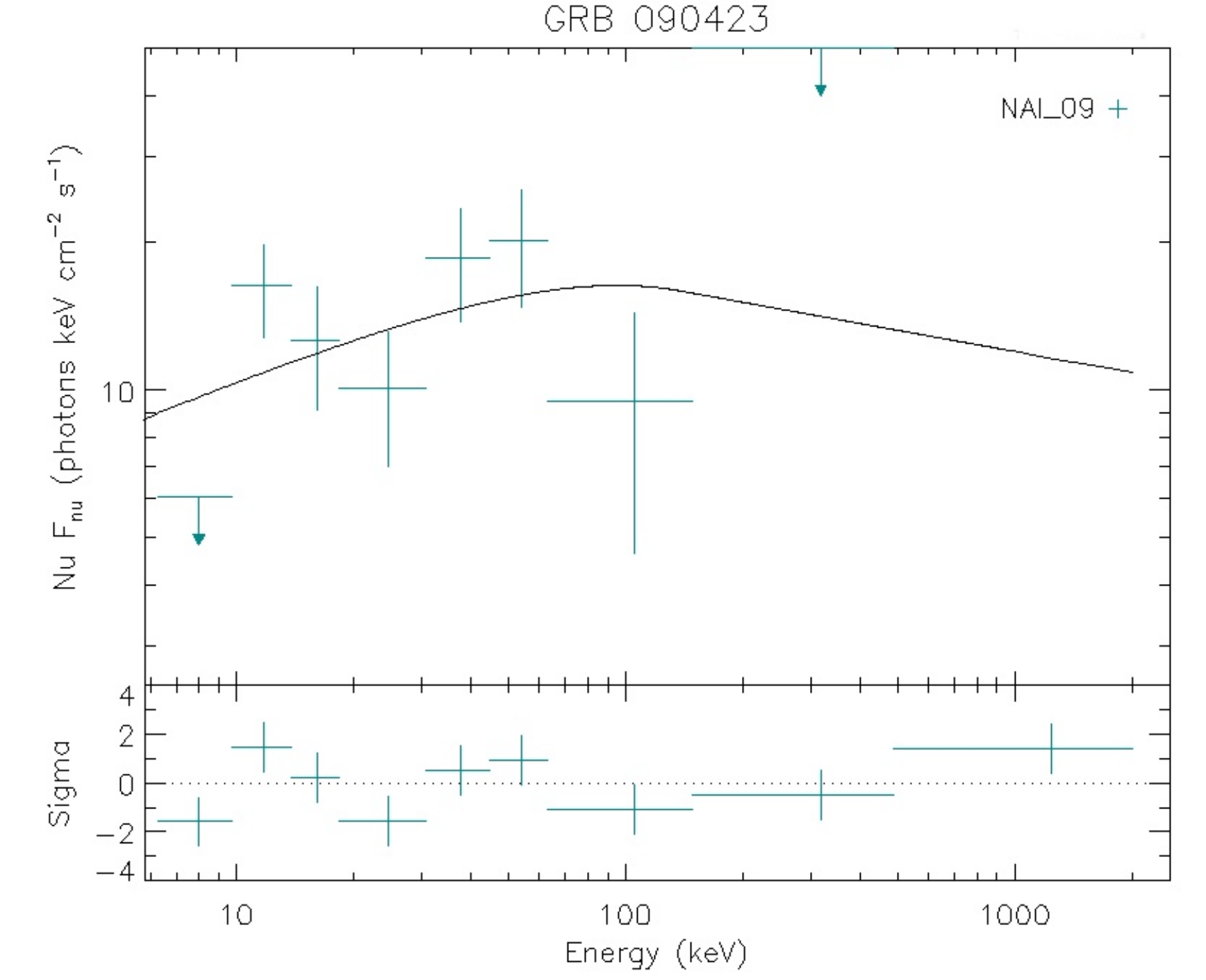} 
\caption{Top: GRB~090423 light-curve, shadow part indicates the period of episode 1. Bottom: the spectrum of episode 1, fitted by a cutoff power-law function.}\label{fig:090423-lc} 
\end{figure}

\begin{figure}
\centering
\includegraphics[width=\hsize,clip]{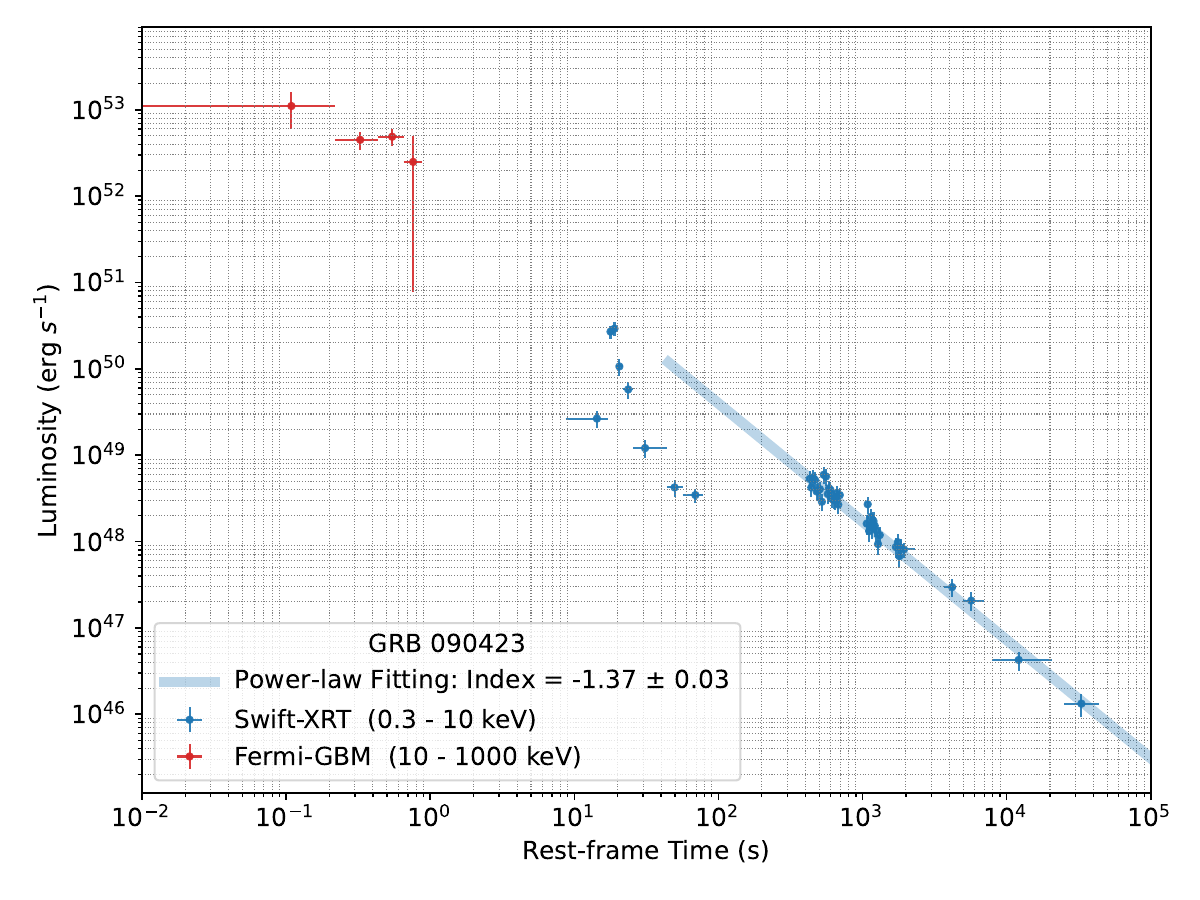}
\caption{Luminosity light-curve of GRB~090423, including the prompt emission and the afterglow, observed by Fermi-GBM and Swift-XRT, respectively.}\label{fig:090423-lc2} 
\end{figure}

Swift-BAT triggered on GRB~090423 at 07:55:19 UT. The event had a double-peaked structure with a duration of about 20 seconds and a peak count rate of ~2000 counts/sec in the 15-350 keV range (see top panel of Fig.~\ref{fig:090423-lc}). Swift's X-Ray Telescope (XRT) began observations 72.5 seconds post-trigger, identifying a fading X-ray source \citep{2009GCN..9198....1K} (see Fig.~\ref{fig:090423-lc2}). The \textit{Fermi} Gamma-Ray Burst Monitor (GBM) triggered on GRB~090423 as well, with a light curve showing a single structured peak with a duration $(T_{90})$ of about 12 seconds \citep{2009GCN..9229....1V}.

The initial discussion on the redshift of GRB~090423 was approximately $z\sim9$ based on early NIR observations and spectral analysis \citep{2009GCN..9209....1C,2009GCN..9213....1C}. However, more refined measurements and analyses have determined the redshift of GRB~090423 more accurately to be $z = 8.2$ through spectroscopic observations \citep{2009GCN..9219....1T,2009GCN..9322....1R}. This makes GRB~090423 one of the most distant cosmic explosions ever observed and a highly significant object for understanding the early universe. The isotropic equivalent energy $(E_{\text{iso}})$ in the $8-1000$ keV band was calculated as $(1.0 \pm 0.3)E+53$ ergs \citep{2009GCN..9229....1V,2009GCN..9251....1V}

Due to the high redshift, it fell outside the capacity of LAT for high energy observation and the optical telescopes for the confirmation of supernova. However, due to the high brightness of early X-rays, the Swift-XRT remains capable of detecting the radiation from high-redshift GRBs. Additionally, because the universe's expansion stretches the timescale of high-redshift GRBs in the observer's frame, Swift has sufficient time to reorient and capture the initial tens of seconds radiation measured in the rest-frame, see more examples in \citet{2023arXiv230605855B}. 

A first analysis of GRB~090423 within the BdHN model was presented in \citet{2014A&A...569A..39R}.

GRB 090423 exhibits its first SN-rise episode takes place from approximately $-5.5$ to $7.4$ seconds in the observed frame, which corresponds to $-0.6$ to $0.8$ seconds in the cosmological rest frame, lasting $1.4$ seconds. The spectrum is best fitted by a cutoff power-law function peaking at $ 80$~keV. The isotropic equivalent energy released during the supernova rise is $1.6 \times 10^{53}$ ergs. The black body temperature is not recognized due to its high redshift (see Fig.~\ref{fig:090423-lc}).

%The Swift satellite was the first to detect this burst. Swift's Burst Alert Telescope (BAT) triggered and located the source, enabling a swift reaction from other observatories around the world including UK Infrared Telescope (UKIRT), Gemini North Telescope, Very Large Telescope (VLT), Hubble Space Telescope (HST). Swift's X-Ray Telescope (XRT) then observed the fading X-ray afterglow of the burst and its location was further refined. The Fermi satellite also detected this burst. The GBM, which is sensitive to hard X-rays and gamma rays with energies ranging from 8 keV to 40 MeV, observed the burst. Unfortunately, due to the high redshift, it possibly fell outside the capacity of LAT.

\subsection{GRB~090429B}

GRB 090429B was detected on April 29, 2009, by the Swift satellite. The burst duration was short among the long GRBs, just about 5 seconds, yet it was extremely bright \citep{2009GCN..9281....1U}. Although the optical afterglow was faint, the Gemini-North telescope was able to capture its near-infrared spectrum \citep{2009GCN..9286....1C,2009GCN..9306....1L}. The GROND (Gamma-Ray burst Optical/Near-Infrared Detector) at the MPG/ESO 2.2-meter telescope in La Silla, Chile, carried out multicolor imaging \citep{2009GCN..9306....1L}. This burst was not detected by Fermi.

\begin{figure}
\centering
\includegraphics[width=\hsize,clip]{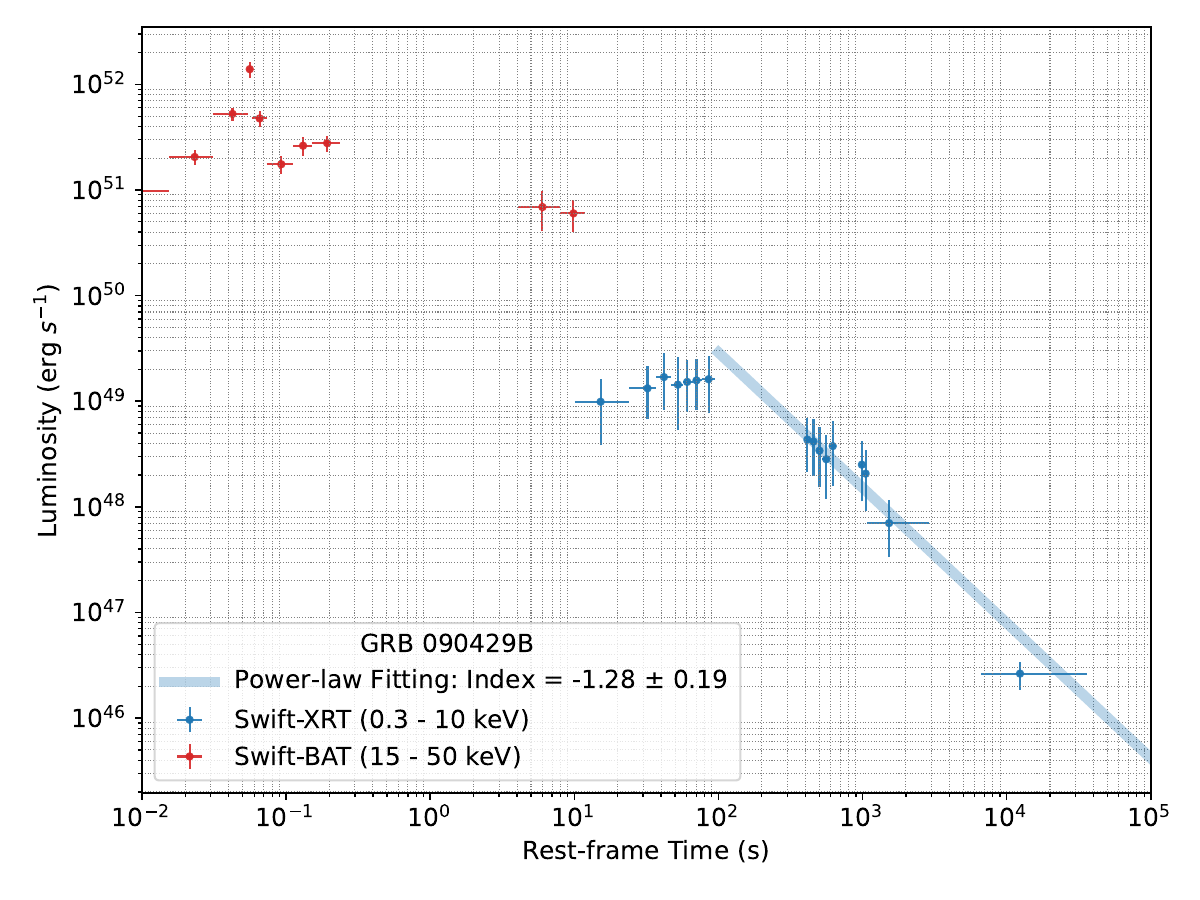}
\caption{Luminosity light-curve of GRB~090429B, including the prompt emission and the afterglow, observed by Swift-BAT and Swift-XRT, respectively.}\label{fig:090429B-lc} 
\end{figure}

Analysis of 8.9 ks of XRT data, covering from 104 s to 29.9 ks post the BAT trigger, reveals an initial light curve ascent fitted by a power-law index of $0.89^{+0.36}_{-0.46}$, transitioning at $T+642$ s to a decay phase with an index of $1.20^{+0.11}_{-0.10}$ (see Fig.~\ref{fig:090429B-lc}). Spectral fitting indicates an absorbed power-law photon index of $2.00^{+0.15}_{-0.24}$, with absorption exceeding Galactic levels \citep{2009GCN..9298....1R}.

The key discovery related to GRB~090429B was its estimated redshift. Due to the faintness of the afterglow and the lack of spectral features, it was impossible to determine the redshift directly through spectroscopic methods. Instead, the photometric redshift estimation is applied, a technique that uses the broad-band colors to estimate its redshift. The photometric data strongly suggested a high redshift for this burst, with an initial estimation of z = 9.4, which would place the GRB's origin just 520 million years after the Big Bang \citep{2011ApJ...736....7C}. This high redshift implies that GRB~090429B can provide valuable insights into the early universe, including the formation of the first stars and galaxies.

This GRB shows its SN-rise episode lasting $0.96$ seconds in the rest frame, spanning $0$ to $10$ seconds observed. It's spectrum is fitted by a cutoff powerlaw function and isotropic energy released is $3.5 \times 10^{52}$ ergs. The black body component has not been resolved due to its high redshift.

\subsection{GRB~090618}

\begin{figure}
\centering
\includegraphics[width=\hsize,clip]{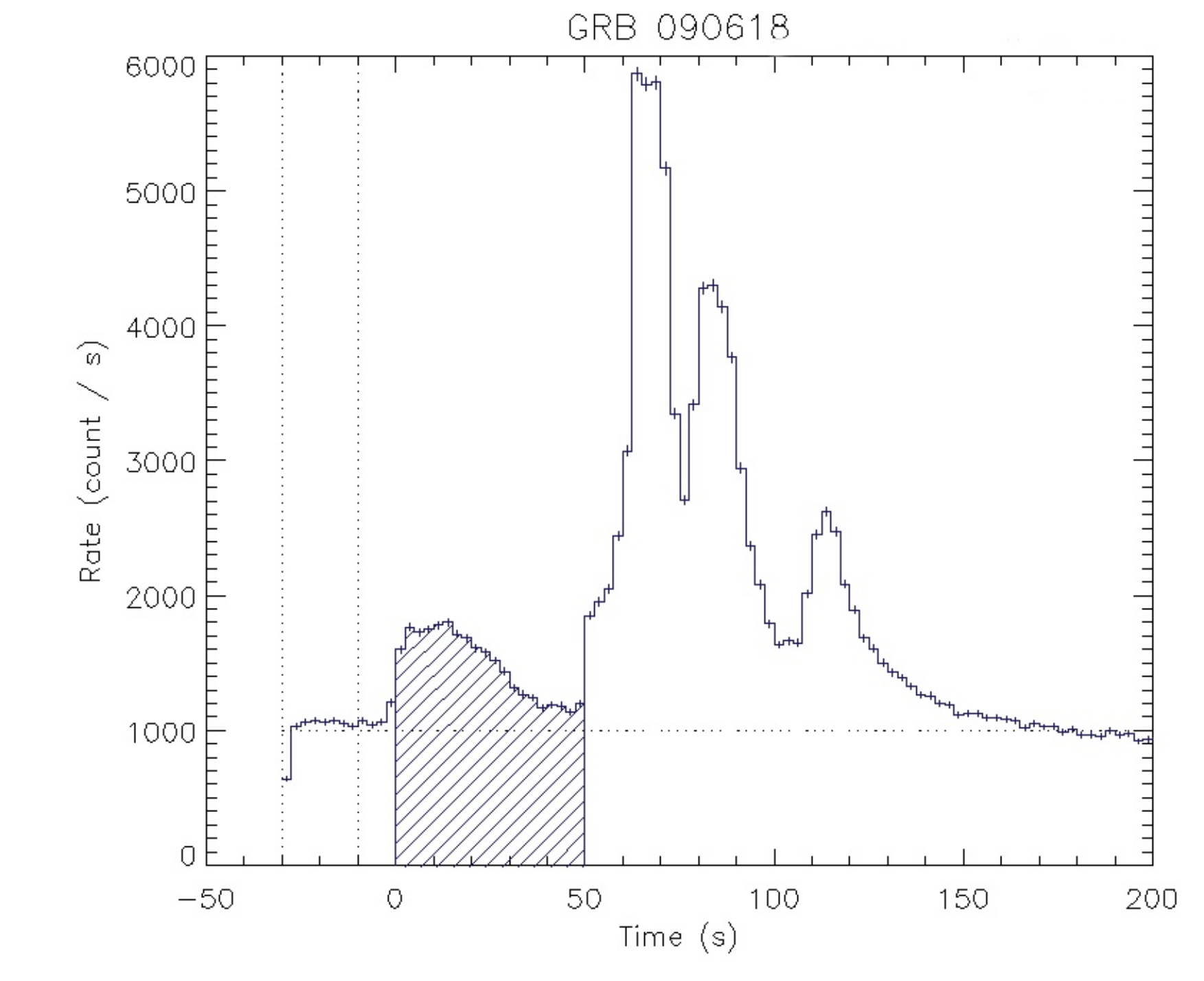} 
\includegraphics[width=\hsize,clip]{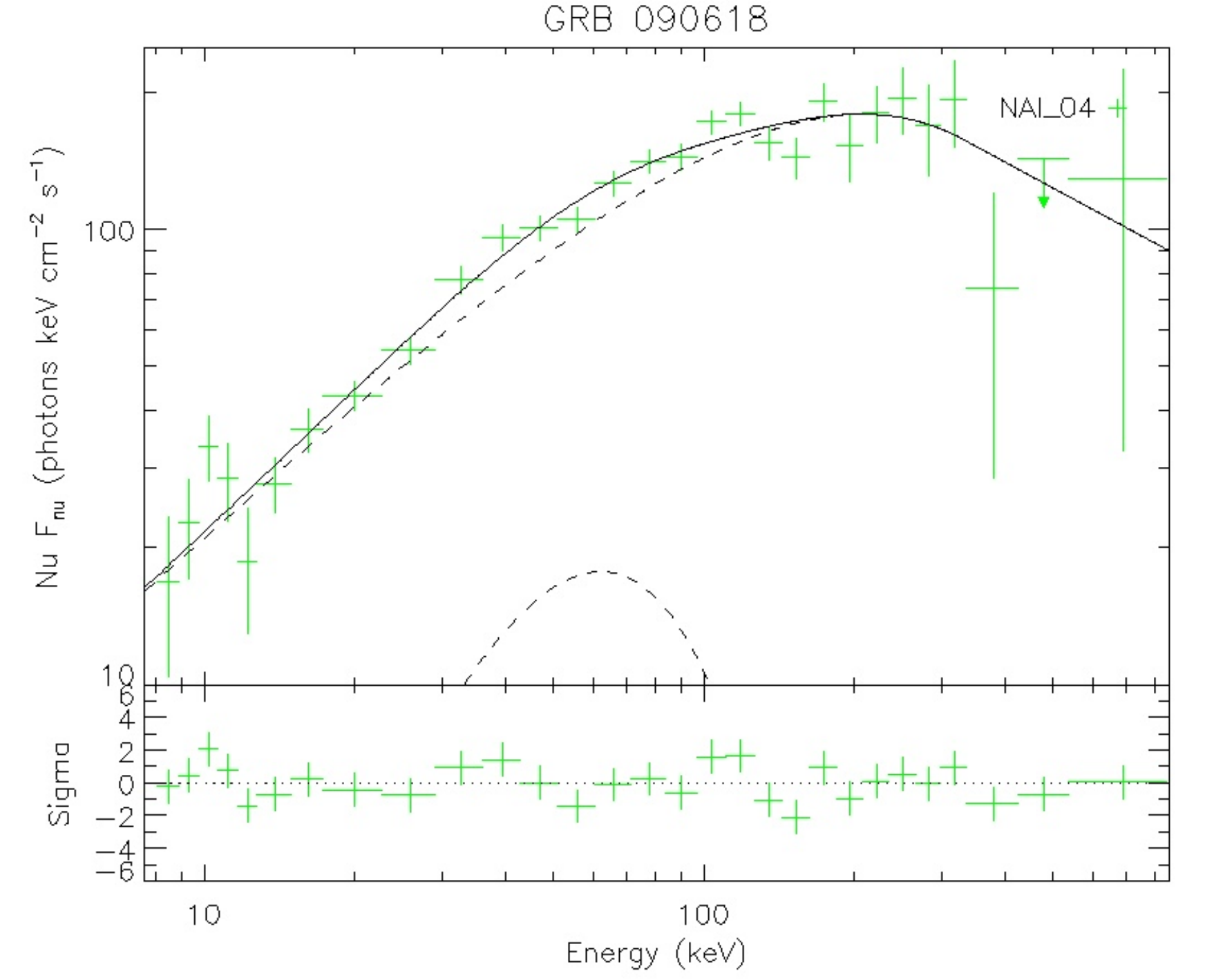}
\caption{Top: GRB~090618 light-curve, shadow part indicates the period of episode 1. Bottom: the spectrum of episode 1, fitted by the Band function plus a blackbody component, the observed temperature is 15.86 keV.}\label{fig:090618-lc} 
\end{figure}

\begin{figure}
\centering
\includegraphics[width=\hsize,clip]{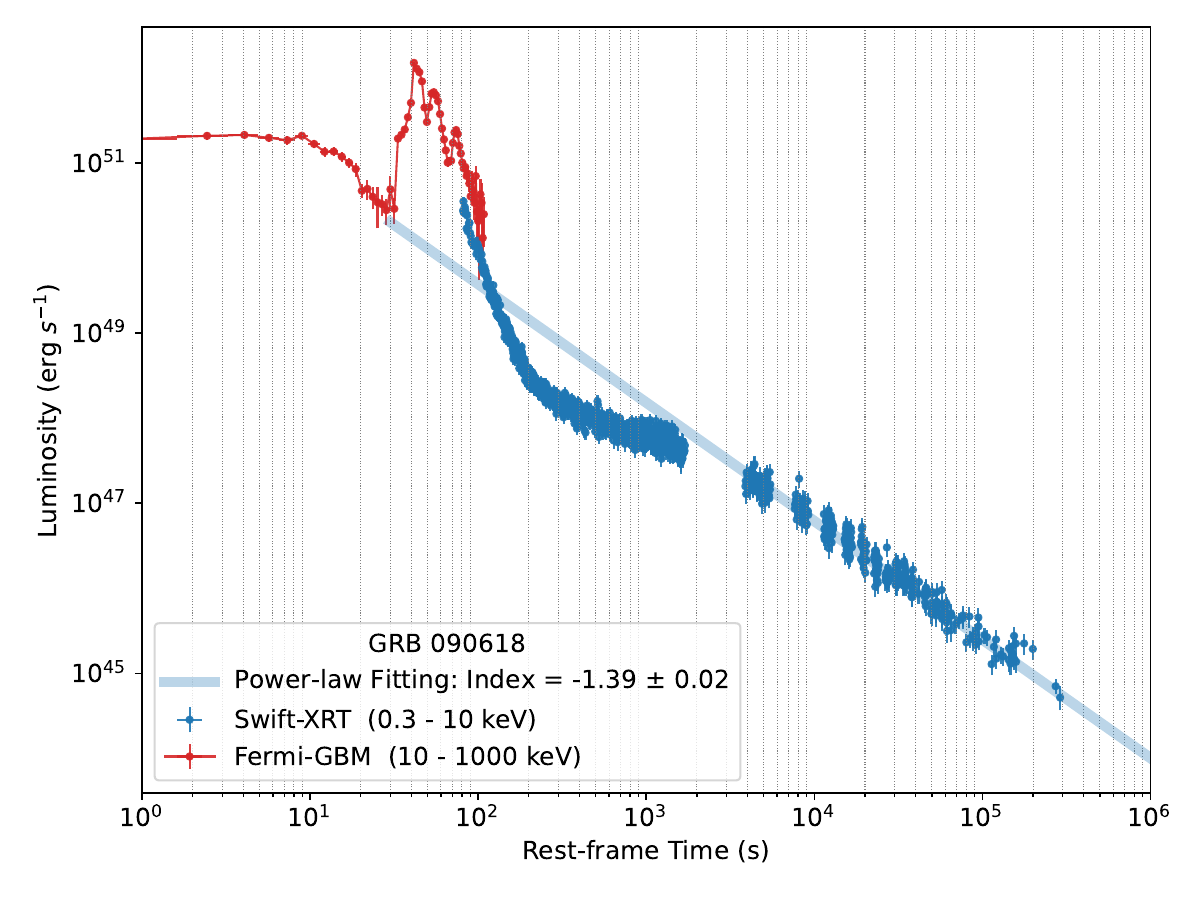}
\caption{Luminosity light-curve of GRB~090618, including the prompt emission and the afterglow, observed by Fermi-GBM and Swift-XRT, respectively.}\label{fig:090618-lc2} 
\end{figure}

GRB~090618 was initially detected by the Swift satellite, Swift-BAT observation shows the multi-peak structure of its light curve, indicating a duration of about 130 seconds with a peak count rate of $\sim 40000$ counts/s (15-350 keV) at 80 sec after the trigger (see top panel of Fig.~\ref{fig:090618-lc}). Swift-XRT observed this source at 120.9 seconds after the BAT trigger (see Fig. \ref{fig:090618-lc2}). Swift-UVOT observations identified the optical afterglow with an estimated magnitude of 14.36 in the white filter, 128 seconds post-trigger \citep{2009GCN..9512....1S}.

The time-averaged spectrum from \(T_0\) to \(T_0+140\) s from Fermi observation can be adequately fit by a Band function, with a peak energy \(E_{\text{peak}} = 155.5\) keV, a low-energy photon index \(\alpha = -1.26\), and a high-energy photon index \(\beta = -2.50\)  \citep{2009GCN..9535....1M} (see Fig.~\ref{fig:090618-lc2}). This suggests significant spectral evolution within the observed time interval. The isotropic energy (\(E_{\text{iso}}\)) estimated from the GBM data is \(2.0E+53\) erg in the 8-1000 keV band, positioning GRB~090618 as one of the energetically significant bursts observed by Fermi. The boresight angle of Fermi-LAT is 133 degrees, out of capacity of detecting GeV photons \citep{2009GCN..9535....1M}.

The AGILE satellite detected GRB~090618 with its MCAL instrument, providing a complementary spectral analysis: The total MCAL spectrum between \(T_0\) and \(T_0+120\) sec could be fit by a simple power law with a photon index of 3.16, in the 0.5-10 MeV energy range, highlighting the burst's brightness below a few MeV and the lack of significant gamma-ray emission above 30 MeV \citep{2009GCN..9524....1L}.

This burst has been extensively studied within the context of the BdHN model, as elucidated by \citet{2012A&A...543A..10I,2012A&A...548L...5I,2013IJMPS..23..202I}. Their analysis divides the GRB into two distinctive episodes: an initial spike due to the supernova explosion and neutron star formation, a later phase associated with the formation of the black hole.

This GRB's first episode spans $0$ to $47.7$ seconds in the observed frame, which translates to $0$ to $31$ seconds in the rest frame and lasts $31$ seconds, of which the spectrum is best fitted by the Band function plus a blackbody component, the integrated isotropic energy is $3.61 \times 10^{52}$ ergs, and the observed black body temperature is $15.86$ keV (see Fig.~\ref{fig:090618-lc}).

\subsection{GRB~130427A}

GRB~130427A, detected on April 27, 2013, was one of the most energetic and brightest GRBs ever observed as of 2021. It was initially detected by the Neil Gehrels Swift Gamma-Ray Burst Mission \citep{2013GCN.14448....1M}, Fermi Gamma-ray Space Telescope \citep{2013GCN.14473....1V}, and INTEGRAL (INTErnational Gamma-Ray Astrophysics Laboratory) \citep{2013GCN.14484....1P}, and subsequently observed across multiple wavelengths by numerous ground and space-based observatories around the world. Using the Gemini-North / GMOS telescope, the observed high-quality spectra revealed absorption lines for Ca H and K, Mg I, and the Mg II doublet at a redshift of z=0.34 \citep{2013GCN.14455....1L}.

\begin{figure}
\centering
\includegraphics[width=\hsize,clip]{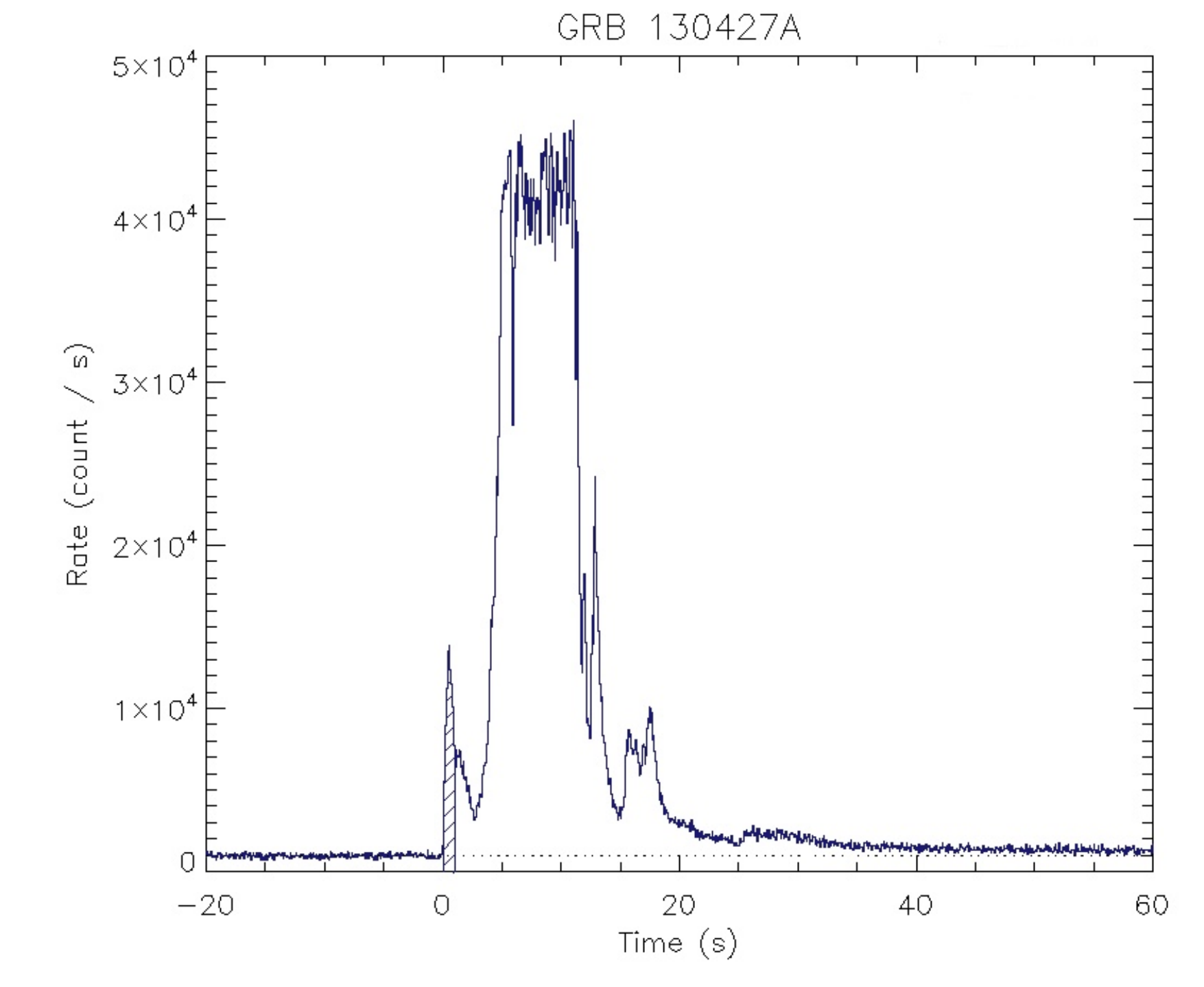} 
\includegraphics[width=\hsize,clip]{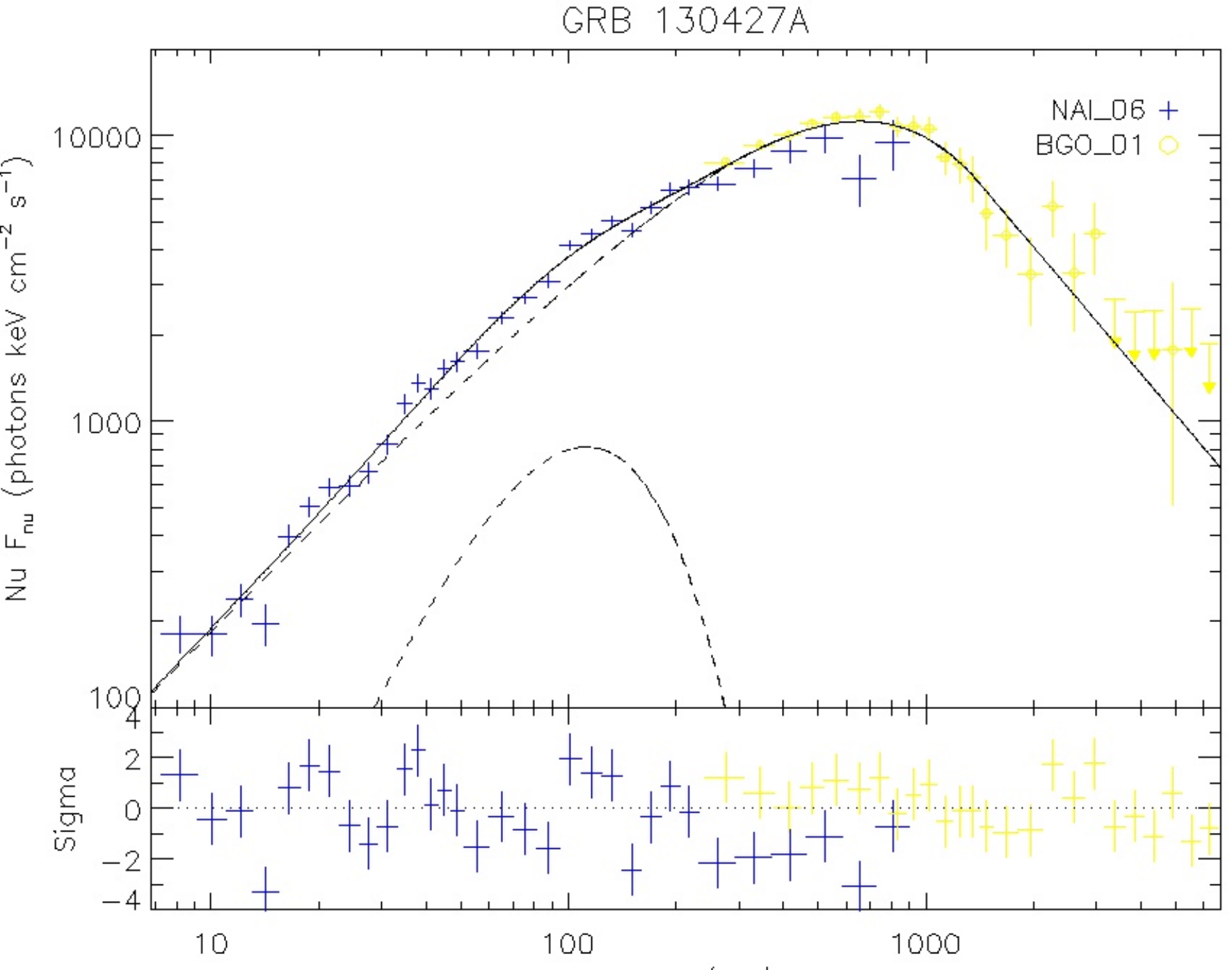}
\caption{Top: GRB~130427A light-curve, shadow part indicates the period of episode 1. Bottom: the spectrum of episode 1, fitted by the Band function plus a blackbody component, the observed temperature is 28.66 keV.}\label{fig:130427A-lc} 
\end{figure}

\begin{figure}
\centering
\includegraphics[width=\hsize,clip]{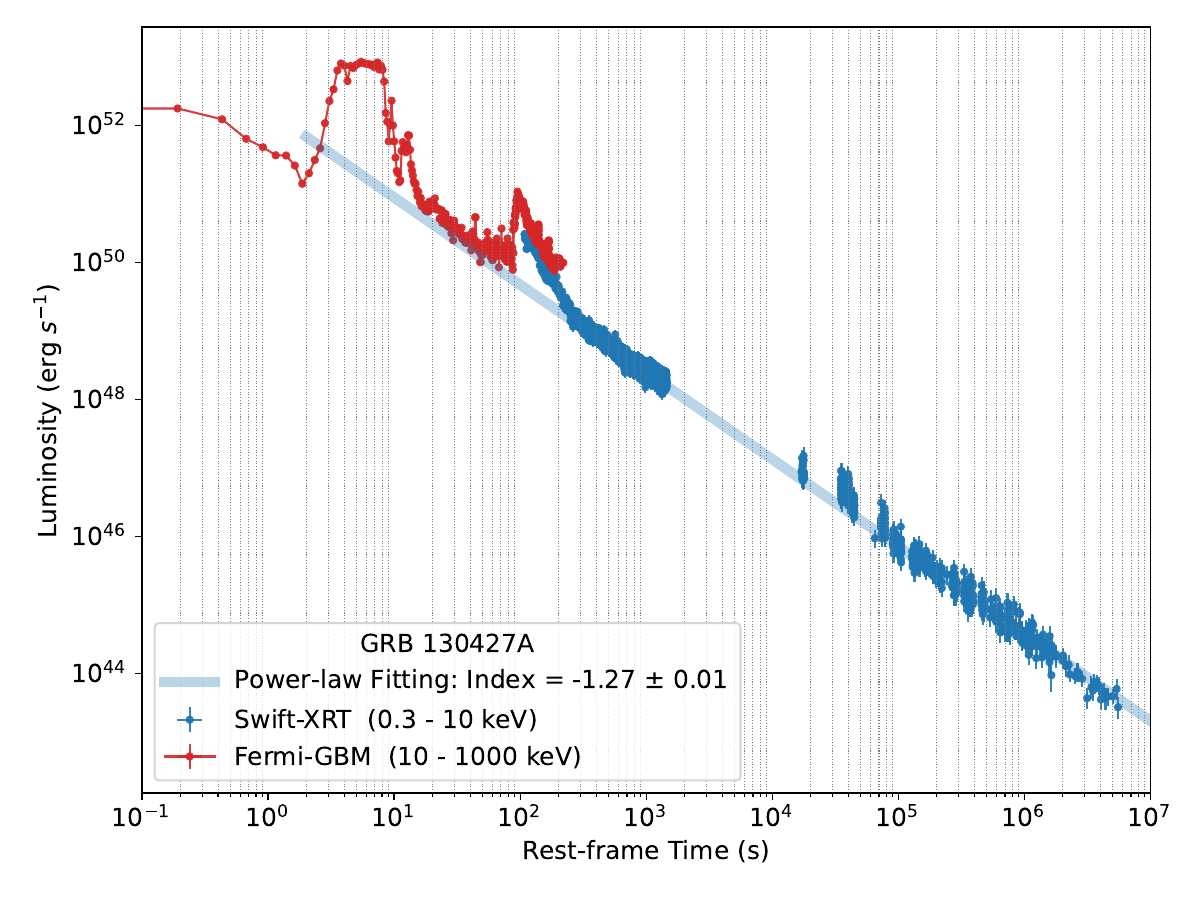}
\includegraphics[width=\hsize,clip]{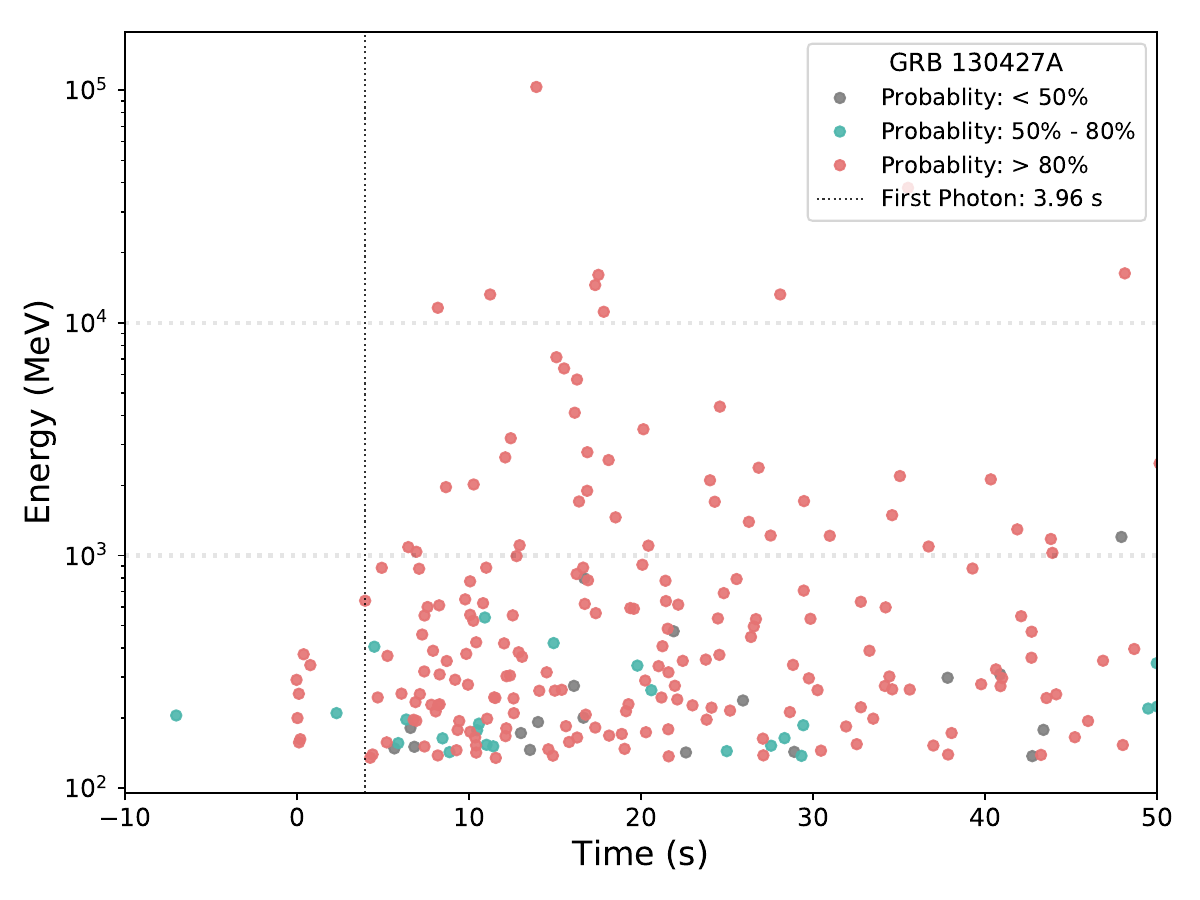} 
\caption{Top: Luminosity light-curve of GRB~130427A, including the prompt emission and the afterglow, observed by Fermi-GBM and Swift-XRT, respectively. Bottom: GeV photons observed by Fermi-LAT.}\label{fig:130427A-lc2} 
\end{figure}

The GBM light curve features a bright structured peak and a FRED-like (Fast Rise Exponential Decay) pulse approximately 120 seconds after the trigger. The overall duration ($T_{90}$) of the burst, covering the energy range 50-300 keV, is about 138 seconds (see top panel of Fig.~\ref{fig:130427A-lc} and top panel of Fig.~\ref{fig:130427A-lc2})). Spectral analysis using a Band function fit for the interval from $T_0+0.002$ s to $T_0+18.432$ s provided parameters of $E_{\text{peak}} = 830 \pm 5$ keV, $\alpha = -0.789 \pm 0.003$, and $\beta = -3.06 \pm 0.02$. However, due to the burst's brightness, systematic effects were significant, and no single model was found to adequately fit the data. The fluence in the 10-1000 keV range for the specified time interval is reported as $(1.975 \pm 0.003) \times 10^{-3}$ erg/cm$^2$, and the 1.024-second peak photon flux starting from $T_0+7.48808$ s in the 8-1000 keV band is $10^{52} \pm 2$ photons/s/cm$^2$. An isotropic-equivalent radiated energy of $(1.05 \pm 0.15) \times 10^{54}$ erg (1-10000 keV cosmological rest-frame) is obtained (see bottom panel of Fig.~\ref{fig:130427A-lc}). These measurements indicate that GRB~130427A is the most intense and fluent GRB detected by Fermi GBM up to that point \citep{2013GCN.14473....1V,2013GCN.14503....1A}.

At the time of the trigger, the burst was approximately 47 degrees from the LAT boresight but remained within the LAT's field of view for the subsequent 700 seconds. The Fermi LAT data revealed a multi-peaked light curve that aligns with the GBM trigger, with more than 200 photons above 100 MeV observed within the first 100 seconds, boasting a Test Statistic (TS) of greater than 1000. Remarkably, the highest energy photon recorded by the LAT for this burst had an energy of 94 GeV \citep{2013GCN.14471....1Z} (see bottom panel of Fig.~\ref{fig:130427A-lc2}).

These comprehensive observations across the electromagnetic spectrum, from radio to gamma-rays, have made GRB~130427A one of the best-studied GRBs, providing a wealth of data for theoretical models like the binary-driven hypernova (BdHN) model \citep[see, e.g.,][and references therein]{2015ApJ...798...10R,2019ApJ...874...39W,2020ApJ...893..148R,2023ApJ...945...10L}, to explain the complex processes during and after the explosion.

The SN-rise episode of this GRB lasts only $0.65$ seconds in the rest frame, occurring from $0$ to $0.9$ seconds observed, in which the spectrum is best fitted by a Band function plus a blackbody component. It releases a modest $0.65 \times 10^{52}$ ergs of energy, with a notably high black body temperature of $28.66$ keV (see Fig.~\ref{fig:130427A-lc}).

\subsection{GRB~160509A}

\begin{figure}
\centering
\includegraphics[width=\hsize,clip]{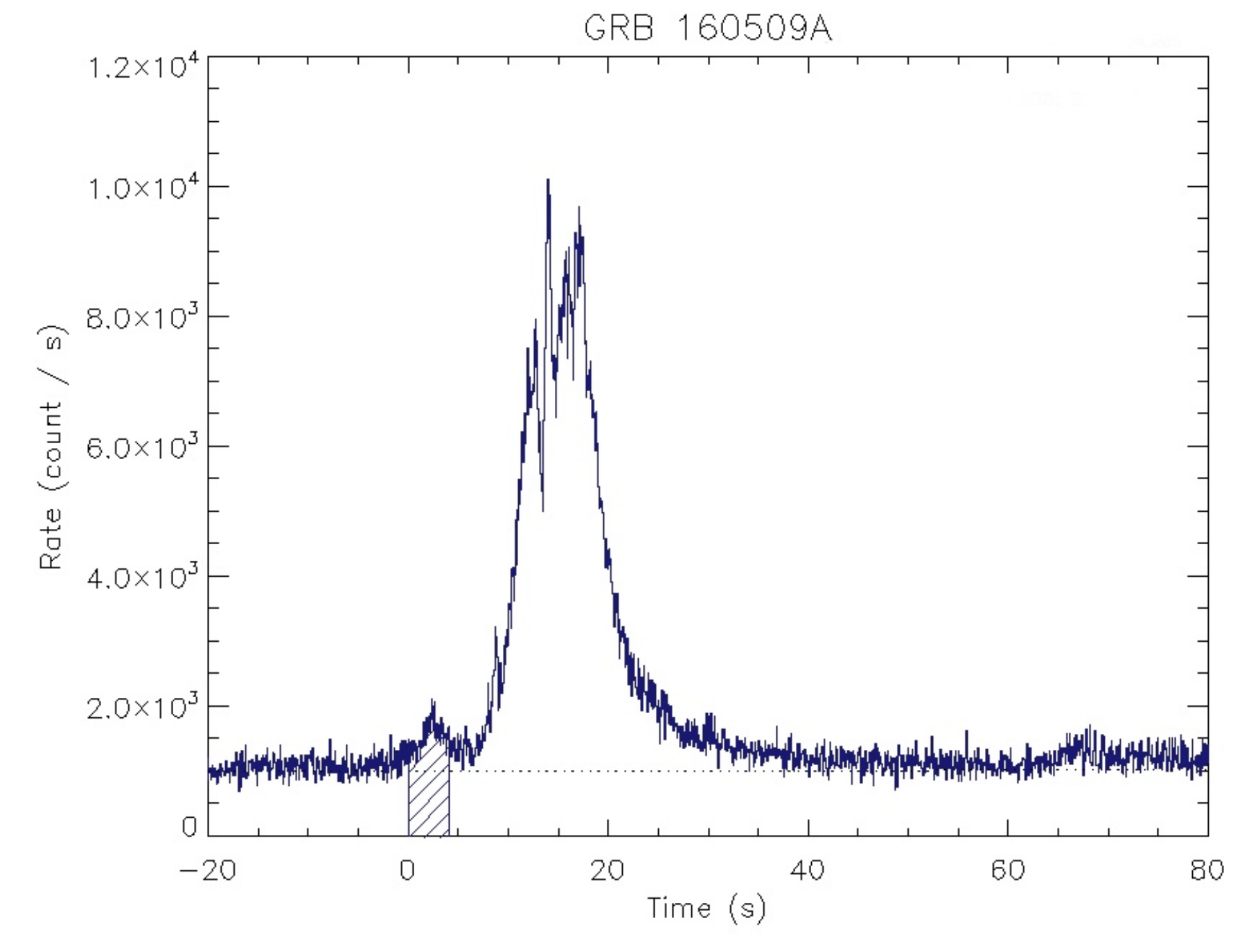} 
\includegraphics[width=\hsize,clip]{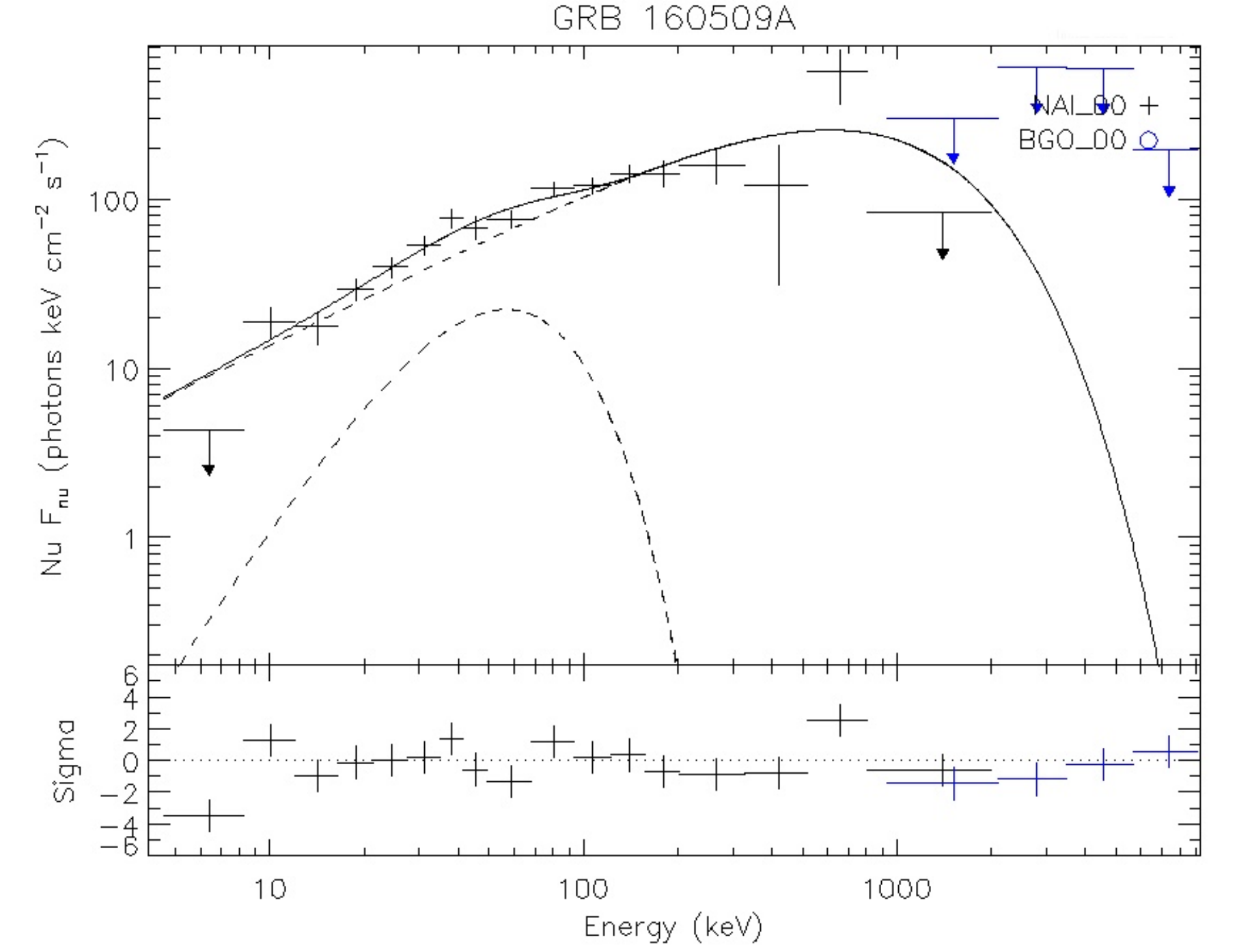}
\caption{Top: GRB~160509A light-curve, shadow part indicates the period of episode 1. Bottom: the spectrum of episode 1, fitted by the cutoff power-law plus a blackbody component, the observed temperature is 14.43 keV.}\label{fig:160509A-lc} 
\end{figure}

GRB~160509A is a long gamma-ray burst detected at 08:59:04.36 UTC on May 09, 2016 by Fermi-LAT \citep{2016GCN.19403....1L}. It also triggered Fermi-GBM \citep{2016GCN.19411....1R} and Swift \citep{2016GCN.19408....1K} (see top panel of Fig.~\ref{fig:160509A-lc}). It shows two gamma-ray pulses separated at about 3 second and has an afterglow lasting weeks. 

The Fermi GBM light curve displayed multiple peaks over a duration (T90) of about 371 seconds. The time-averaged spectrum of the initial bright 40~s was best fit by a Band function, with $E_{\text{peak}} = 370 \pm 7$ keV, $alpha = -0.89 \pm 0.01$, and $beta = -2.11 \pm 0.02$. The event's fluence (10-1000 keV) was $(1.51 \pm 0.01) \times 10^{-4} \text{ erg cm}^{-2}$, and the peak photon flux was $75.5 \pm 0.6 \text{s cm}^{-2}$ \citep{2016GCN.19411....1R}. 

\begin{figure}
\centering
\includegraphics[width=\hsize,clip]{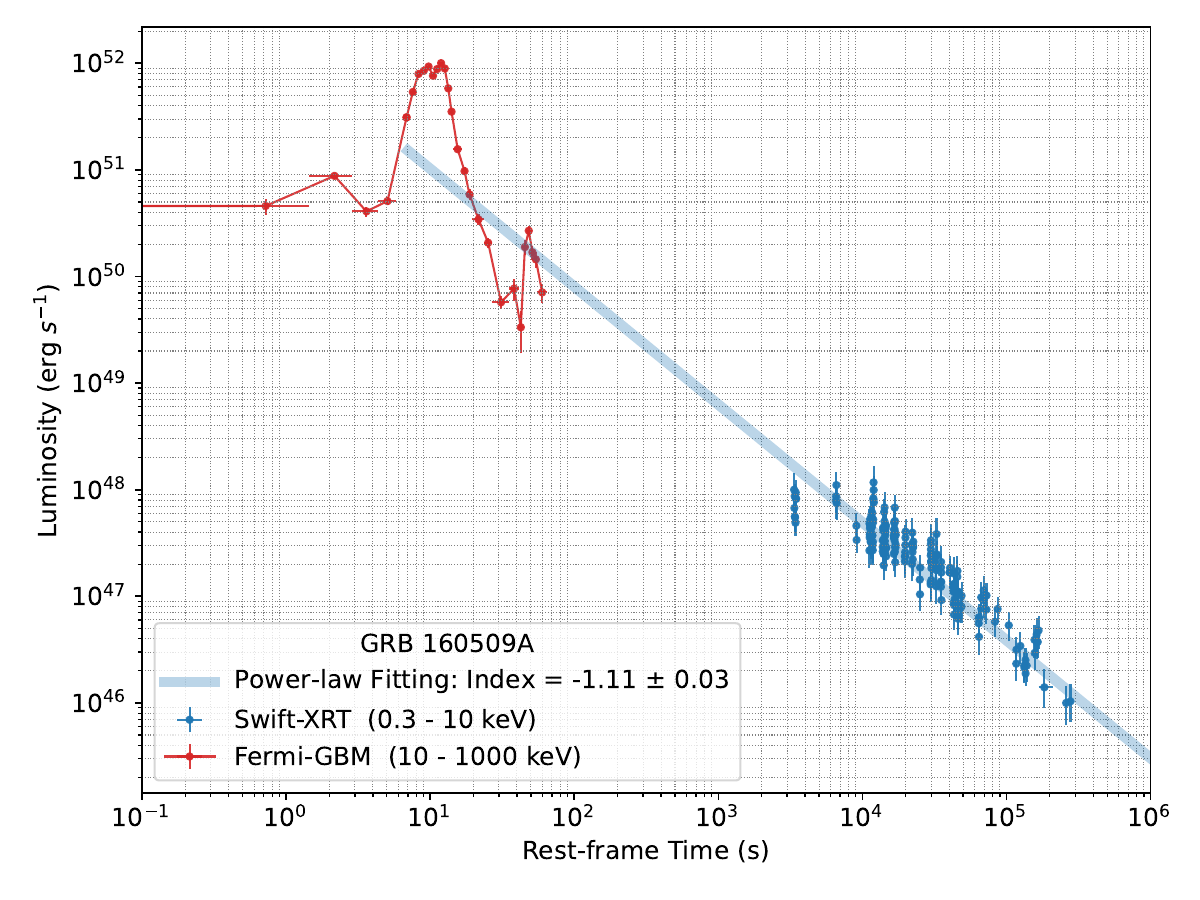}
\includegraphics[width=\hsize,clip]{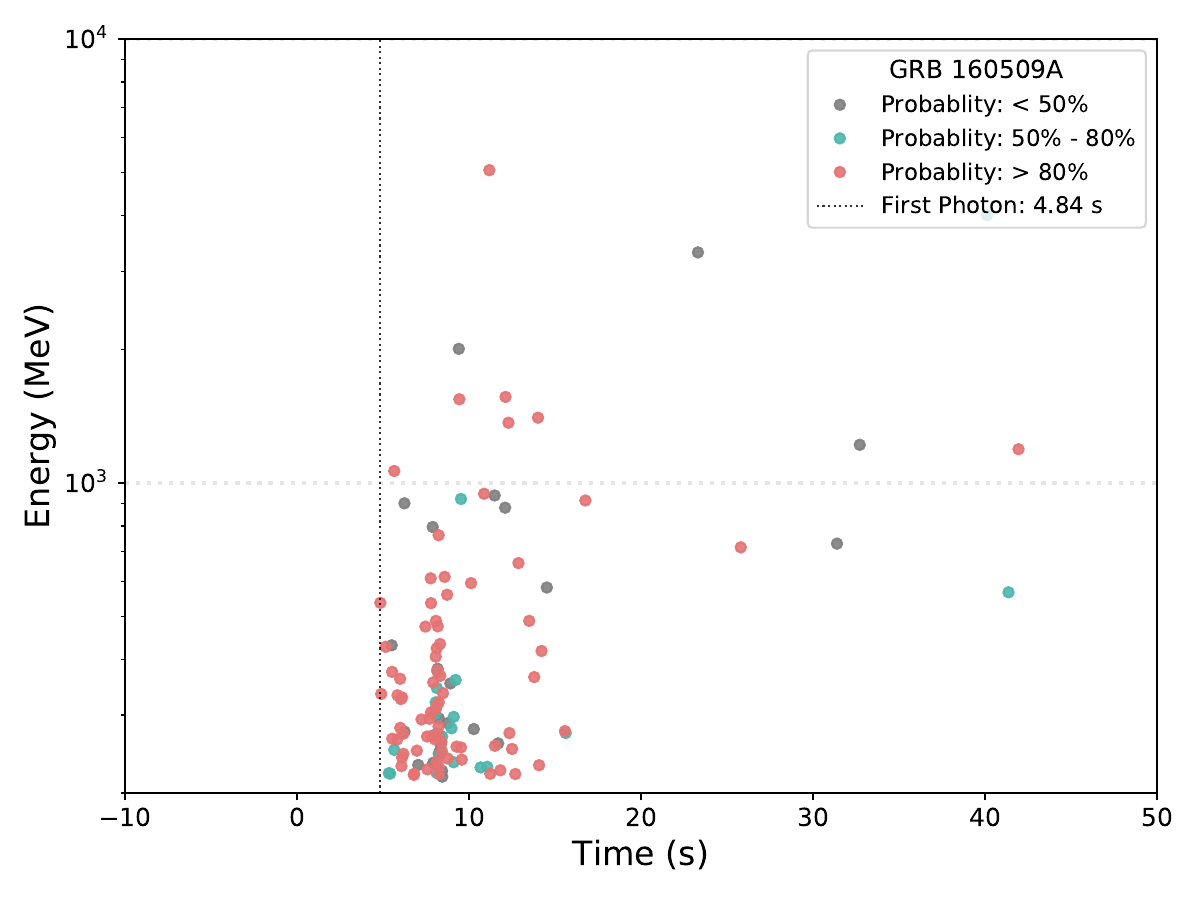} 
\caption{Top: Luminosity light-curve of GRB~160509A, including the prompt emission and the afterglow, observed by Fermi-GBM and Swift-XRT, respectively. Bottom: GeV photons observed by Fermi-LAT.}\label{fig:160509A-lc2} 
\end{figure}

Swift-XRT conducted follow-up observations on GRB~160509A in a series of 7 tiles totaling 1.7 ks of exposure time, with the longest single exposure being 560 seconds (see top panel of Fig.~\ref{fig:160509A-lc2}). The spectrum obtained from PC mode data fits an absorbed power-law model with a photon spectral index of $1.7 \pm 0.3$ and an absorption column density of $4.6^{+2.3}_{-1.8} \times 10^{21} \text{ cm}^{-2}$ \citep{2016GCN.19408....1K}.

Utilizing Fermi-LAT events of greater than 100 MeV from 0 until 2660 seconds after the trigger, refined the burst's localization and detected a 52 GeV photon, 77 seconds post-trigger (see bottom panel of Fig.~\ref{fig:160509A-lc2}). The high-energy emission was characterized by a soft power-law with an index of $-3.4 \pm 0.2$ during the main GBM emission episode, transitioning to a harder index of $-2.0 \pm 0.1$ afterwards \citep{2016GCN.19413....1L}.

The analysis of GRB~160509A within the BdHN model has been presented in \citet{2020ApJ...893..148R,2023ApJ...945...10L}.

The first episode of this burst is observed from $0$ to $4.0$ seconds , translating to $0$ to $1.84$ seconds in the rest frame, with a duration of $1.84$ seconds. The spectrum of this episode is fitted by a cutoff power-law plus a thermal component. It releases $1.47 \times 10^{52}$ ergs of energy, and the observed black body temperature is $14.43$ keV (see Fig.~\ref{fig:160509A-lc}).

\subsection{GRB~160625B}

\begin{figure}
\centering
\includegraphics[width=\hsize,clip]{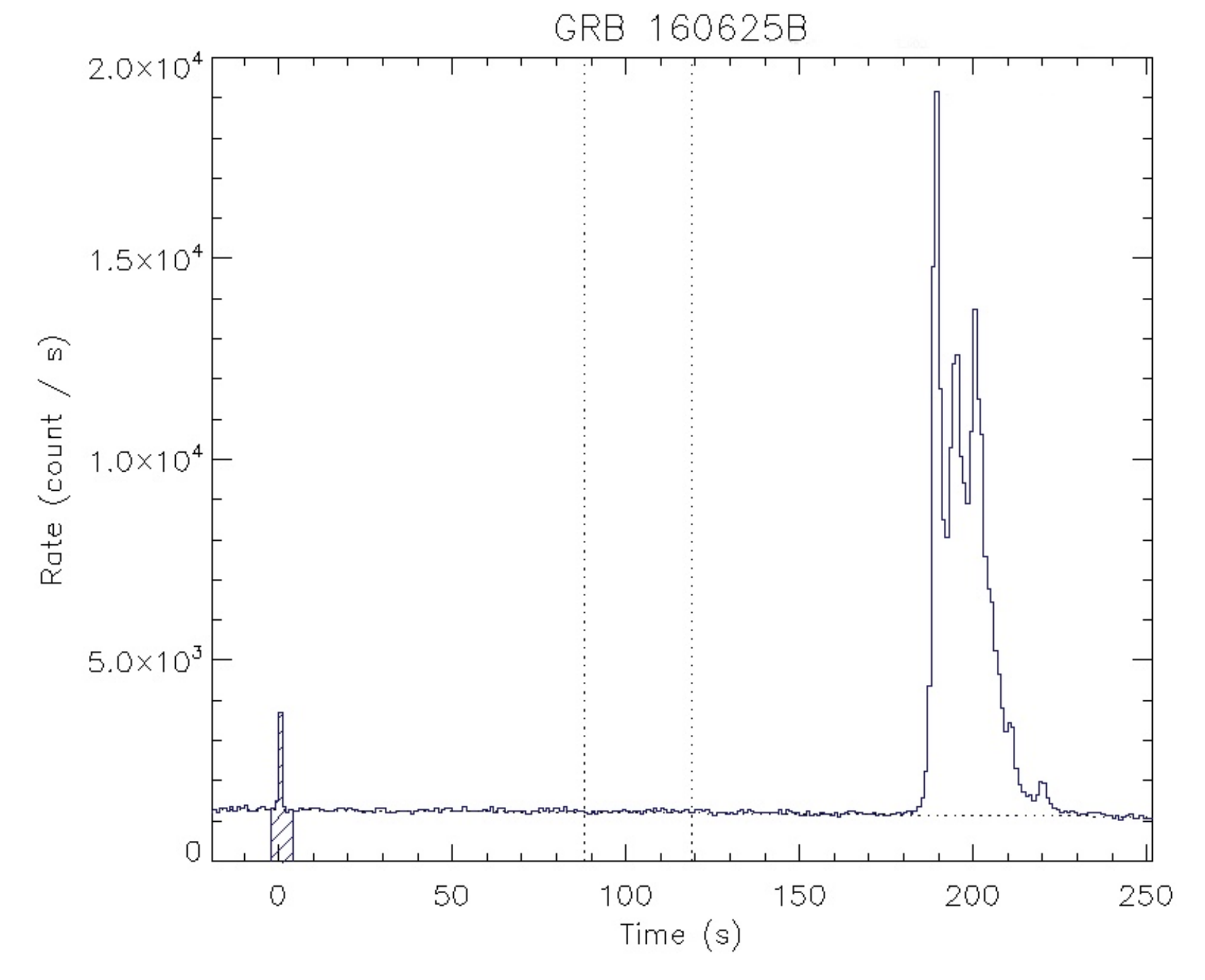} 
\includegraphics[width=\hsize,clip]{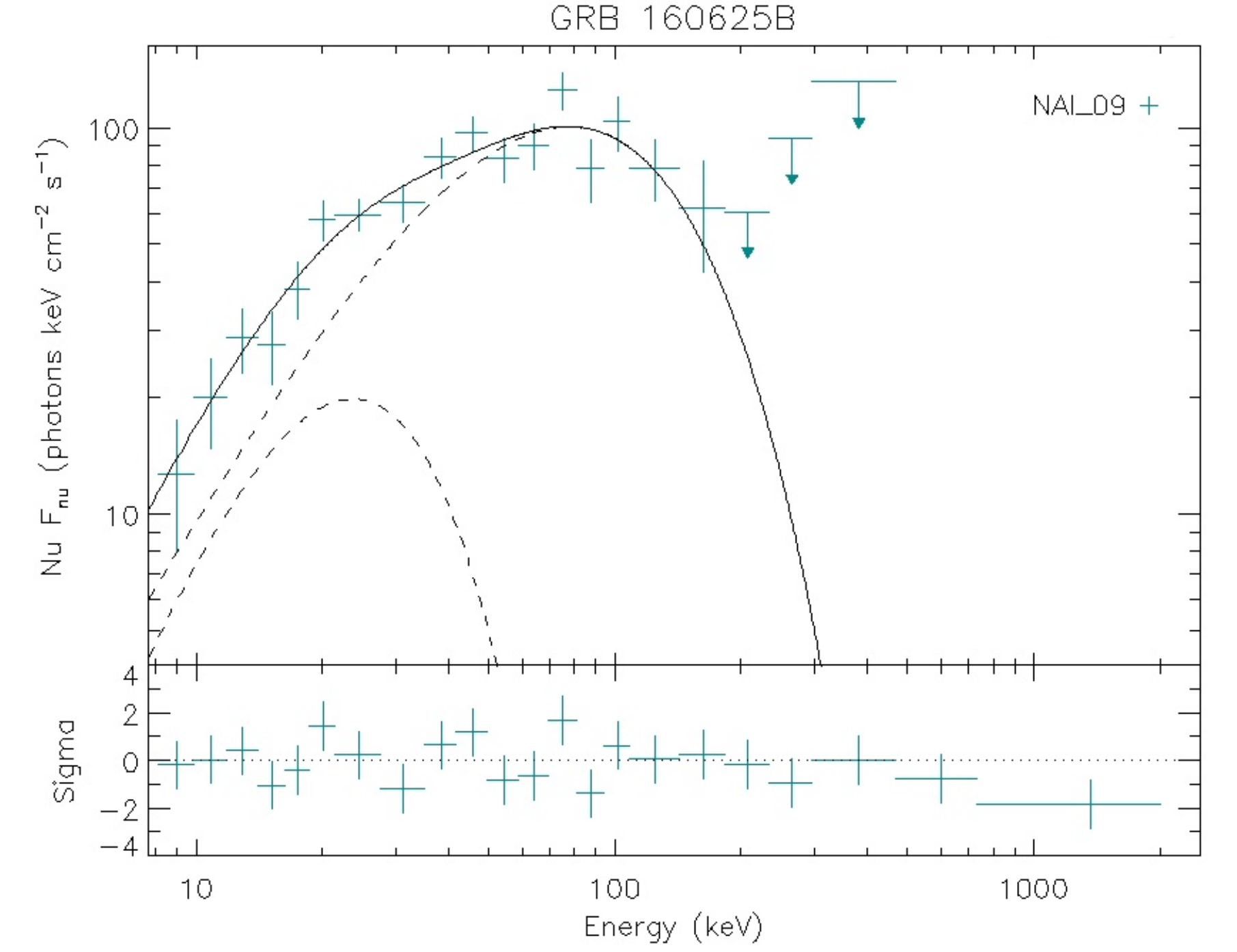}
\caption{Top: GRB~160625B light-curve, shadow part indicates the period of episode 1. Bottom: the spectrum of episode 1, fitted by the cutoff power-law plus a blackbody component, the observed temperature is 6.025 keV.}\label{fig:160625B-lc} 
\end{figure}

\begin{figure}
\centering
\includegraphics[width=\hsize,clip]{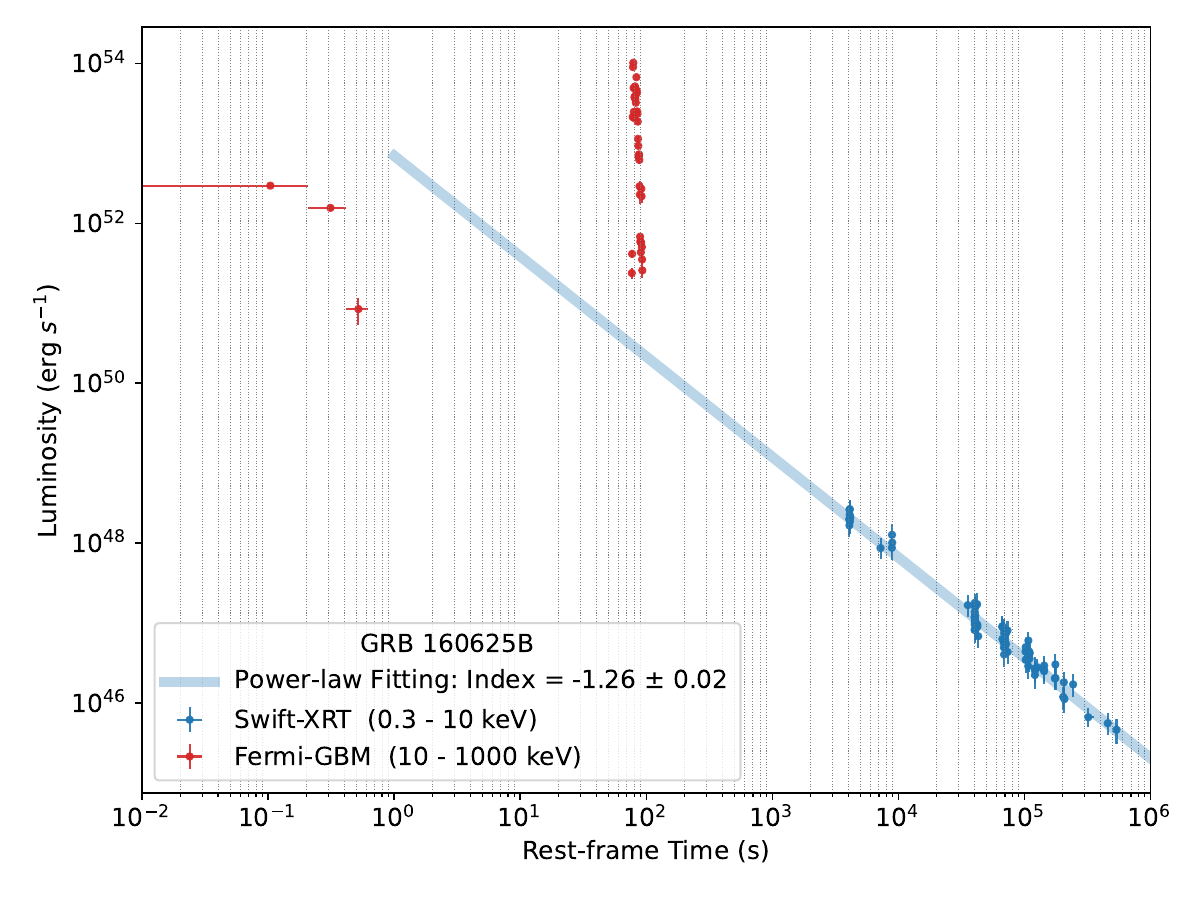}
\includegraphics[width=\hsize,clip]{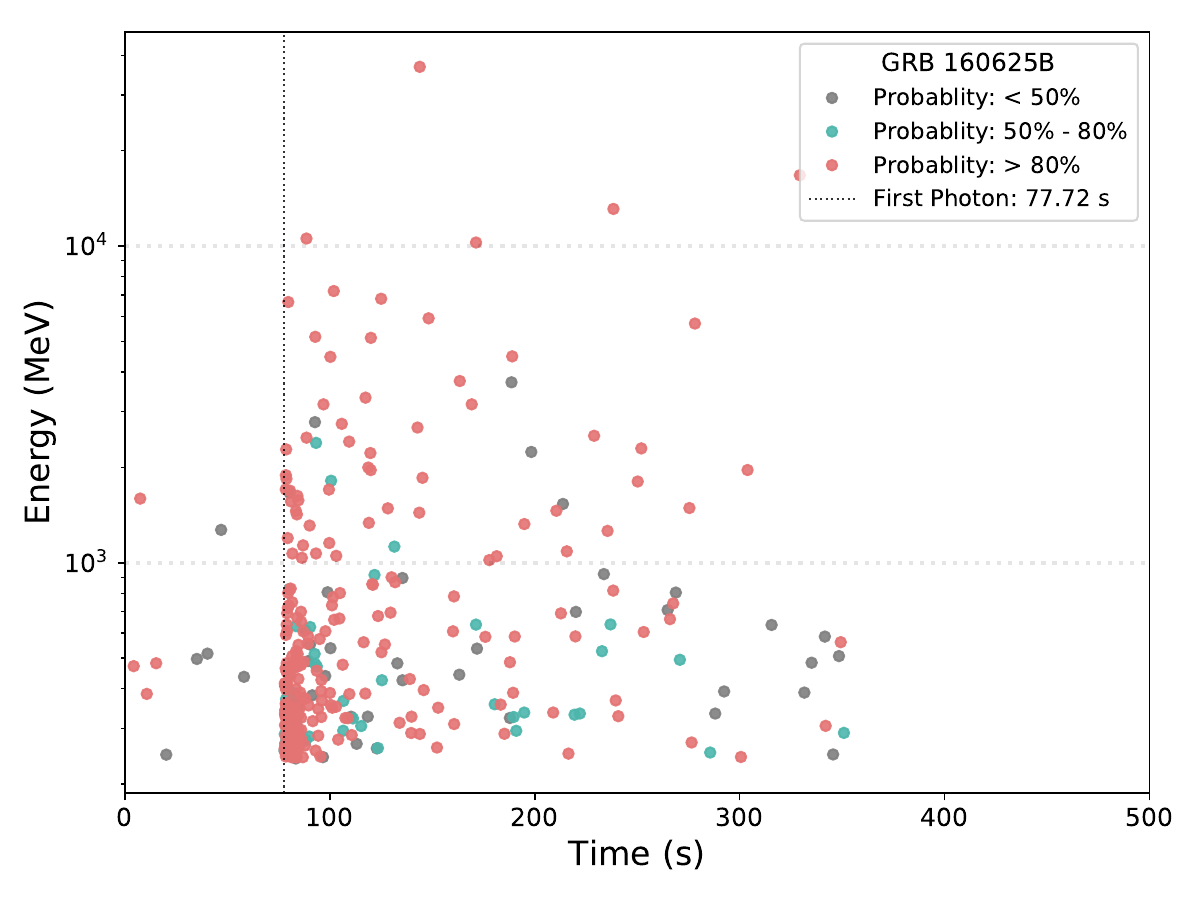} 
\caption{Top: Luminosity light-curve of GRB~160625B, including the prompt emission and the afterglow, observed by Fermi-GBM and Swift-XRT, respectively. Bottom: GeV photons observed by Fermi-LAT.}\label{fig:160625B-lc2} 
\end{figure}

On June 25, 2016, at 22:40:16.28 UT, the NASA Fermi Gamma-ray Space Telescope's GBM was triggered by GRB~160625B \citep{2016GCN.19581....1B} (see top panel of Fig.~\ref{fig:160625B-lc}). The Fermi-LAT started its observation 188.54 seconds post-trigger, detecting over 300 photons with energy exceeding 100 MeV, with the highest photon energy around 15 GeV \citep{2016GCN.19586....1D} (see bottom panel of Fig.~\ref{fig:160625B-lc2}). The Swift-XRT began its observation later, discovering a power-law behavior with a decaying index of approximately -1.25 \citep{2016GCN.19585....1M}. The redshift $z=1.406$ is reported \citep{2016GCN.19600....1X,2016GCN.19601....1D} (see top panel of Fig.~\ref{fig:160625B-lc2}). GRB~160625B is among the most energetic GRBs, with an isotropic energy of about 3$\times$10$^{54}$ erg \citep{2016GCN.19600....1X}. GRB~160625B is a bright GRB with detectable polarization \citep{2017Natur.547..425T}. Due to its high redshift, $z>1$, there is no associated supernova confirmation.

The GBM light curve of GRB~160625B showcases multiple peaks, with a total $T_{90}$ 

duration of approximately 460 seconds in the 50-300 keV range (see top panel of Fig.~\ref{fig:160625B-lc}). The initial GBM trigger was due to a soft peak lasting about 1 second. Spectral analysis of this phase reveals a power-law function with an exponential cutoff, characterized by a power-law index of $-0.2 \pm 0.1$ and a cutoff energy ($E_{\text{peak}}$) of $68 \pm 1$ keV. The fluence during this interval was $(1.65 \pm 0.03) \times 10^{-6}$ erg/cm$^2$. The main peak, responsible for the LAT trigger, extended for about 25 seconds. Its spectrum fits well with a Band function, showing an $E_{\text{peak}}$ of $657 \pm 5$ keV, an alpha of $-0.74$, and a beta of $2.36 \pm 0.01$. The fluence in this period was $(5.00 \pm 0.01) \times 10^{-4}$ erg/cm$^2$ \citep{2016GCN.19587....1B} (see bottom panel of Fig.~\ref{fig:160625B-lc}).

The early emission can be categorized into three episodes as suggested by several independent studies: a short precursor ($G_1$), a main burst ($G_2$), and a long-lasting tail ($G_3$). A significant and variable linear optical polarization in $G_2$ was detected, and it was inferred that the GRB outflows might be dominated by Poynting flux, where the magnetic energy dissipates quickly before the magnetic reconnection, resulting in bright gamma rays. A meticulous time-resolved analysis revealed an evolution of the thermal component in $G_1$. The bright $G_2$ episode was divided into 71 slices, each with at least 2500 net counts, for a detailed time-resolved spectral analysis. All slices fit a Band function; no thermal component was determined. $G_3$ is faint, and its time-resolved spectra were fitted by a single power law or cutoff power laws. The spectral evolution from thermal to nonthermal suggests a transition of the outflow from fireball to Poynting-flux-dominated.

The analysis of GRB~160625B within the BdHN model has been presented in \citet{2020ApJ...893..148R,2023ApJ...945...10L}.

The SN-rise episode of this GRB lasts from $-1.2$ to $3.1$ seconds observed ($-0.5$ to $1.3$ seconds in the rest frame), lasting $0.75$ seconds. The burst emits $1.04 \times 10^{52}$ ergs and has a lower black body temperature of $6.025$ keV (see Fig.~\ref{fig:160625B-lc}).

\subsection{GRB~180720B}

\begin{figure}
\centering
\includegraphics[width=\hsize,clip]{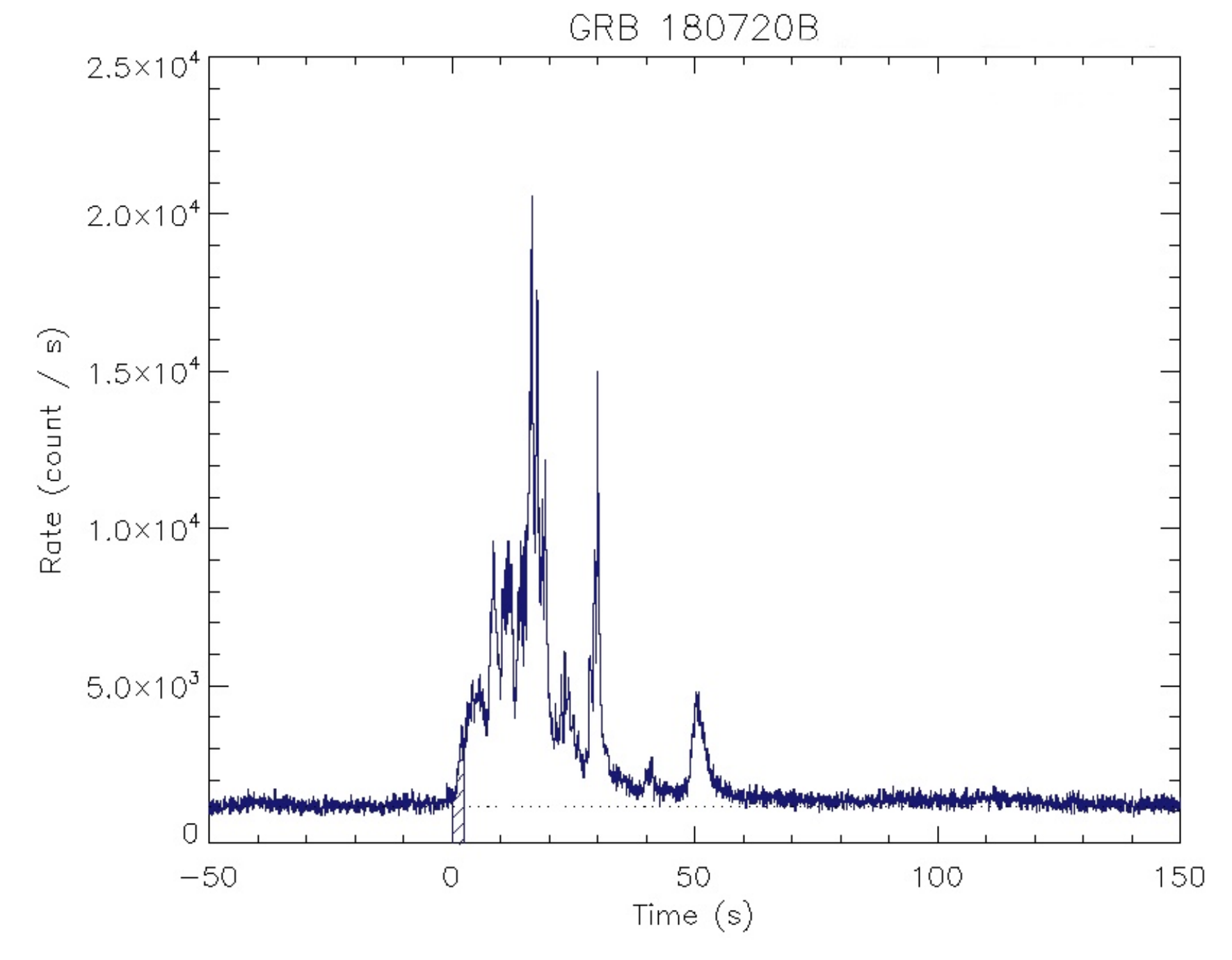} 
\includegraphics[width=\hsize,clip]{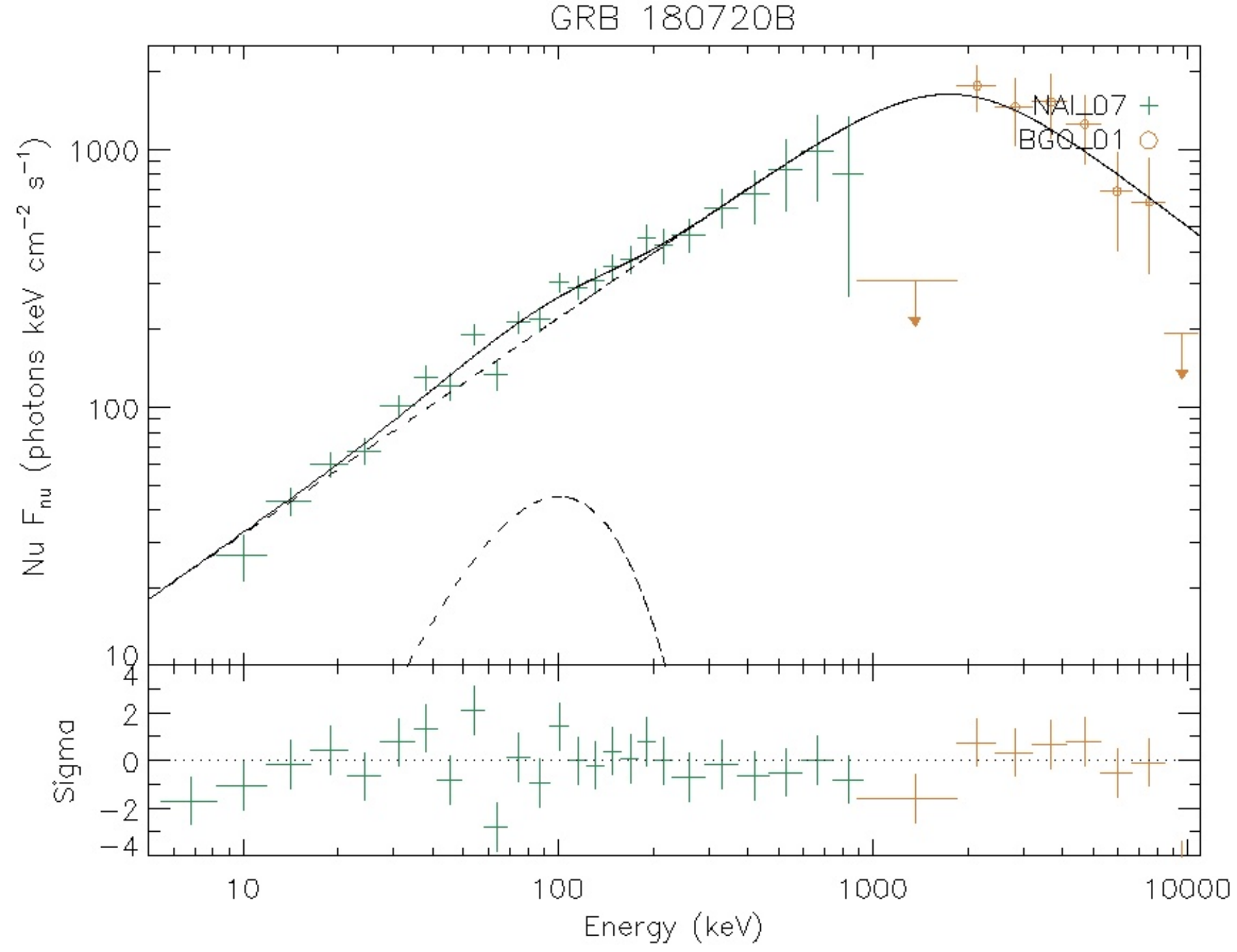}
\caption{Top: GRB~180720B light-curve, shadow part indicates the period of episode 1. Bottom: the spectrum of episode 1, fitted by the Band function plus a blackbody component, the observed temperature is 17.67 keV.}\label{fig:180720B-lc} 
\end{figure}

\begin{figure}
\centering
\includegraphics[width=\hsize,clip]{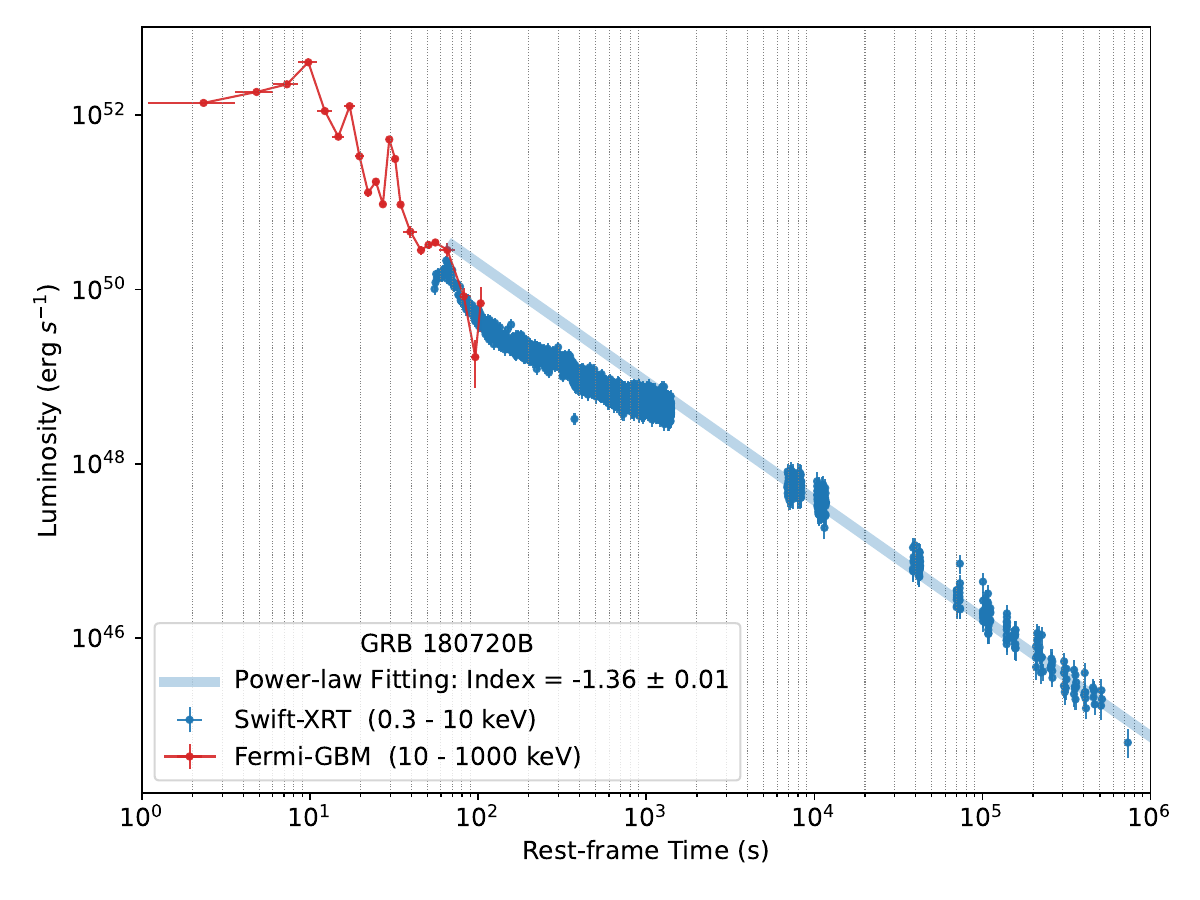}
\includegraphics[width=\hsize,clip]{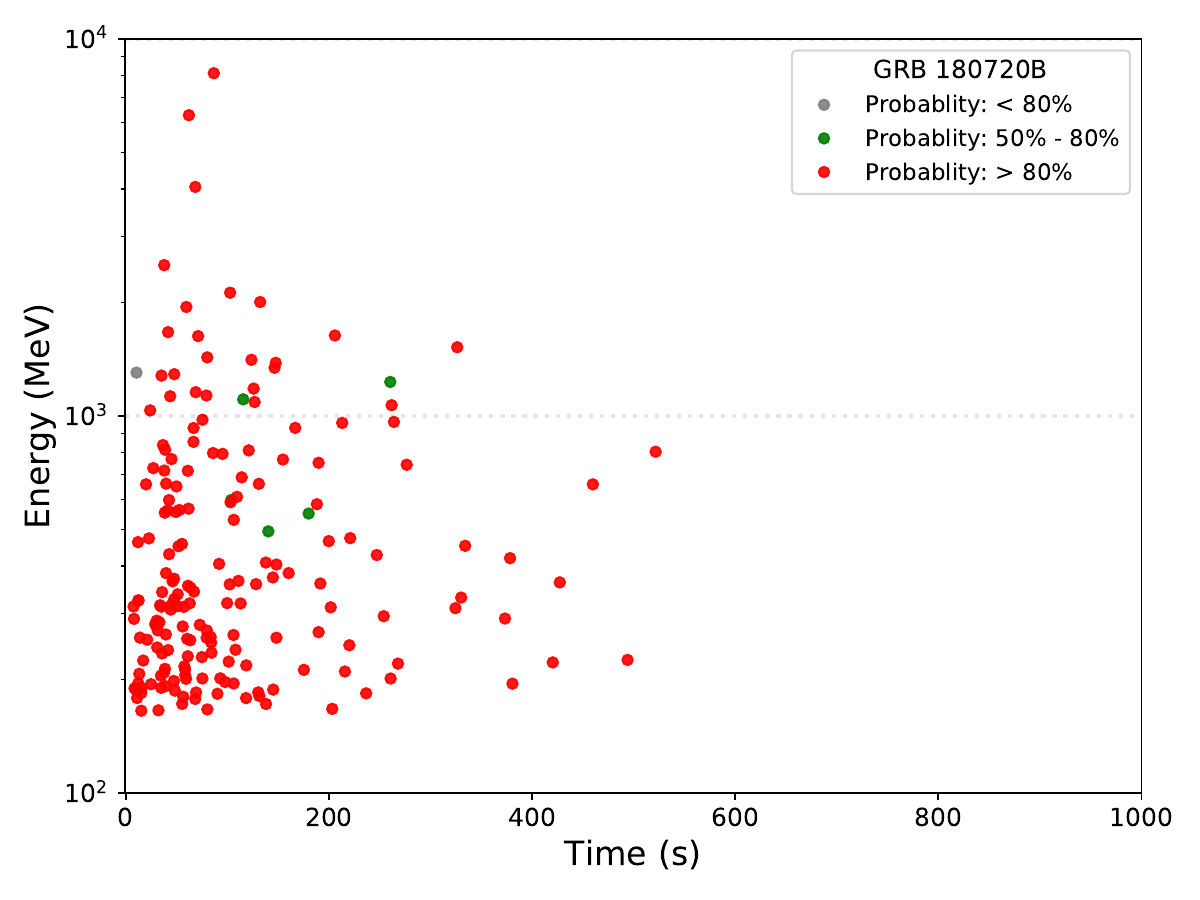} 
\caption{Top: Luminosity light-curve of GRB~180720B, including the prompt emission and the afterglow, observed by Fermi-GBM and Swift-XRT, respectively. Bottom: GeV photons observed by Fermi-LAT.}\label{fig:180720B-lc2}
\end{figure}

GRB~180720B was detected by Swift BAT on July 20, 2018, at 14:21:44 UT. BAT  shows a multi-peaked light curve with a duration of $\sim 150$~s and a peak count rate of $\sim 5\times 10^4$ counts/s (see top panel of Fig.~\ref{fig:180720B-lc}).  XRT observation began $86.5$~s after BAT trigger, locating the X-ray afterglow at RA = 00h 02m 6.70s, Dec = -02d 55' 01.2", with an uncertainty of 5.0 arcseconds \citep{2018GCN.22973....1S} (see top panel of Fig.~\ref{fig:180720B-lc2}). An optical counterpart within the XRT error circle with a magnitude of $R\sim9.4$ mag 73~s after the trigger by Kanata telescope at Higashi-Hiroshima Observatory, indicating a bright optical afterglow \citep{2018GCN.22977....1S}. The absorption features were identified corresponding to a redshift of $z = 0.654$ by ESO's VLT/X-shooter \citep{2018GCN.22996....1V}.

The Fermi GBM triggered almost simultaneously with Swift BAT, providing a detailed light curve and spectral analysis of the burst \citep{2018GCN.22981....1R}. The GBM light curve displayed a very bright, FRED-like peak with numerous overlapping pulses, lasting a duration (T90) of 49 s in the 50-300 keV range.
Spectral analysis from GBM data revealed a time-averaged spectrum best fit by a Band function, with peak energy $E_p = 631 \pm 10$ keV, lower photon index $\alpha = -1.11 \pm  0.01$, and higher photon index $\beta = -2.30 \pm  0.03$. The event's fluence (10-1000 keV) over the T90 interval was reported as $(2.985 \pm 0.001) 10^{-4} erg~cm^{-2}$. The 1-sec peak photon flux measured starting from T0+4.4 s in the 10-1000 keV band was $125 \pm 1 s^{-1} cm^{-2}$. The isotropic energy released  is $E_{iso} = 6.82(-0.22, 0.24) \times 10^{53}$~erg \citep{2018GCN.23042....1C}.

The Very-High-Energy (VHE) emission from GRB~180720B was first announced by Ruiz-Velasco at the 1st International CTA Symposium in May 2019. The High Energy Stereoscopic System (H.E.S.S.) observations revealed a new gamma-ray source ranging from 100-440 GeV at time about 10 hours post-burst, pinpointing the emission's origin close to the identified  GRB location at other wavelengths. Follow-up observations under similar conditions 18 days later showed a background-consistent sky map, effectively ruling out associations with steady gamma ray emitters like active galactic nuclei or persistent systematic effects, thereby reinforcing the link between the VHE emission and GRB~180720B \citep{2019ApJ...884..117W,2019Natur.575..464A} (see bottom panel of Fig.~\ref{fig:180720B-lc2}).

The analysis of GRB~180720B within the BdHN model has been presented in \citet{2022ApJ...939...62R,2022A&AT...33..191R}.

This GRB's first episode spans $0$ to $2.5$ seconds observed, translating to $0$ to $1.51$ seconds in the rest frame, and lasts $1.51$ seconds. Its spectrum is fitted by a Band function plus a blackbody. It releases $1.6 \pm 0.3 \times 10^{52}$ ergs of energy, with a black body temperature of $17.67$ keV (see Fig.~\ref{fig:180720B-lc}).

\subsection{GRB~190114C}

GRB~190114C was detected on January 14, 2019 \citep{2019GCN.23688....1G}, and sparked significant interest due to its brightness and the detection of high-energy emissions, including VHE gamma-ray emissions observed by the MAGIC telescope \citep{2019GCN.23701....1M}. 

The Swift Burst Alert Telescope (BAT) triggered on the burst at 20:57:03 UT, locating it with coordinates RA = 03h 38m 02s, Dec = -26d 56' 18" (J2000). Swift's X-Ray Telescope (XRT) began observing 64.0 seconds after the BAT trigger, identifying a bright X-ray source within the BAT error circle. The Ultraviolet/Optical Telescope (UVOT) detected a candidate afterglow in its imaging \citep{2019GCN.23688....1G}.  Optical telescopes, including MASTER, Nordic Optical Telescope (NOT), and others, provided early-time optical observations and spectroscopy, identifying the optical counterpart and determining a redshift of $z = 0.42$ \citep{2019GCN.23695....1S,2019GCN.23708....1C}. Observations in the radio, infrared, and sub-millimeter bands were conducted, including with the VLA, ALMA, and ATCA, providing multi-wavelength coverage of the afterglow.

\begin{figure}
\centering
\includegraphics[width=\hsize,clip]{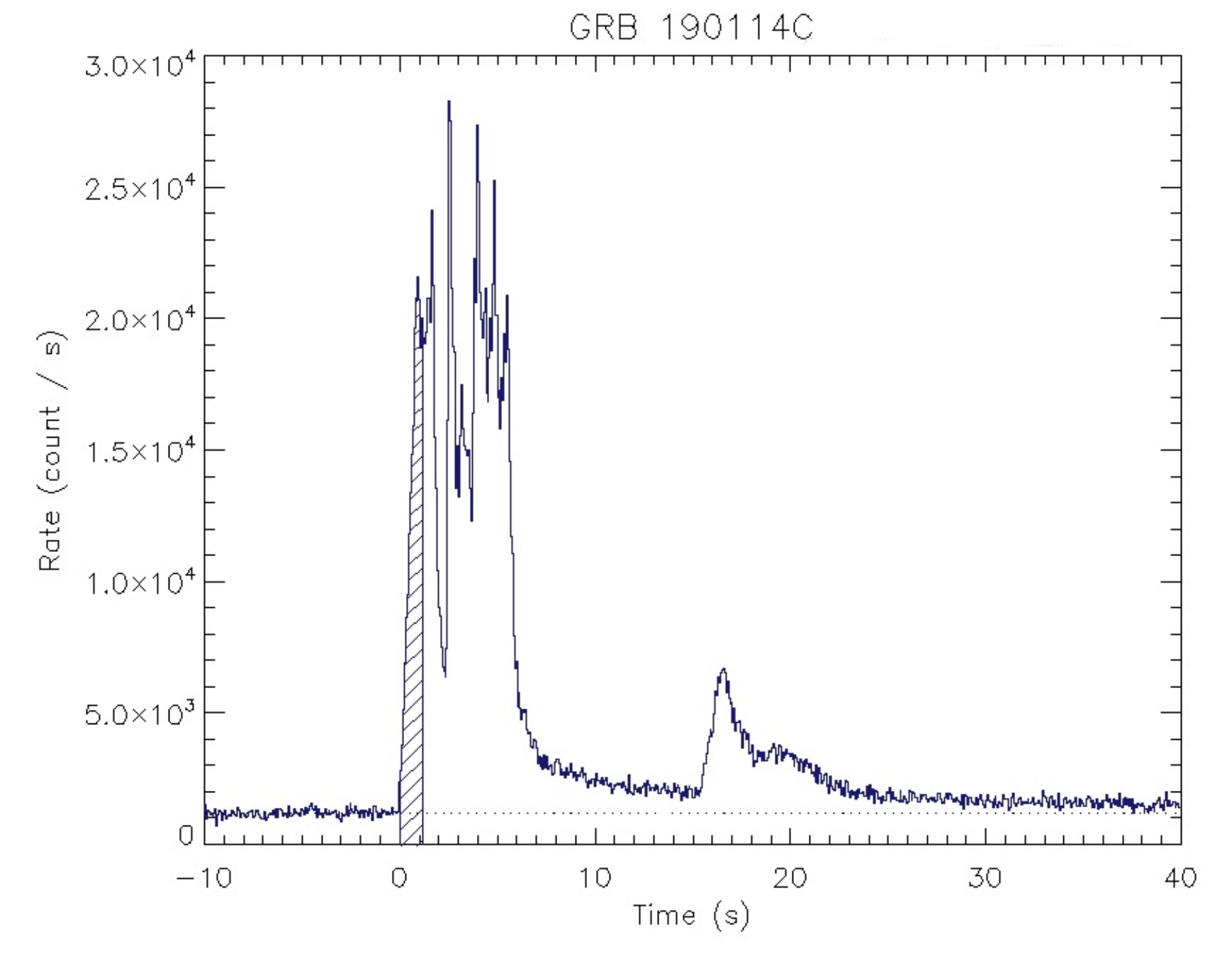} 
\includegraphics[width=\hsize,clip]{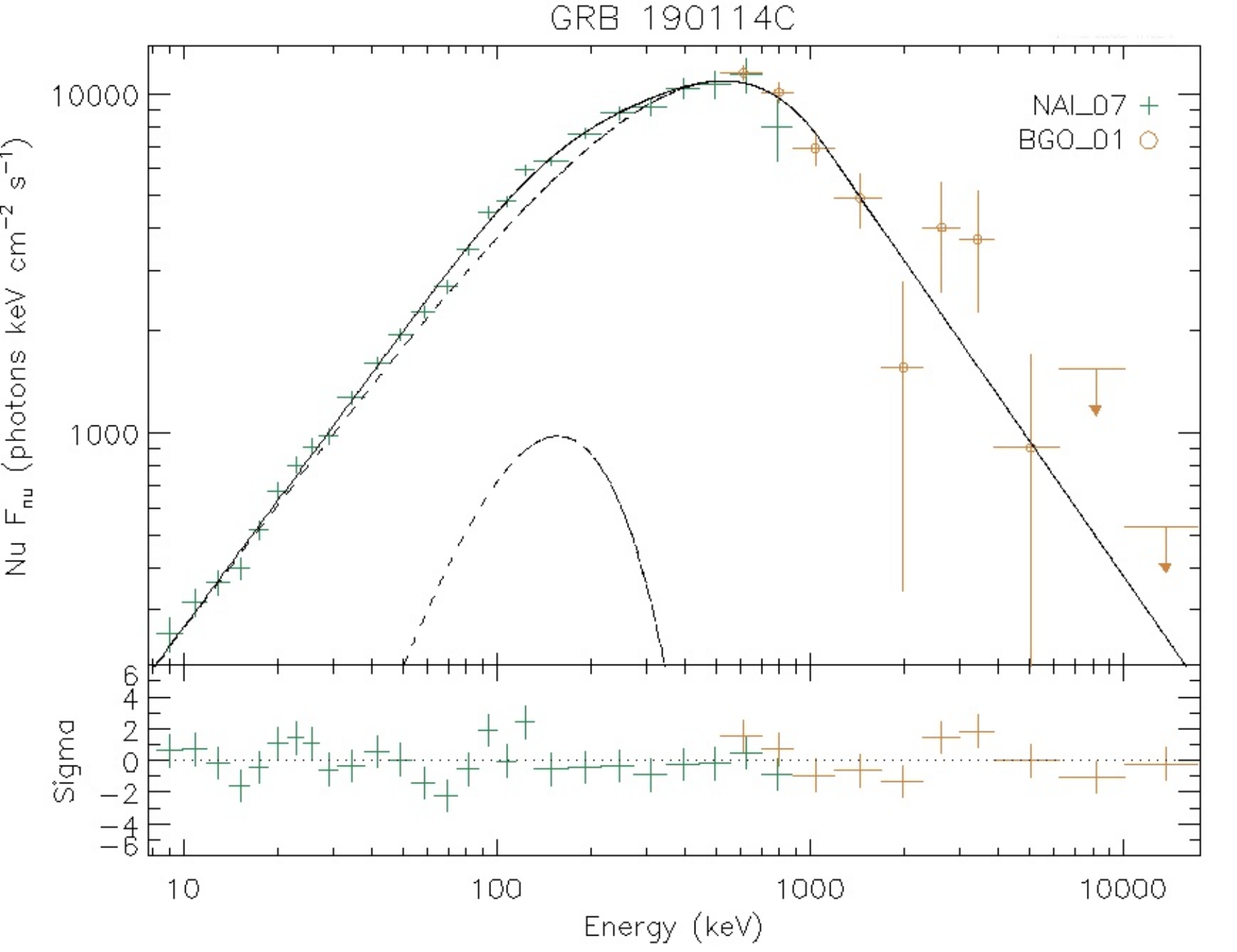}s 
\caption{Top: GRB~190114C light-curve, shadow part indicates the period of episode 1. Bottom: the spectrum of episode 1, fitted by the Band function plus a blackbody component, the observed temperature is 39.99 keV.}\label{fig:190114C-lc} 
\end{figure}

The GBM light curve for GRB~190114C reveals a highly luminous, multi-peaked structure extending to $15$ s, succeeded by a less intense pulse from roughly $15$ s to $25$ s (see top panel of Fig.~\ref{fig:190114C-lc}). Fainter emissions are discernible up to about $200$ s. The duration, $T_{90}$, is estimated at approximately 116 s. The time-averaged spectrum from $0$ s to $38.59$ s can be accurately modeled by a Band function, with $E_{\text{p}} = 998.6 \pm 11.9$ keV, $\alpha = -1.058 \pm 0.003$, and $\beta = -3.18 \pm 0.07$. The fluence for the event, within the 10-1000 keV energy band for this time interval, is $(3.99 \pm 0.00081) \times 10^{-4} erg~cm^{-2}$. The peak photon flux measured starting from $T_0+3.84$ seconds in the 10-1000 keV band is $246.86 \pm 0.86 \,  s^{-1}cm^{-2}$. Utilizing the Band spectral fit and the redshift measurement of $z = 0.42$ \citep{2019GCN.23695....1S}, the isotropic energy release in gamma-rays, $E_{\text{iso}}$, is estimated to be $3 \times 10^{53}$ erg, and the isotropic peak luminosity, $L_{\text{iso}}$, is calculated to be $\sim 10^{53} \text{erg s}^{-1}$ , within the 1 keV to 10 MeV energy band \citep{2019GCN.23707....1H}. Fermi/LAT detected a significant increase in the event rate that is spatially correlated with the GBM trigger with high significance. The highest-energy photon is a $22.9$ GeV event which is observed 15 s after the GBM trigger \citep{2019GCN.23709....1K}.

\begin{figure}
\centering 
\includegraphics[width=\hsize,clip]{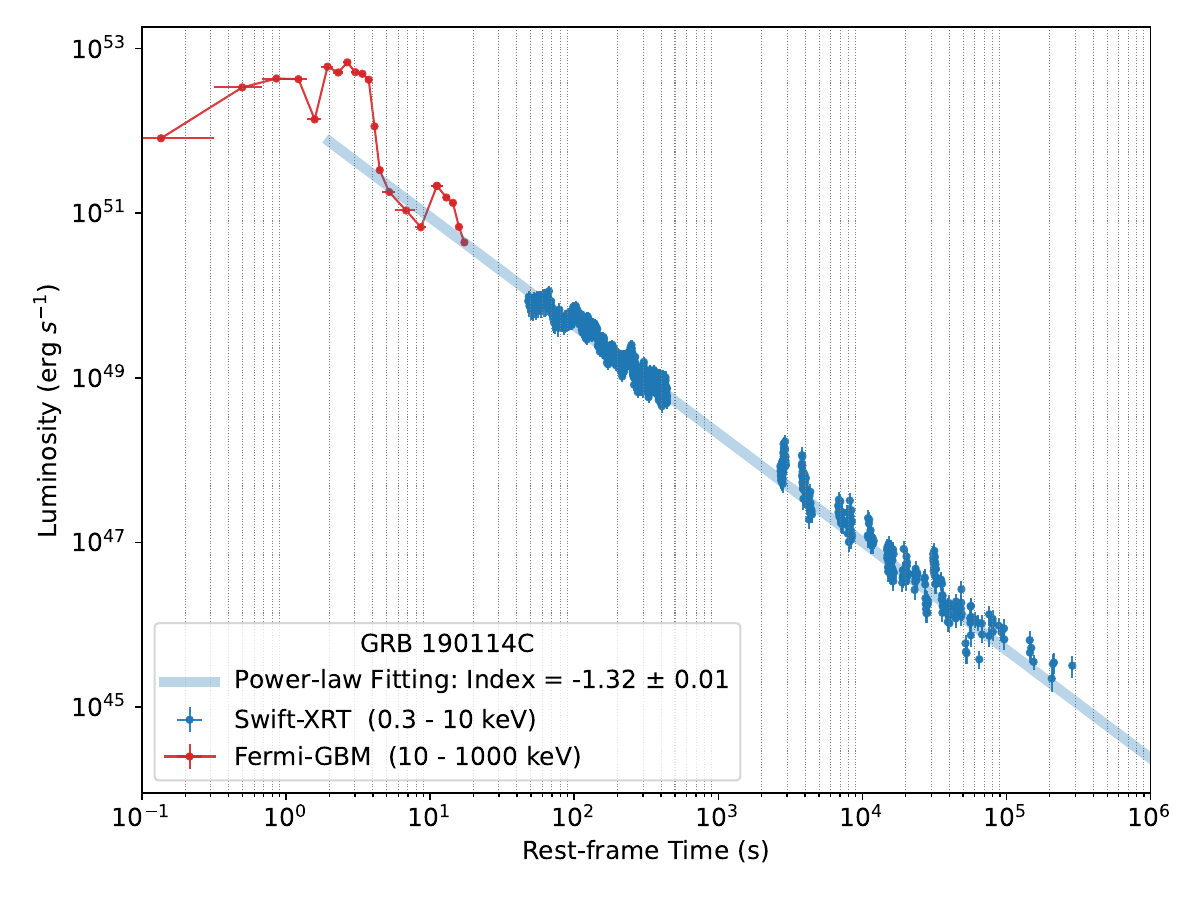}
\includegraphics[width=\hsize,clip]{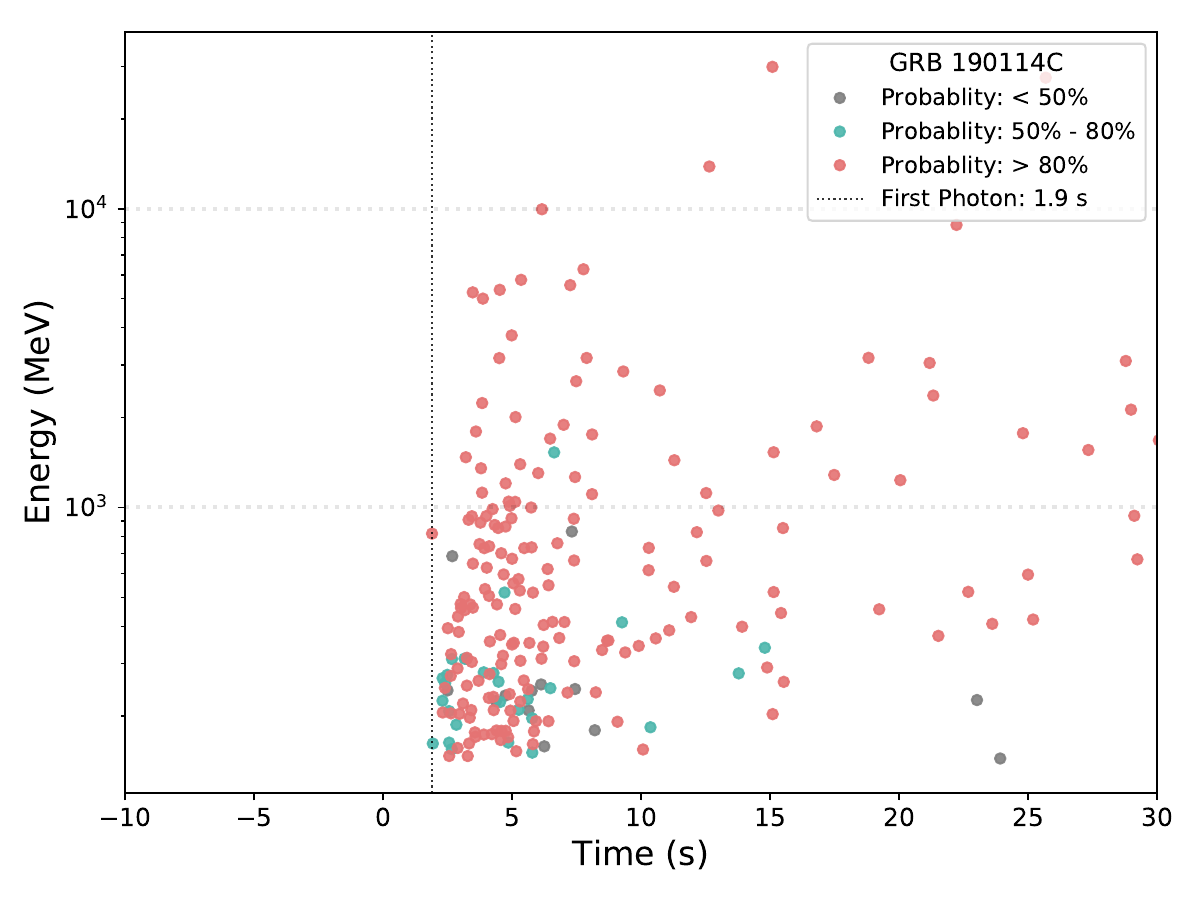}
\caption{Top: Luminosity light-curve of GRB~190114C, including the prompt emission and the afterglow, observed by Fermi-GBM and Swift-XRT, respectively. Bottom: GeV photons observed by Fermi-LAT.}\label{fig:190114C-lc2} 
\end{figure}

The afterglow of GRB~190114C exhibited a power-law decay across X-ray, optical, and radio wavelengths, indicative of synchrotron radiation origination. X-ray observations by Swift-XRT identified a fading afterglow with a decay index of $\alpha_X \approx 1.34 \pm 0.01$, fitting the typical afterglow emission model. The initial X-ray flux, measured 64 seconds post-trigger, was $7.39 \times 10^{-8} \, \text{erg cm}^{-2} \text{s}^{-1}$ in the 0.3-10 keV range \citep{2019GCN.23706....1D} (see top panel of Fig.~\ref{fig:190114C-lc2}). Optical follow-up revealed that late-time light curve was influenced by an emerging supernova component \citep{2019GCN.23983....1M}. This supernova, associated with the GRB, peaked at approximately $r \approx 23.9$ mag (AB) in the rest frame, about 15 days post-burst.

GRB~190114C has been extensively studied within the BdHN model \citep[see, e.g.,][and references therein]{2019ApJ...883..191R,2020ApJ...893..148R,2021PhRvD.104f3043M,2022A&AT...33..191R,2023ApJ...945...10L}.

The SN-rise of this burst spanning $0$ to $1.1$ seconds observed ($0$ to $0.79$ seconds in the rest frame), and lasts $0.79$ seconds, described by the Band function plus a blackbody component. It emits $3.5 \pm 0.2 \times 10^{52}$ ergs of energy, with a particularly high black body temperature of $39.99$ keV (see Fig.~\ref{fig:190114C-lc}).

\subsection{GRB~220101A}

\begin{figure}
\centering
\includegraphics[width=\hsize,clip]{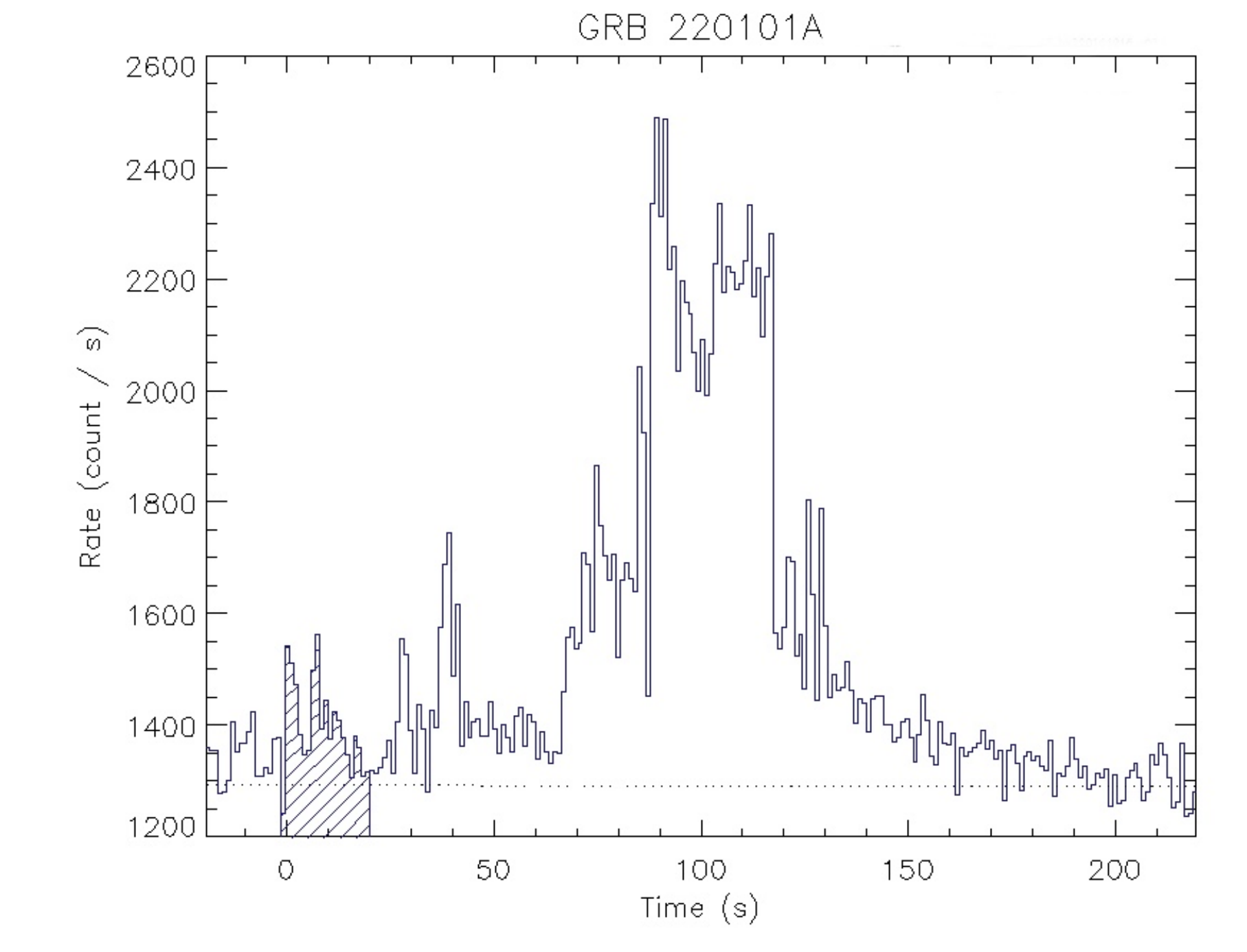} 
\includegraphics[width=\hsize,clip]{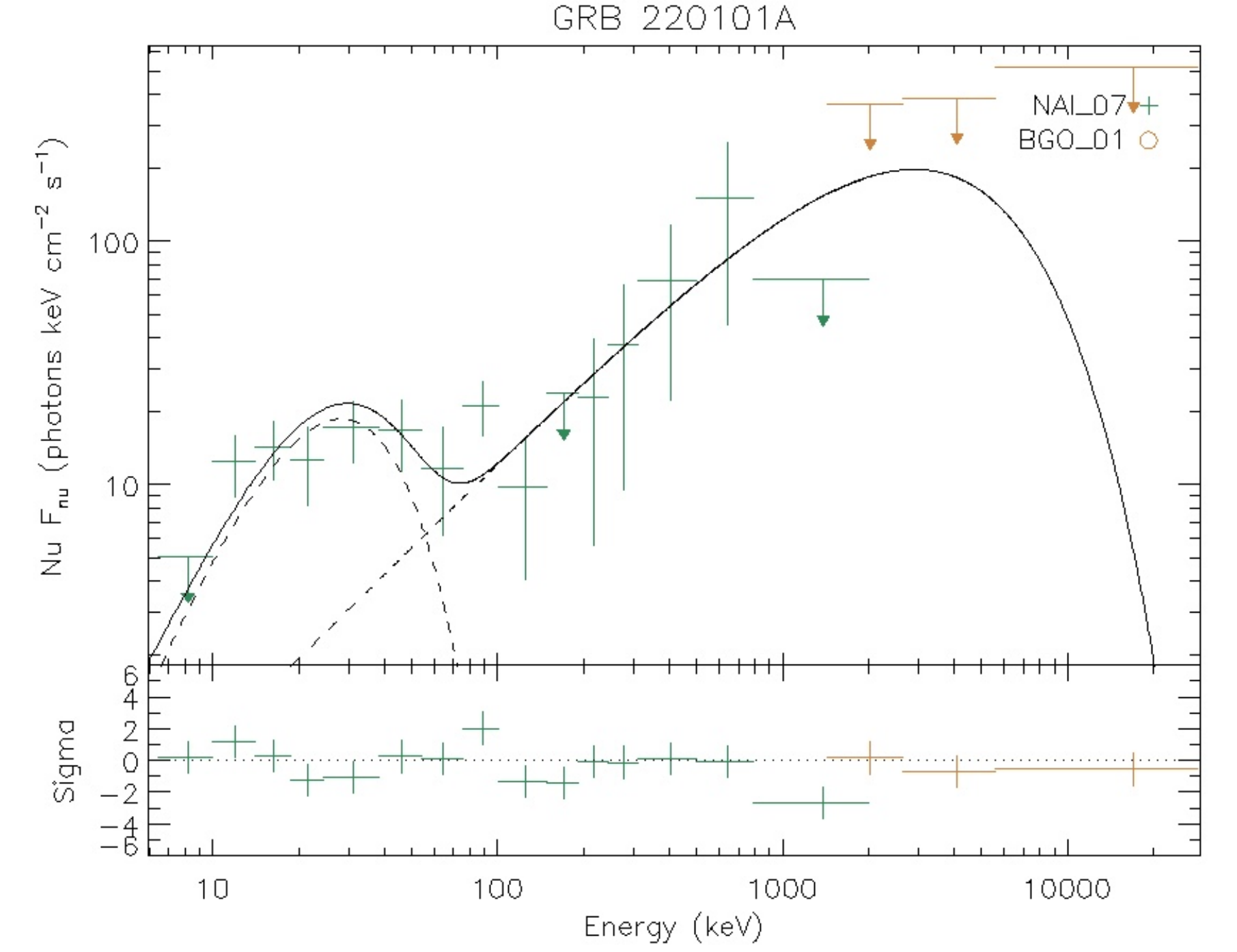}
\caption{Top: GRB~220101A light-curve, shadow part indicates the period of episode 1. Bottom: the spectrum of episode 1, fitted by the cutoff power-law plus a blackbody component, the observed temperature is 7.215 keV.}\label{fig:220101A-lc} 
\end{figure}

\begin{figure}
\centering
\includegraphics[width=\hsize,clip]{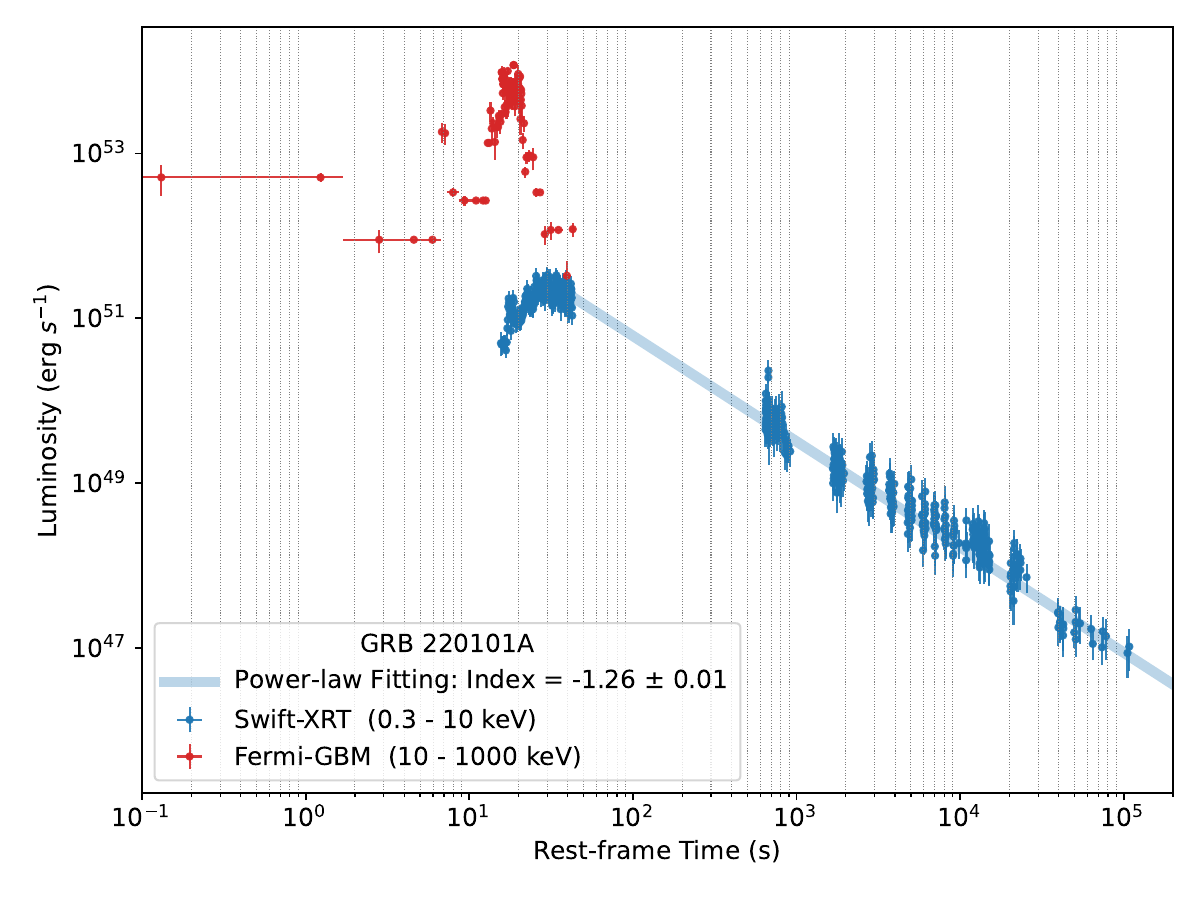}
\includegraphics[width=\hsize,clip]{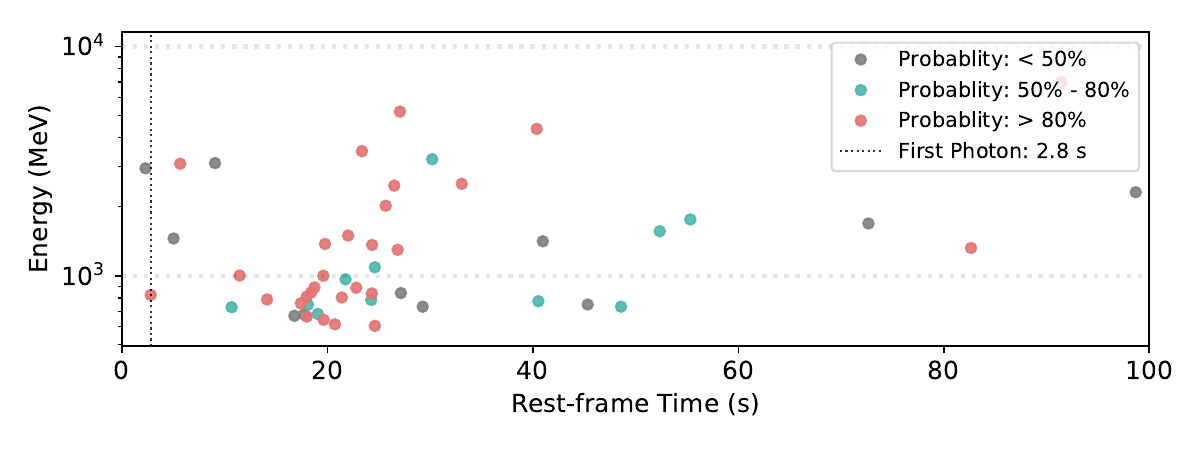} 
\caption{Top: Luminosity light-curve of GRB~220101A, including the prompt emission and the afterglow, observed by Fermi-GBM and Swift-XRT, respectively. Bottom: GeV photons observed by Fermi-LAT.}\label{fig:220101A-lc2} 
\end{figure}

The Swift-BAT detected GRB~220101A at 05:09:55, January 1, 2022 UT. The peak count rate was 7000 counts/s in the 15-350 keV range, approximately 89 seconds after the trigger (see top panel of Fig.~\ref{fig:220101A-lc}). The X-Ray Telescope (XRT) began observations 80.8 seconds after the trigger, identifying a bright, uncatalogued X-ray source at RA = 00h 05m 25.46s, Dec = +31d 46' 12.7" with an uncertainty of 4.7 arcseconds (see top panel of Fig.~\ref{fig:220101A-lc2}). The Ultraviolet/Optical Telescope (UVOT) identified a candidate afterglow with a magnitude of 14.60 in the white filter \citep{2022GCN.31347....1T}.

The spectral analysis by Fermi-GBM indicated a best fit with a power law function with an exponential high-energy cutoff, where the power law index was $-1.09 \pm 0.02$, and the cutoff energy ($E_{\text{c}}$) was $330 \pm 15$ keV. A Band function fit was equally well with parameters $E_{\text{p}}= 290 \pm 18$ keV, $\alpha = -1.06 \pm 0.02$, and $\beta = -2.3 \pm 0.2$ \citep{2022GCN.31360....1L} (see bottom panel of Fig.~\ref{fig:220101A-lc}). The GRB has been detected also by AGILE \citep{2022GCN.31354....1U}. The isotropic energy ($E_{\text{iso}}$) was calculated to be approximately $3.7 \times 10^{54}$ ergs, equating to one of the highest measured isotropic energies for GRBs to date \citep{2022GCN.31365....1A,2022GCN.31433....1T,2022GCN.31436....1T}. 

The \textit{Fermi-LAT} detected high-energy emission from GRB~220101A with a photon flux above 100 MeV of $2.5 \times 10^{-5} \pm 5 \times 10^{-6} \text{ph/cm}^2/\text{s}$ in the time interval 0-600s after the Swift trigger. The estimated photon index was $-2.46 \pm 0.25$ \citep{2022GCN.31350....1A} (see bottom panel of Fig.~\ref{fig:220101A-lc2}).

The redshift of GRB~220101A was determined to be $z = 4.61$, based on spectroscopic observations using the Xinglong-2.16m telescope and the Nordic Optical Telescope (NOT), identifying a broad absorption feature likely corresponding to Lyman-alpha absorption \citep{2022GCN.31353....1F,2022GCN.31359....1F}.

For a preliminary analysis of GRB~220101A within the BdHN model see \citet{2023arXiv230605855B}.

The first episode of this GRB ranges from $-1.0$ to $20.0$ seconds observed ($-0.18$ to $3.57$ seconds in the rest frame), lasting $3.75$ seconds. The spectrum of this episode is fitted by the Band function plus a blackbody component. It releases $1.2 \times 10^{53}$ ergs of isotropic energy, with a black body temperature of $7.215$ keV (see Fig.~\ref{fig:220101A-lc}).

\subsection{GRB~221009A}

\begin{figure}
\centering
\includegraphics[width=\hsize,clip]{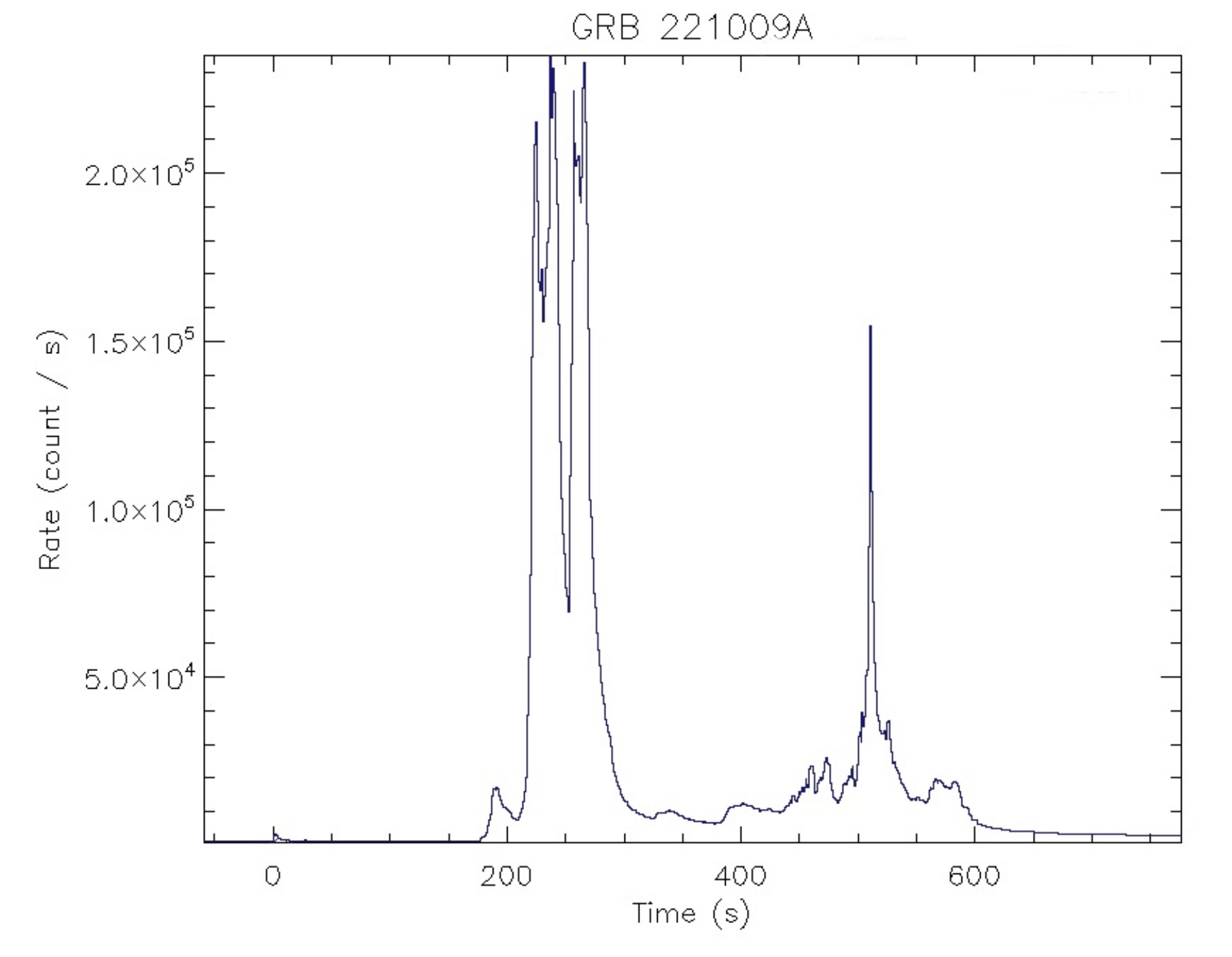} 
\includegraphics[width=\hsize,clip]{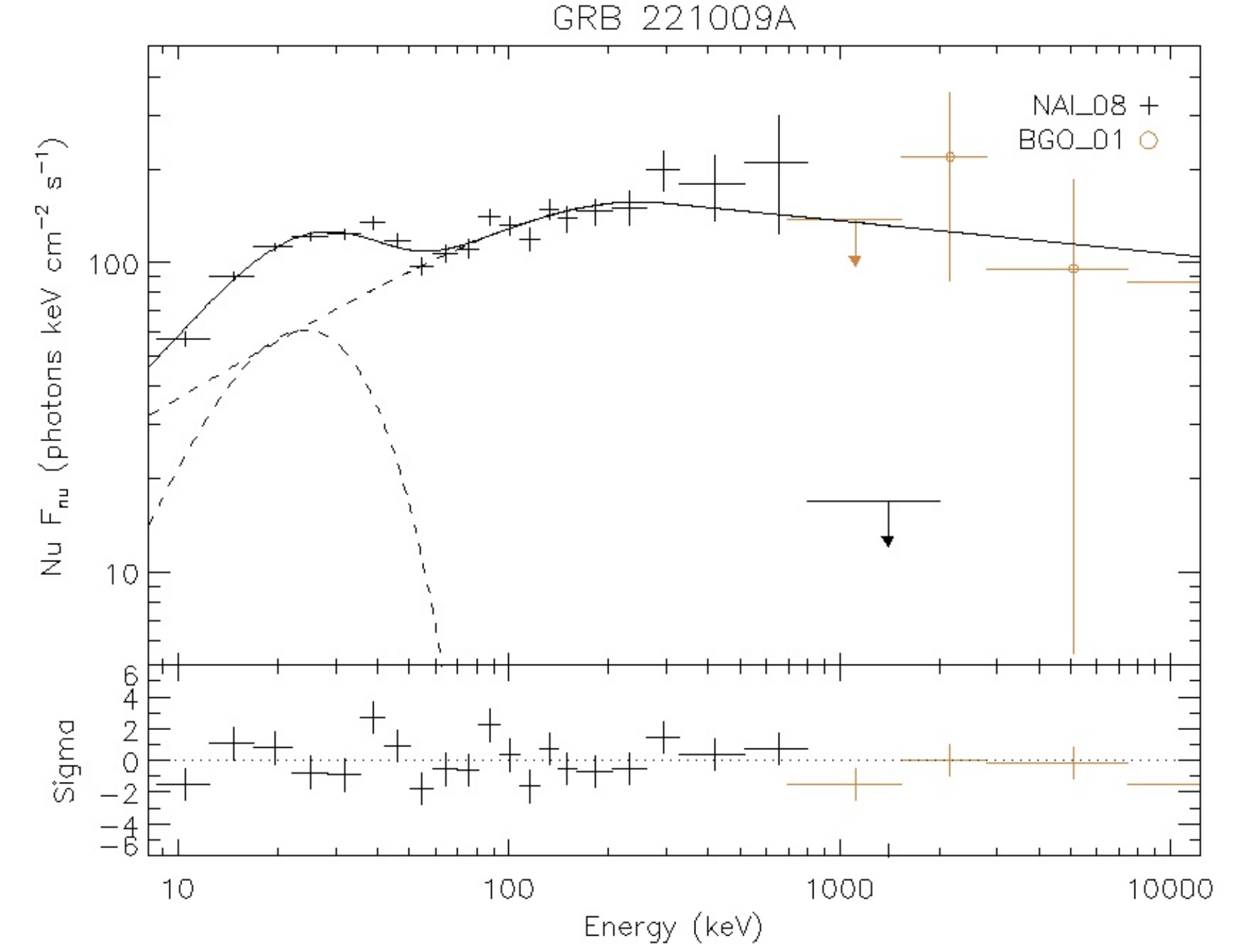}
\caption{Top: GRB~221009A light-curve, shadow part indicates the period of episode 1. Bottom: the spectrum of episode 1, fitted by the Band function plus a blackbody component, the observed temperature is 6.169 keV.}\label{fig:221009A-lc} 
\end{figure}

\begin{figure}
\centering
\includegraphics[width=\hsize,clip]{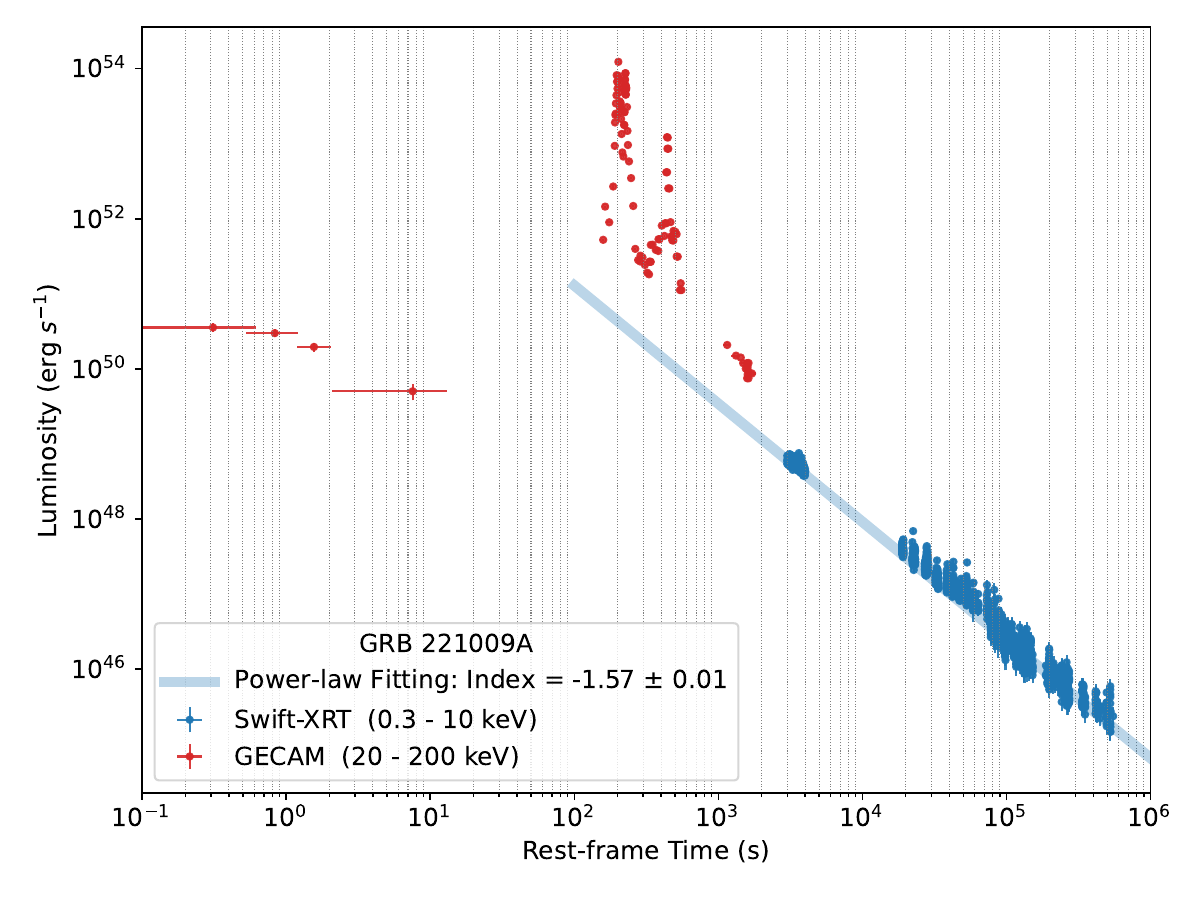}s 
\caption{Luminosity light-curve of GRB~221009A, including the prompt emission and the afterglow, observed by GECAM and Swift-XRT, respectively.}\label{fig:221009A-lc2} 
\end{figure}

The Swift-BAT detected GRB~221009A, initially named Swift J1913.1+1946, on October 9 of 2022 at 14:10:17 UT (see top panel of Fig.~\ref{fig:221009A-lc}). Swift-XRT began observing the field 143 seconds after the BAT trigger, identifying a bright, fading X-ray source at RA = 19h 13m 3.43s, Dec = +19d 46' 16.3" with an uncertainty of 5.6 arcseconds (see Fig.~\ref{fig:221009A-lc2}). The Swift-UVOT took an exposure 179 seconds after the BAT trigger, identifying a candidate counterpart with an estimated magnitude of 16.63 \citep{2022GCN.32632....1D}.

The Fermi-GBM team reported \citep{2022GCN.32642....1L} that the time-averaged spectrum for the first emission episode is best fitted by a power-law function with an exponential high-energy cutoff. The photon index was found to be $-1.70 \pm 0.02$, the exponential cutoff is at $375 \pm 87$ keV. The fluence in the 10-1000 keV range for this time interval was reported as $(2.12 \pm 0.05) \times 10^{-5} \text{ erg/cm}^2$. 

For the high-energy observations made by Fermi LAT, \citet{2022GCN.32637....1B} detailed the detection. The initial boresight angle was 94 degrees then the satellite rotated to point to the source. The spectral fitting from LAT observations revealed that in the time interval 500-3500 s after the Swift trigger, the photon flux in the range of 100 MeV - 1 GeV is $(1.27 \pm 0.16) \times 10^{-5} \text{ cm}^2/\text{s}$. The estimated photon index above 100 MeV is $-2.12 \pm 0.11$. LAT observed a highest-energy photon of $7.8$ GeV 766 seconds after the Swift trigger. 

\begin{table*}
\centering
\caption{GRB SN-rise properties: For each GRB Episode 1 is given its Trigger number, its redshift, its starting time $t_s$ and ending time $t_e$ both in the observed frame and in the GRB cosmological rest frame, its duration in the GRB cosmological rest frame, its isotropic equivalent energy $E_{iso}^{SN-rise}$, and its observed black body temperature.}
\label{table:GRBSN}
\begin{tabular}{l|l|c|c|c|c|c|c|c}
\hline
\textbf{GRB}      & \textbf{Trigger} & \textbf{z} & \textbf{Episode} & \textbf{$t_{s} \sim t_{e}$} & \textbf{$t_{s} \sim t_{e}$} & \textbf{Duration} & \textbf{$E_{iso}^{SN-rise}$} & \textbf{BB Temp.} \\
\textbf{name} & \textbf{number} & & & \textbf{(s, obs.)} & \textbf{(s, rest)} & \textbf{(s, rest)} & \textbf{($10^{52}$erg)} & \textbf{(keV, obs.)}\\
\hline\hline
090423   & 090423330      & 8.2      & 1 (SN-rise)  & -5.5 $\sim$ 7.4  & -0.6$\sim$0.8           & 1.4          & $16$   &   --- \\ \hline
090429B  & 350854         & 9.4      & 1 (SN-rise)  & 0 $\sim$ 10      & 0$\sim$0.96             & 0.96         & $3.5$    &  --- \\ \hline
090618   & 090618353      & 0.54     & 1 (SN-rise)  & 0 $\sim$ 47.7    & 0$\sim$31               & 31           & $3.61$    & 15.86 \\ \hline
130427A  & 130427324      & 0.34     & 1 (SN-rise)  & 0 $\sim$ 0.9     & 0$\sim$0.65             & 0.65         & $0.65$    &  28.66 \\ \hline
160509A  & 160509374      & 1.17     & 1 (SN-rise)  & 0 $\sim$ 4.0     & 0$\sim$1.84             & 1.84         & $1.47$    & 14.43 \\ \hline
160625B  & 160625945      & 1.406    & 1 (SN-rise)  & -1.2 $\sim$ 3.1  & -0.5$\sim$1.3           & 0.75         & $1.04$   &  6.025 \\ \hline
180720B  & 180720598      & 0.653    & 1 (SN-rise)  & 0 $\sim$ 2.5     & 0$\sim$1.51             & 1.51         & $1.6 \pm 0.3$ & 17.67 \\ \hline
190114C  & 190114873      & 0.425    & 1 (SN-rise)  & 0 $\sim$ 1.1     & 0$\sim$0.79             & 0.79         & $3.5 \pm 0.2$ & 39.99 \\ \hline
220101A  & 220101215      & 4.61     & 1 (SN-rise)  & -1.0 $\sim$ 20.0 & -0.18$\sim$3.57         & 3.75         & $12$   &   7.215 \\ \hline
221009A  & 221009553      & 0.151    & 1 (SN-rise)  & 0 $\sim$  15.0   & 0$\sim$13               & 13           & $0.18 \pm 0.01$ & 6.169 \\ \hline \hline
\end{tabular}
\end{table*}

Observations of X-shooter instrument at ESO's Very Large Telescope began 11.55 hours after the Fermi GBM trigger and 10.66 hours after the Swift BAT trigger. The spectrum showed a very red continuum with absorption features corresponding to CaII, CaI, and NaID. These absorption features were used to determine a redshift of $z = 0.151$. Using this redshift and the GBM fluence, an \(E_{\text{iso}}\) value of \(2 \times 10^{54}\) erg is determined, placing GRB~221009A within the upper end of GRB energetics \citet{2022GCN.32648....1D}.

LHAASO detected GRB~221009A with significant findings that by LHAASO-WCDA above 500 GeV and LHAASO-KM2A, marking the first detection of photons above 10 TeV from GRBs. Above 100 standard deviations (s.d.) for LHAASO-WCDA and about 10 s.d. for LHAASO-KM2A. The highest photon energy reached 18 TeV \citep{2022GCN.32677....1H}.

Spectra of the afterglow of GRB~221009A were also obtained with JWST/NIRSpec under DDT program 2784 (P.I. Blanchard) on 2023 April 20, 193 observer-frame days after the burst. The spectrum, differing significantly from the earlier power-law continuum observed, indicates a considerable contribution from the SN/host galaxy. JWST/NIRSpec detected a feature centered at ~1 micron, consistent with the Ca II IR triplet from a SN. Prominent narrow lines were also observed \citep{2023GCN.33676....1B}.

This GRB's SN-rise episode lasts from $0$ to $15.0$ seconds observed ($0$ to $13$ seconds in the rest frame), with a duration of $13$ seconds, of which the spectrum is best fitted by the Band function plus a blackbody component. It releases a much lower $0.18 \pm 0.01 \times 10^{52}$ ergs of energy, with a black body temperature of $6.169$ keV (see Fig.~\ref{fig:221009A-lc}).

\subsection{Challenges in Detecting Thermal Components of High Redshift GRBs}

For GRBs at high redshifts, the detection of their thermal components becomes notably challenging compared to those at lower redshifts. This phenomenon can be attributed to two primary factors:

\begin{enumerate}
\item \textbf{Data Quality and Model Complexity}

 Introducing a thermal component to the fitting models introduces two additional degrees of freedom. Consequently, constraining the models with an additional  thermal component mandates high-quality observational data.
        
 Considering GRBs at significant redshifts, like those at \(z=5\), the luminosity distance is magnified to be seven times that of GRBs situated at \(z=1\). This translates to a drastic reduction, about 50-fold, in the number of photons that can be observed. Moreover, if the selected time bin is small, there will be even fewer accumulated photons, making it more difficult to constrain a model that includes a thermal component.

\item \textbf{Cosmic Expansion and Energy Range Limitations}

The universe's ongoing expansion plays a role in redshifting the thermal component of the GRB. This redshifting can push the thermal component towards, or even beyond, the peripheries of a telescope's bandwidth. 

Typically, the temperature of a GRB's thermal component hovers around tens of keV. Several GRBs in this article have temperatures ranging between 20-50 keV in their rest frames. Such temperatures, when emitting from a high redshift location like \(z=5\), undergo redshifting to fall between 3-8 keV by the time they reach Earth. The peak of a blackbody spectrum is situated at \(2.82\) times its temperature. This means the most discernible region of the thermal component lies between 10-25 keV. Given the Fermi-GBM NaI detector's energy bandwidth of 8-900 keV, the thermal component is located at the spectrum's lower energy edge. The blackbody spectrum, especially its ascent, might not be wholly observable. Adding another layer of difficulty to detect and fit the thermal component.

\end{enumerate}

An illustrative example from this article is GRB~090423, which has a redshift of $8.2$. The spectral data from its SN-rise phase (depicted in Figure \ref{fig:090423-lc}) clearly reflects the aforementioned challenges. The spectrum has a limited number of data points, with a mere 4 points lying below 30 keV, which is not adequate to constrain an additional thermal component. While the spectrum fitting in Figure \ref{fig:090423-lc} employs the CPL model, this is more of a compromise due to sparse data. We cannot justify the presence of a thermal component in the intrinsic spectrum.

\section{Conclusion}

The duration, energetics and thermal component of the SN-rise for the ten sources are summarized in Table~\ref{table:GRBSN}. The thermal component identified in the analysis of GRBs associated with supernovae is a significant aspect of these observations. This thermal component has been detected in all the 8 sources at redshifts less than 5. The temperatures of these thermal components range between $6.23$ keV and $39.99$ keV. These observations suggest a possible signature of pair-driven supernovae, indicative of the immense energy and the high-density environment in which these bursts occur. Such thermal emissions provide crucial insights into the physical conditions prevailing during the explosive events, contributing to our understanding of the mechanisms driving supernovae and associated GRBs.

\bibliography{hypernova}

\begin{thebibliography}{89}
\expandafter\ifx\csname natexlab\endcsname\relax\def\natexlab#1{#1}\fi

\bibitem[{{Abdalla} {et~al.}(2019){Abdalla}, {Adam}, {Aharonian}, {Ait
  Benkhali}, {Ang{\"u}ner}, {Arakawa}, {Arcaro}, {Armand}, {Ashkar}, {Backes},
  {Barbosa Martins}, {Barnard}, {Becherini}, {Berge}, {Bernl{\"o}hr},
  {Bissaldi}, {Blackwell}, {B{\"o}ttcher}, {Boisson}, {Bolmont}, {Bonnefoy},
  {Bregeon}, {Breuhaus}, {Brun}, {Brun}, {Bryan}, {B{\"u}chele}, {Bulik},
  {Bylund}, {Capasso}, {Caroff}, {Carosi}, {Casanova}, {Cerruti}, {Chand},
  {Chandra}, {Chen}, {Colafrancesco}, {Cury{\l}o}, {Davids}, {Deil}, {Devin},
  {deWilt}, {Dirson}, {Djannati-Ata{\"\i}}, {Dmytriiev}, {Donath},
  {Doroshenko}, {Dyks}, {Egberts}, {Emery}, {Ernenwein}, {Eschbach}, {Feijen},
  {Fegan}, {Fiasson}, {Fontaine}, {Funk}, {F{\"u}{\ss}ling}, {Gabici},
  {Gallant}, {Gat{\'e}}, {Giavitto}, {Giunti}, {Glawion}, {Glicenstein},
  {Gottschall}, {Grondin}, {Hahn}, {Haupt}, {Heinzelmann}, {Henri}, {Hermann},
  {Hinton}, {Hofmann}, {Hoischen}, {Holch}, {Holler}, {Horns}, {Huber},
  {Iwasaki}, {Jamrozy}, {Jankowsky}, {Jankowsky}, {Jardin-Blicq},
  {Jung-Richardt}, {Kastendieck}, {Katarzy{\'n}ski}, {Katsuragawa}, {Katz},
  {Khangulyan}, {Kh{\'e}lifi}, {King}, {Klepser}, {Klu{\'z}niak}, {Komin},
  {Kosack}, {Kostunin}, {Kreter}, {Lamanna}, {Lemi{\`e}re}, {Lemoine-Goumard},
  {Lenain}, {Leser}, {Levy}, {Lohse}, {Lypova}, {Mackey}, {Majumdar},
  {Malyshev}, {Marandon}, {Marcowith}, {Mares}, {Mariaud}, {Mart{\'\i}-Devesa},
  {Marx}, {Maurin}, {Meintjes}, {Mitchell}, {Moderski}, {Mohamed}, {Mohrmann},
  {Moore}, {Moulin}, {Muller}, {Murach}, {Nakashima}, {de Naurois},
  {Ndiyavala}, {Niederwanger}, {Niemiec}, {Oakes}, {O'Brien}, {Odaka}, {Ohm},
  {de Ona Wilhelmi}, {Ostrowski}, {Oya}, {Panter}, {Parsons}, {Perennes},
  {Petrucci}, {Peyaud}, {Piel}, {Pita}, {Poireau}, {Priyana Noel}, {Prokhorov},
  {Prokoph}, {P{\"u}hlhofer}, {Punch}, {Quirrenbach}, {Raab}, {Rauth},
  {Reimer}, {Reimer}, {Remy}, {Renaud}, {Rieger}, {Rinchiuso}, {Romoli},
  {Rowell}, {Rudak}, {Ruiz-Velasco}, {Sahakian}, {Sailer}, {Saito}, {Sanchez},
  {Santangelo}, {Sasaki}, {Schlickeiser}, {Sch{\"u}ssler}, {Schulz}, {Schutte},
  {Schwanke}, {Schwemmer}, {Seglar-Arroyo}, {Senniappan}, {Seyffert}, {Shafi},
  {Shiningayamwe}, {Simoni}, {Sinha}, {Sol}, {Specovius}, {Spir-Jacob},
  {Stawarz}, {Steenkamp}, {Stegmann}, {Steppa}, {Takahashi}, {Tavernier},
  {Taylor}, {Terrier}, {Tiziani}, {Tluczykont}, {Trichard}, {Tsirou}, {Tsuji},
  {Tuffs}, {Uchiyama}, {van der Walt}, {van Eldik}, {van Rensburg}, {van
  Soelen}, {Vasileiadis}, {Veh}, {Venter}, {Vincent}, {Vink}, {V{\"o}lk},
  {Vuillaume}, {Wadiasingh}, {Wagner}, {White}, {Wierzcholska}, {Yang},
  {Yoneda}, {Zacharias}, {Zanin}, {Zdziarski}, {Zech}, {Ziegler}, {Zorn},
  {{\.Z}ywucka}, {de Palma}, {Axelsson}, \& {Roberts}}]{2019Natur.575..464A}
{Abdalla}, H., {Adam}, R., {Aharonian}, F., {et~al.} 2019, \nat, 575, 464

\bibitem[{{Aimuratov} {et~al.}(2023){Aimuratov}, {Becerra}, {Bianco},
  {Cherubini}, {Della Valle}, {Filippi}, {Li}, {Moradi}, {Rastegarnia},
  {Rueda}, {Ruffini}, {Sahakyan}, {Wang}, \& {Zhang}}]{2023ApJ...955...93A}
{Aimuratov}, Y., {Becerra}, L.~M., {Bianco}, C.~L., {et~al.} 2023, \apj, 955,
  93

\bibitem[{{Amati} {et~al.}(2013){Amati}, {Dichiara}, {Frontera}, \&
  {Guidorzi}}]{2013GCN.14503....1A}
{Amati}, L., {Dichiara}, S., {Frontera}, F., \& {Guidorzi}, C. 2013, GRB
  Coordinates Network, 14503, 1

\bibitem[{{Arimoto} {et~al.}(2022){Arimoto}, {Scotton}, {Longo}, \& {Fermi-LAT
  Collaboration}}]{2022GCN.31350....1A}
{Arimoto}, M., {Scotton}, L., {Longo}, F., \& {Fermi-LAT Collaboration}. 2022,
  GRB Coordinates Network, 31350, 1

\bibitem[{{Atteia}(2022)}]{2022GCN.31365....1A}
{Atteia}, J.~L. 2022, GRB Coordinates Network, 31365, 1

\bibitem[{{Becerra} {et~al.}(2016){Becerra}, {Bianco}, {Fryer}, {Rueda}, \&
  {Ruffini}}]{2016ApJ...833..107B}
{Becerra}, L., {Bianco}, C.~L., {Fryer}, C.~L., {Rueda}, J.~A., \& {Ruffini},
  R. 2016, \apj, 833, 107

\bibitem[{{Becerra} {et~al.}(2015){Becerra}, {Cipolletta}, {Fryer}, {Rueda}, \&
  {Ruffini}}]{2015ApJ...812..100B}
{Becerra}, L., {Cipolletta}, F., {Fryer}, C.~L., {Rueda}, J.~A., \& {Ruffini},
  R. 2015, \apj, 812, 100

\bibitem[{{Becerra} {et~al.}(2019){Becerra}, {Ellinger}, {Fryer}, {Rueda}, \&
  {Ruffini}}]{2019ApJ...871...14B}
{Becerra}, L., {Ellinger}, C.~L., {Fryer}, C.~L., {Rueda}, J.~A., \& {Ruffini},
  R. 2019, \apj, 871, 14

\bibitem[{{Becerra} {et~al.}(2018){Becerra}, {Guzzo}, {Rossi-Torres}, {Rueda},
  {Ruffini}, \& {Uribe}}]{2018ApJ...852..120B}
{Becerra}, L., {Guzzo}, M.~M., {Rossi-Torres}, F., {et~al.} 2018, \apj, 852,
  120

\bibitem[{{Bianco} {et~al.}(2023){Bianco}, {Mirtorabi}, {Moradi},
  {Rastegarnia}, {Rueda}, {Ruffini}, {Wang}, {Della Valle}, {Li}, \&
  {Zhang}}]{2023arXiv230605855B}
{Bianco}, C.~L., {Mirtorabi}, M.~T., {Moradi}, R., {et~al.} 2023, arXiv
  e-prints, arXiv:2306.05855

\bibitem[{{Bissaldi} {et~al.}(2022){Bissaldi}, {Omodei}, {Kerr}, \& {Fermi-LAT
  Team}}]{2022GCN.32637....1B}
{Bissaldi}, E., {Omodei}, N., {Kerr}, M., \& {Fermi-LAT Team}. 2022, GRB
  Coordinates Network, 32637, 1

\bibitem[{{Blanchard} {et~al.}(2023){Blanchard}, {Villar}, {Chornock}, {Sears},
  {LeBaron}, {Yadavalli}, {Laskar}, {Alexander}, {Margutti}, {Berger},
  {Barnes}, {Siegel}, {Metzger}, {Kasen}, {Cendes}, {Eftekhari}, \&
  {Leja}}]{2023GCN.33676....1B}
{Blanchard}, P.~K., {Villar}, V.~A., {Chornock}, R., {et~al.} 2023, GRB
  Coordinates Network, 33676, 1

\bibitem[{{Burns}(2016{\natexlab{a}})}]{2016GCN.19581....1B}
{Burns}, E. 2016{\natexlab{a}}, GRB Coordinates Network, 19581, 1

\bibitem[{{Burns}(2016{\natexlab{b}})}]{2016GCN.19587....1B}
{Burns}, E. 2016{\natexlab{b}}, GRB Coordinates Network, 19587, 1

\bibitem[{{Castro-Tirado} {et~al.}(2019){Castro-Tirado}, {Hu},
  {Fernandez-Garcia}, {Valeev}, {Sokolov}, {Guziy}, {Oates}, {Jeong}, {Pandey},
  {Carrasco}, \& {Reverte-Paya}}]{2019GCN.23708....1C}
{Castro-Tirado}, A.~J., {Hu}, Y., {Fernandez-Garcia}, E., {et~al.} 2019, GRB
  Coordinates Network, 23708, 1

\bibitem[{{Cherry} {et~al.}(2018){Cherry}, {Yoshida}, {Sakamoto}, {Sugita},
  {Kawakubo}, {Tezuka}, {Matsukawa}, {Onozawa}, {Ito}, {Morita}, {Sone},
  {Yamaoka}, {Nakahira}, {Takahashi}, {Asaoka}, {Ozawa}, {Torii}, {Shimizu},
  {Tamura}, {Ishizaki}, {Ricciarini}, {Penacchioni}, \&
  {Marrocchesi}}]{2018GCN.23042....1C}
{Cherry}, M.~L., {Yoshida}, A., {Sakamoto}, T., {et~al.} 2018, GRB Coordinates
  Network, 23042, 1

\bibitem[{{Cipolletta} {et~al.}(2017){Cipolletta}, {Cherubini}, {Filippi},
  {Rueda}, \& {Ruffini}}]{2017PhRvD..96b4046C}
{Cipolletta}, F., {Cherubini}, C., {Filippi}, S., {Rueda}, J.~A., \& {Ruffini},
  R. 2017, \prd, 96, 024046

\bibitem[{{Cucchiara} {et~al.}(2009{\natexlab{a}}){Cucchiara}, {Fox}, \&
  {Berger}}]{2009GCN..9213....1C}
{Cucchiara}, A., {Fox}, D.~B., \& {Berger}, E. 2009{\natexlab{a}}, GRB
  Coordinates Network, 9213, 1

\bibitem[{{Cucchiara} {et~al.}(2009{\natexlab{b}}){Cucchiara}, {Fox}, \&
  {Berger}}]{2009GCN..9209....1C}
{Cucchiara}, A., {Fox}, D.~B., \& {Berger}, E. 2009{\natexlab{b}}, GRB
  Coordinates Network, 9209, 1

\bibitem[{{Cucchiara} {et~al.}(2009{\natexlab{c}}){Cucchiara}, {Levan},
  {Tanvir}, {Fox}, \& {Berger}}]{2009GCN..9286....1C}
{Cucchiara}, A., {Levan}, A., {Tanvir}, N., {Fox}, D.~B., \& {Berger}, E.
  2009{\natexlab{c}}, GRB Coordinates Network, 9286, 1

\bibitem[{{Cucchiara} {et~al.}(2011){Cucchiara}, {Levan}, {Fox}, {Tanvir},
  {Ukwatta}, {Berger}, {Kr{\"u}hler}, {K{\"u}pc{\"u} Yolda{\c{s}}}, {Wu},
  {Toma}, {Greiner}, {Olivares}, {Rowlinson}, {Amati}, {Sakamoto}, {Roth},
  {Stephens}, {Fritz}, {Fynbo}, {Hjorth}, {Malesani}, {Jakobsson}, {Wiersema},
  {O'Brien}, {Soderberg}, {Foley}, {Fruchter}, {Rhoads}, {Rutledge}, {Schmidt},
  {Dopita}, {Podsiadlowski}, {Willingale}, {Wolf}, {Kulkarni}, \&
  {D'Avanzo}}]{2011ApJ...736....7C}
{Cucchiara}, A., {Levan}, A.~J., {Fox}, D.~B., {et~al.} 2011, \apj, 736, 7

\bibitem[{{de Ugarte Postigo} {et~al.}(2022){de Ugarte Postigo}, {Izzo},
  {Pugliese}, {Xu}, {Schneider}, {Fynbo}, {Tanvir}, {Malesani}, {Saccardi},
  {Kann}, {Wiersema}, {Gompertz}, {Thoene}, {Levan}, \& {Stargate
  Collaboration}}]{2022GCN.32648....1D}
{de Ugarte Postigo}, A., {Izzo}, L., {Pugliese}, G., {et~al.} 2022, GRB
  Coordinates Network, 32648, 1

\bibitem[{{D'Elia} {et~al.}(2019){D'Elia}, {D'Ai}, {Sbarufatti}, {Burrows},
  {Tohuvavohu}, {Page}, {Beardmore}, {Evans}, {Melandri}, \&
  {Gropp}}]{2019GCN.23706....1D}
{D'Elia}, V., {D'Ai}, A., {Sbarufatti}, B., {et~al.} 2019, GRB Coordinates
  Network, 23706, 1

\bibitem[{{D'Elia} {et~al.}(2016){D'Elia}, {Melandri}, \&
  {Malesani}}]{2016GCN.19601....1D}
{D'Elia}, V., {Melandri}, A., \& {Malesani}, D. 2016, GRB Coordinates Network,
  19601, 1

\bibitem[{{Dichiara} {et~al.}(2022){Dichiara}, {Gropp}, {Kennea}, {Kuin},
  {Lien}, {Marshall}, {Tohuvavohu}, {Williams}, \& {Neil Gehrels Swift
  Observatory Team}}]{2022GCN.32632....1D}
{Dichiara}, S., {Gropp}, J.~D., {Kennea}, J.~A., {et~al.} 2022, GRB Coordinates
  Network, 32632, 1

\bibitem[{{Dirirsa} {et~al.}(2016){Dirirsa}, {Vianello}, {Racusin}, \&
  {Axelsson}}]{2016GCN.19586....1D}
{Dirirsa}, F., {Vianello}, G., {Racusin}, J., \& {Axelsson}, M. 2016, GRB
  Coordinates Network, 19586, 1

\bibitem[{{Fryer} {et~al.}(2015){Fryer}, {Oliveira}, {Rueda}, \&
  {Ruffini}}]{2015PhRvL.115w1102F}
{Fryer}, C.~L., {Oliveira}, F.~G., {Rueda}, J.~A., \& {Ruffini}, R. 2015,
  Physical Review Letters, 115, 231102

\bibitem[{{Fryer} {et~al.}(2014){Fryer}, {Rueda}, \&
  {Ruffini}}]{2014ApJ...793L..36F}
{Fryer}, C.~L., {Rueda}, J.~A., \& {Ruffini}, R. 2014, \apjl, 793, L36

\bibitem[{{Fu} {et~al.}(2022){Fu}, {Zhu}, {Xu}, {Liu}, \&
  {Jiang}}]{2022GCN.31353....1F}
{Fu}, S.~Y., {Zhu}, Z.~P., {Xu}, D., {Liu}, X., \& {Jiang}, S.~Q. 2022, GRB
  Coordinates Network, 31353, 1

\bibitem[{{Fynbo} {et~al.}(2022){Fynbo}, {de Ugarte Postigo}, {Xu}, {Malesani},
  {Milvang-Jensen}, \& {Viuho}}]{2022GCN.31359....1F}
{Fynbo}, J.~P.~U., {de Ugarte Postigo}, A., {Xu}, D., {et~al.} 2022, GRB
  Coordinates Network, 31359, 1

\bibitem[{{Galama} {et~al.}(1998){Galama}, {Vreeswijk}, {van Paradijs},
  {Kouveliotou}, {Augusteijn}, {B{\"o}hnhardt}, {Brewer}, {Doublier},
  {Gonzalez}, {Leibundgut}, {Lidman}, {Hainaut}, {Patat}, {Heise}, {in't Zand},
  {Hurley}, {Groot}, {Strom}, {Mazzali}, {Iwamoto}, {Nomoto}, {Umeda},
  {Nakamura}, {Young}, {Suzuki}, {Shigeyama}, {Koshut}, {Kippen}, {Robinson},
  {de Wildt}, {Wijers}, {Tanvir}, {Greiner}, {Pian}, {Palazzi}, {Frontera},
  {Masetti}, {Nicastro}, {Feroci}, {Costa}, {Piro}, {Peterson}, {Tinney},
  {Boyle}, {Cannon}, {Stathakis}, {Sadler}, {Begam}, \&
  {Ianna}}]{1998Natur.395..670G}
{Galama}, T.~J., {Vreeswijk}, P.~M., {van Paradijs}, J., {et~al.} 1998, \nat,
  395, 670

\bibitem[{{Gropp} {et~al.}(2019){Gropp}, {Kennea}, {Klingler}, {Krimm},
  {Laporte}, {Lien}, {Moss}, {Palmer}, {Sbarufatti}, \&
  {Siegel}}]{2019GCN.23688....1G}
{Gropp}, J.~D., {Kennea}, J.~A., {Klingler}, N.~J., {et~al.} 2019, GRB
  Coordinates Network, 23688, 1

\bibitem[{{Hamburg} {et~al.}(2019){Hamburg}, {Veres}, {Meegan}, {Burns},
  {Connaughton}, {Goldstein}, {Kocevski}, \& {Roberts}}]{2019GCN.23707....1H}
{Hamburg}, R., {Veres}, P., {Meegan}, C., {et~al.} 2019, GRB Coordinates
  Network, 23707, 1

\bibitem[{{Huang} {et~al.}(2022){Huang}, {Hu}, {Chen}, {Zha}, {Liu}, {Yao},
  {Cao}, \& {Experiment}}]{2022GCN.32677....1H}
{Huang}, Y., {Hu}, S., {Chen}, S., {et~al.} 2022, GRB Coordinates Network,
  32677, 1

\bibitem[{{Izzo} {et~al.}(2013){Izzo}, {Bianco}, {Muccino}, {Penacchioni}, \&
  {Ruffini}}]{2013IJMPS..23..202I}
{Izzo}, L., {Bianco}, C.~L., {Muccino}, M., {Penacchioni}, A.~V., \& {Ruffini},
  R. 2013, International Journal of Modern Physics Conference Series, 23, 202

\bibitem[{{Izzo} {et~al.}(2012{\natexlab{a}}){Izzo}, {Rueda}, \&
  {Ruffini}}]{2012A&A...548L...5I}
{Izzo}, L., {Rueda}, J.~A., \& {Ruffini}, R. 2012{\natexlab{a}}, \aap, 548, L5

\bibitem[{{Izzo} {et~al.}(2012{\natexlab{b}}){Izzo}, {Ruffini}, {Penacchioni},
  {Bianco}, {Caito}, {Chakrabarti}, {Rueda}, {Nandi}, \&
  {Patricelli}}]{2012A&A...543A..10I}
{Izzo}, L., {Ruffini}, R., {Penacchioni}, A.~V., {et~al.} 2012{\natexlab{b}},
  \aap, 543, A10

\bibitem[{{Kennea} {et~al.}(2016){Kennea}, {Roegiers}, {Osborne}, {Page},
  {Melandri}, {D'Avanzo}, {D'Elia}, {Burrows}, {McCauley}, {Pagani}, \&
  {Evans}}]{2016GCN.19408....1K}
{Kennea}, J.~A., {Roegiers}, T.~G.~R., {Osborne}, J.~P., {et~al.} 2016, GRB
  Coordinates Network, 19408, 1

\bibitem[{{Kocevski} {et~al.}(2019){Kocevski}, {Omodei}, {Axelsson}, {Burns},
  {Vianello}, {Bissaldi}, \& {Longo}}]{2019GCN.23709....1K}
{Kocevski}, D., {Omodei}, N., {Axelsson}, M., {et~al.} 2019, GRB Coordinates
  Network, 23709, 1

\bibitem[{{Krimm} {et~al.}(2009){Krimm}, {Beardmore}, {Evans}, {Godet},
  {Gronwall}, {Guidorzi}, {O'Brien}, {Page}, {Palmer}, {Perri}, {Sbarufatti},
  {Schady}, {Stratta}, {Tagliaferri}, \& {Ukwatta}}]{2009GCN..9198....1K}
{Krimm}, H.~A., {Beardmore}, A.~P., {Evans}, P.~A., {et~al.} 2009, GRB
  Coordinates Network, 9198, 1

\bibitem[{{Lesage} {et~al.}(2022{\natexlab{a}}){Lesage}, {Meegan}, \& {Fermi
  Gamma-ray Burst Monitor Team}}]{2022GCN.31360....1L}
{Lesage}, S., {Meegan}, C., \& {Fermi Gamma-ray Burst Monitor Team}.
  2022{\natexlab{a}}, GRB Coordinates Network, 31360, 1

\bibitem[{{Lesage} {et~al.}(2022{\natexlab{b}}){Lesage}, {Veres}, {Roberts},
  {Burns}, {Bissaldi}, \& {Fermi GBM Team}}]{2022GCN.32642....1L}
{Lesage}, S., {Veres}, P., {Roberts}, O.~J., {et~al.} 2022{\natexlab{b}}, GRB
  Coordinates Network, 32642, 1

\bibitem[{{Levan} {et~al.}(2009){Levan}, {Warwick}, {Cucchiara}, {Tanvir},
  {Fox}, \& {Berger}}]{2009GCN..9306....1L}
{Levan}, A., {Warwick}, U., {Cucchiara}, A., {et~al.} 2009, GRB Coordinates
  Network, 9306, 1

\bibitem[{{Levan} {et~al.}(2013){Levan}, {Cenko}, {Perley}, \&
  {Tanvir}}]{2013GCN.14455....1L}
{Levan}, A.~J., {Cenko}, S.~B., {Perley}, D.~A., \& {Tanvir}, N.~R. 2013, GRB
  Coordinates Network, 14455, 1

\bibitem[{{Li} {et~al.}(2023){Li}, {Rueda}, {Moradi}, {Wang}, {Xue}, \&
  {Ruffini}}]{2023ApJ...945...10L}
{Li}, L., {Rueda}, J.~A., {Moradi}, R., {et~al.} 2023, \apj, 945, 10

\bibitem[{{Longo} {et~al.}(2016{\natexlab{a}}){Longo}, {Bissaldi}, {Bregeon},
  {McEnery}, {Ohno}, \& {Zhu}}]{2016GCN.19403....1L}
{Longo}, F., {Bissaldi}, E., {Bregeon}, J., {et~al.} 2016{\natexlab{a}}, GRB
  Coordinates Network, 19403, 1

\bibitem[{{Longo} {et~al.}(2016{\natexlab{b}}){Longo}, {Bissaldi}, {Vianello},
  {Moretti}, {Omodei}, {Bregeon}, {Dirirsa}, {Yassine}, {Kocevski}, {Racusin},
  {McEnery}, \& {Ohno}}]{2016GCN.19413....1L}
{Longo}, F., {Bissaldi}, E., {Vianello}, G., {et~al.} 2016{\natexlab{b}}, GRB
  Coordinates Network, 19413, 1

\bibitem[{{Longo} {et~al.}(2009){Longo}, {Moretti}, {Barbiellini}, {Vallazza},
  {Trifoglio}, {Bulgarelli}, {Gianotti}, {Fuschino}, {Marisaldi}, {Labanti},
  {Galli}, {Di Cocco}, {Cutini}, {Pittori}, {Tavani}, {Striani}, {Pucella},
  {D'Ammando}, {Vittorini}, {Argan}, {Trois}, {Piano}, {Sabatini}, {Del Monte},
  {Feroci}, {Evangelista}, {Donnarumma}, {Pacciani}, {Soffitta}, {Costa},
  {Lazzarotto}, {Lapshov}, {Rapisarda}, {Giuliani}, {Chen}, {Mereghetti},
  {Perotti}, {Caraveo}, {Pellizzoni}, {Pilia}, {Vercellone}, {Picozza},
  {Morselli}, {Prest}, {Lipari}, {Zanello}, {Rappoldi}, {Cattaneo}, {Giommi},
  {Santolamazza}, {Verrecchia}, \& {Salotti}}]{2009GCN..9524....1L}
{Longo}, F., {Moretti}, E., {Barbiellini}, G., {et~al.} 2009, GRB Coordinates
  Network, 9524, 1

\bibitem[{{Maselli} {et~al.}(2013){Maselli}, {Beardmore}, {Lien}, {Mangano},
  {Mountford}, {Page}, {Palmer}, \& {Siegel}}]{2013GCN.14448....1M}
{Maselli}, A., {Beardmore}, A.~P., {Lien}, A.~Y., {et~al.} 2013, GRB
  Coordinates Network, 14448, 1

\bibitem[{{McBreen}(2009)}]{2009GCN..9535....1M}
{McBreen}, S. 2009, GRB Coordinates Network, 9535, 1

\bibitem[{{Melandri} {et~al.}(2016){Melandri}, {D'Avanzo}, {D'Elia}, {Burrows},
  {Roegiers}, {McCauley}, {Gibson}, {Osborne}, \&
  {Evans}}]{2016GCN.19585....1M}
{Melandri}, A., {D'Avanzo}, P., {D'Elia}, V., {et~al.} 2016, GRB Coordinates
  Network, 19585, 1

\bibitem[{{Melandri} {et~al.}(2019){Melandri}, {Izzo}, {D'Avanzo}, {Malesani},
  {Della Valle}, {Pian}, {Tanvir}, {Ragosta}, {Olivares}, {Carini}, {Palazzi},
  {Piranomonte}, {Jonker}, {Rossi}, {Kann}, {Hartmann}, {Inserra}, {Kankare},
  {Maguire}, {Smartt}, {Yaron}, {Young}, \& {Manulis}}]{2019GCN.23983....1M}
{Melandri}, A., {Izzo}, L., {D'Avanzo}, P., {et~al.} 2019, GRB Coordinates
  Network, 23983, 1

\bibitem[{{Mirzoyan} {et~al.}(2019){Mirzoyan}, {Noda}, {Moretti}, {Berti},
  {Nigro}, {Hoang}, {Micanovic}, {Takahashi}, {Chai}, {Moralejo}, \& {MAGIC
  Collaboration}}]{2019GCN.23701....1M}
{Mirzoyan}, R., {Noda}, K., {Moretti}, E., {et~al.} 2019, GRB Coordinates
  Network, 23701, 1

\bibitem[{{Moradi} {et~al.}(2021){Moradi}, {Rueda}, {Ruffini}, {Li}, {Bianco},
  {Campion}, {Cherubini}, {Filippi}, {Wang}, \& {Xue}}]{2021PhRvD.104f3043M}
{Moradi}, R., {Rueda}, J.~A., {Ruffini}, R., {et~al.} 2021, \prd, 104, 063043

\bibitem[{{Pozanenko} {et~al.}(2013){Pozanenko}, {Minaev}, \&
  {Volnova}}]{2013GCN.14484....1P}
{Pozanenko}, A., {Minaev}, P., \& {Volnova}, A. 2013, GRB Coordinates Network,
  14484, 1

\bibitem[{{Riechers} {et~al.}(2009){Riechers}, {Walter}, {Bertoldi}, {Carilli},
  {Cox}, {Kramer}, \& {Riquelme}}]{2009GCN..9322....1R}
{Riechers}, D.~A., {Walter}, F., {Bertoldi}, F., {et~al.} 2009, GRB Coordinates
  Network, 9322, 1

\bibitem[{{Roberts} {et~al.}(2016){Roberts}, {Fitzpatrick}, \&
  {Veres}}]{2016GCN.19411....1R}
{Roberts}, O.~J., {Fitzpatrick}, G., \& {Veres}, P. 2016, GRB Coordinates
  Network, 19411, 1

\bibitem[{{Roberts} \& {Meegan}(2018)}]{2018GCN.22981....1R}
{Roberts}, O.~J. \& {Meegan}, C. 2018, GRB Coordinates Network, 22981, 1

\bibitem[{{Rowlinson} \& {Ukwatta}(2009)}]{2009GCN..9298....1R}
{Rowlinson}, A. \& {Ukwatta}, T.~N. 2009, GRB Coordinates Network, 9298, 1

\bibitem[{{Rueda} {et~al.}(2022){Rueda}, {Li}, {Moradi}, {Ruffini}, {Sahakyan},
  \& {Wang}}]{2022ApJ...939...62R}
{Rueda}, J.~A., {Li}, L., {Moradi}, R., {et~al.} 2022, \apj, 939, 62

\bibitem[{{Rueda} \& {Ruffini}(2012)}]{2012ApJ...758L...7R}
{Rueda}, J.~A. \& {Ruffini}, R. 2012, \apjl, 758, L7

\bibitem[{{Rueda} {et~al.}(2020){Rueda}, {Ruffini}, {Karlica}, {Moradi}, \&
  {Wang}}]{2020ApJ...893..148R}
{Rueda}, J.~A., {Ruffini}, R., {Karlica}, M., {Moradi}, R., \& {Wang}, Y. 2020,
  \apj, 893, 148

\bibitem[{{Rueda} {et~al.}(2021){Rueda}, {Ruffini}, {Moradi}, \&
  {Wang}}]{2021IJMPD..3030007R}
{Rueda}, J.~A., {Ruffini}, R., {Moradi}, R., \& {Wang}, Y. 2021, International
  Journal of Modern Physics D, 30, 2130007

\bibitem[{{Ruffini}(2022)}]{2022A&AT...33..191R}
{Ruffini}, R. 2022, Astronomical and Astrophysical Transactions, 33, 191

\bibitem[{{Ruffini} {et~al.}(2014){Ruffini}, {Izzo}, {Muccino}, {Pisani},
  {Rueda}, {Wang}, {Barbarino}, {Bianco}, {Enderli}, \&
  {Kovacevic}}]{2014A&A...569A..39R}
{Ruffini}, R., {Izzo}, L., {Muccino}, M., {et~al.} 2014, \aap, 569, A39

\bibitem[{{Ruffini} {et~al.}(2019){Ruffini}, {Melon Fuksman}, \&
  {Vereshchagin}}]{2019ApJ...883..191R}
{Ruffini}, R., {Melon Fuksman}, J.~D., \& {Vereshchagin}, G.~V. 2019, \apj,
  883, 191

\bibitem[{{Ruffini} {et~al.}(2021){Ruffini}, {Moradi}, {Rueda}, {Li},
  {Sahakyan}, {Chen}, {Wang}, {Aimuratov}, {Becerra}, {Bianco}, {Cherubini},
  {Filippi}, {Karlica}, {Mathews}, {Muccino}, {Pisani}, \&
  {Xue}}]{2021MNRAS.504.5301R}
{Ruffini}, R., {Moradi}, R., {Rueda}, J.~A., {et~al.} 2021, \mnras, 504, 5301

\bibitem[{{Ruffini} {et~al.}(2015){Ruffini}, {Wang}, {Enderli}, {Muccino},
  {Kovacevic}, {Bianco}, {Penacchioni}, {Pisani}, \&
  {Rueda}}]{2015ApJ...798...10R}
{Ruffini}, R., {Wang}, Y., {Enderli}, M., {et~al.} 2015, \apj, 798, 10

\bibitem[{{Sasada} {et~al.}(2018){Sasada}, {Nakaoka}, {Kawabata}, {Uchida},
  {Yamazaki}, \& {Kawabata}}]{2018GCN.22977....1S}
{Sasada}, M., {Nakaoka}, T., {Kawabata}, M., {et~al.} 2018, GRB Coordinates
  Network, 22977, 1

\bibitem[{{Schady} {et~al.}(2009){Schady}, {Baumgartner}, {Beardmore},
  {Campana}, {Curran}, {Guidorzi}, {Kennea}, {Mao}, {Margutti}, {Osborne},
  {Page}, {Romano}, {Siegel}, {Stratta}, \& {Ukwatta}}]{2009GCN..9512....1S}
{Schady}, P., {Baumgartner}, W.~H., {Beardmore}, A.~P., {et~al.} 2009, GRB
  Coordinates Network, 9512, 1

\bibitem[{{Selsing} {et~al.}(2019){Selsing}, {Fynbo}, {Heintz}, \&
  {Watson}}]{2019GCN.23695....1S}
{Selsing}, J., {Fynbo}, J.~P.~U., {Heintz}, K.~E., \& {Watson}, D. 2019, GRB
  Coordinates Network, 23695, 1

\bibitem[{{Siegel} {et~al.}(2018){Siegel}, {Burrows}, {Deich}, {Gropp},
  {Kennea}, {Laporte}, {Lien}, {Moss}, {Page}, {Palmer}, {Sbarufatti}, \&
  {Tohuvavohu}}]{2018GCN.22973....1S}
{Siegel}, M.~H., {Burrows}, D.~N., {Deich}, A., {et~al.} 2018, GRB Coordinates
  Network, 22973, 1

\bibitem[{{Tanvir} {et~al.}(2009){Tanvir}, {Levan}, {Wiersema}, {Fynbo},
  {Hjorth}, \& {Jakobsson}}]{2009GCN..9219....1T}
{Tanvir}, N., {Levan}, A., {Wiersema}, K., {et~al.} 2009, GRB Coordinates
  Network, 9219, 1

\bibitem[{{Tohuvavohu} {et~al.}(2022){Tohuvavohu}, {Gropp}, {Kennea}, {Lien},
  {Palmer}, {Parsotan}, {Sbarufatti}, {Siegel}, \& {Neil Gehrels Swift
  Observatory Team}}]{2022GCN.31347....1T}
{Tohuvavohu}, A., {Gropp}, J.~D., {Kennea}, J.~A., {et~al.} 2022, GRB
  Coordinates Network, 31347, 1

\bibitem[{{Troja} {et~al.}(2017){Troja}, {Lipunov}, {Mundell}, {Butler},
  {Watson}, {Kobayashi}, {Cenko}, {Marshall}, {Ricci}, {Fruchter}, {Wieringa},
  {Gorbovskoy}, {Kornilov}, {Kutyrev}, {Lee}, {Toy}, {Tyurina}, {Budnev},
  {Buckley}, {Gonz{\'a}lez}, {Gress}, {Horesh}, {Panasyuk}, {Prochaska},
  {Ramirez-Ruiz}, {Rebolo Lopez}, {Richer}, {Roman-Zuniga}, {Serra-Ricart},
  {Yurkov}, \& {Gehrels}}]{2017Natur.547..425T}
{Troja}, E., {Lipunov}, V.~M., {Mundell}, C.~G., {et~al.} 2017, \nat, 547, 425

\bibitem[{{Tsvetkova} {et~al.}(2022){Tsvetkova}, {Frederiks}, {Lysenko},
  {Ridnaia}, {Svinkin}, {Ulanov}, {Cline}, \& {Konus-Wind
  Team}}]{2022GCN.31433....1T}
{Tsvetkova}, A., {Frederiks}, D., {Lysenko}, A., {et~al.} 2022, GRB Coordinates
  Network, 31433, 1

\bibitem[{{Tsvetkova} \& {Konus-Wind Team}(2022)}]{2022GCN.31436....1T}
{Tsvetkova}, A. \& {Konus-Wind Team}. 2022, GRB Coordinates Network, 31436, 1

\bibitem[{{Ukwatta} {et~al.}(2009){Ukwatta}, {Barthelmy}, {Evans}, {Gehrels},
  {Markwardt}, {Page}, {Palmer}, {Rowlinson}, {Siegel}, {Stamatikos}, \&
  {Vetere}}]{2009GCN..9281....1U}
{Ukwatta}, T.~N., {Barthelmy}, S.~D., {Evans}, P.~A., {et~al.} 2009, GRB
  Coordinates Network, 9281, 1

\bibitem[{{Ursi} {et~al.}(2022){Ursi}, {Menegoni}, {Longo}, {Pittori},
  {Verrecchia}, {Tempesta}, {Tavani}, {Argan}, {Cardillo}, {Casentini},
  {Evangelista}, {Foffano}, {Piano}, {Lucarelli}, {Bulgarelli}, {di Piano},
  {Fioretti}, {Fuschino}, {Parmiggiani}, {Marisaldi}, {Pilia}, {Trois},
  {Donnarumma}, {Giuliani}, \& {Agile Team}}]{2022GCN.31354....1U}
{Ursi}, A., {Menegoni}, E., {Longo}, F., {et~al.} 2022, GRB Coordinates
  Network, 31354, 1

\bibitem[{{von Kienlin}(2009{\natexlab{a}})}]{2009GCN..9229....1V}
{von Kienlin}, A. 2009{\natexlab{a}}, GRB Coordinates Network, 9229, 1

\bibitem[{{von Kienlin}(2009{\natexlab{b}})}]{2009GCN..9251....1V}
{von Kienlin}, A. 2009{\natexlab{b}}, GRB Coordinates Network, 9251, 1

\bibitem[{{von Kienlin}(2013)}]{2013GCN.14473....1V}
{von Kienlin}, A. 2013, GRB Coordinates Network, 14473, 1

\bibitem[{{Vreeswijk} {et~al.}(2018){Vreeswijk}, {Kann}, {Heintz}, {de Ugarte
  Postigo}, {Milvang-Jensen}, {Malesani}, {Covino}, {Levan}, \&
  {Pugliese}}]{2018GCN.22996....1V}
{Vreeswijk}, P.~M., {Kann}, D.~A., {Heintz}, K.~E., {et~al.} 2018, GRB
  Coordinates Network, 22996, 1

\bibitem[{{Wang} {et~al.}(2019{\natexlab{a}}){Wang}, {Liu}, {Zhang}, {Xi}, \&
  {Zhang}}]{2019ApJ...884..117W}
{Wang}, X.-Y., {Liu}, R.-Y., {Zhang}, H.-M., {Xi}, S.-Q., \& {Zhang}, B.
  2019{\natexlab{a}}, \apj, 884, 117

\bibitem[{{Wang} {et~al.}(2019{\natexlab{b}}){Wang}, {Rueda}, {Ruffini},
  {Becerra}, {Bianco}, {Becerra}, {Li}, \& {Karlica}}]{2019ApJ...874...39W}
{Wang}, Y., {Rueda}, J.~A., {Ruffini}, R., {et~al.} 2019{\natexlab{b}}, \apj,
  874, 39

\bibitem[{{Wang} {et~al.}(2022){Wang}, {Rueda}, {Ruffini}, {Moradi}, {Li},
  {Aimuratov}, {Rastegarnia}, {Eslamzadeh}, {Sahakyan}, \&
  {Zheng}}]{2022ApJ...936..190W}
{Wang}, Y., {Rueda}, J.~A., {Ruffini}, R., {et~al.} 2022, \apj, 936, 190

\bibitem[{{Woosley} \& {Bloom}(2006)}]{2006ARA&A..44..507W}
{Woosley}, S.~E. \& {Bloom}, J.~S. 2006, \araa, 44, 507

\bibitem[{{Xu} {et~al.}(2016){Xu}, {Malesani}, {Fynbo}, {Tanvir}, {Levan}, \&
  {Perley}}]{2016GCN.19600....1X}
{Xu}, D., {Malesani}, D., {Fynbo}, J.~P.~U., {et~al.} 2016, GRB Coordinates
  Network, 19600, 1

\bibitem[{{Zhu} {et~al.}(2013){Zhu}, {Racusin}, {Kocevski}, {McEnery}, {Longo},
  {Chiang}, \& {Vianello}}]{2013GCN.14471....1Z}
{Zhu}, S., {Racusin}, J., {Kocevski}, D., {et~al.} 2013, GRB Coordinates
  Network, 14471, 1

\end{thebibliography}

\end{CJK*}
\end{document}